\documentclass[aps, prd, superscriptaddress, nofootinbib, preprintnumbers]{revtex4}
\usepackage{amsmath}
\usepackage{graphicx}
\usepackage{color}

\newcommand{\lesssim}{\mathrel{\mathpalette\vereq<}}
\newcommand{\gtrsim}{\mathrel{\mathpalette\vereq>}}

\newcommand{\chushi}[1]{}
\newcommand{\xbar}[1]{#1 \hspace{-5.5pt}/}
\newcommand{\beq}{\begin{eqnarray}}
\newcommand{\eeq}{\end{eqnarray}}
\begin{document}

\title{Walking on the Ladder: 125 GeV Technidilaton, or Conformal Higgs\\
\hspace{1cm}
{\it -Dedicated to the late Professor Yoichiro Nambu-}}      
\author{Shinya Matsuzaki}\thanks{\tt synya@hken.phys.nagoya-u.ac.jp}
      \affiliation{ Institute for Advanced Research, Nagoya University, Nagoya 464-8602, Japan.}
      \affiliation{ Department of Physics, Nagoya University, Nagoya 464-8602, Japan.}   
\author{{Koichi Yamawaki}} \thanks{
      {\tt yamawaki@kmi.nagoya-u.ac.jp}}
      \affiliation{ Kobayashi-Maskawa Institute for the Origin of Particles and the Universe (KMI) \\ 
 Nagoya University, Nagoya 464-8602, Japan.}

\begin{abstract} 
The walking technicolor based on the ladder Schwinger-Dyson gap equation is studied, with the scale-invariant coupling being an idealization
of the Caswell-Banks-Zaks infrared  
fixed point in the ``anti-Veneziano limit'', such that $N_C \rightarrow \infty$ with $N_C \cdot \alpha(\mu^2)=$ fixed {\it and} $N_F/N_C=$ fixed ($\gg 1$), of the $SU(N_C)$ gauge theory with massless $N_F$ flavors {\it near criticality}.
We show that the 125 GeV Higgs can be naturally identified with the technidilaton (TD)   
predicted in the walking technicolor, a pseudo Nambu-Goldstone (NG) boson of the spontaneous symmetry breaking of the approximate scale symmetry.
Ladder calculations 
yield the TD mass $M_\phi$ from the trace anomaly as 
 $M_\phi^2 F_\phi^2=  
-4 \langle \theta_\mu^\mu \rangle 
=-  \frac{\beta(\alpha (\mu^2))}{\alpha(\mu^2)}
\, \langle G_{\lambda \nu}^2(\mu^2)\rangle 
\simeq N_C N_F\frac{16 }{\pi^4} m_F^4$, 
independently of the renormalization point $\mu$,
where $m_F$ is the
dynamical mass of the technifermion, and $F_\phi={\cal O} (\sqrt{N_F N_C}\, m_F)$ the TD decay constant.
It reads 
$M_\phi^2\simeq (\frac{v_{\rm EW}}{2} \cdot \frac{5 v_{\rm EW}}{F_\phi})^2 \cdot [\frac{8}{N_F}\frac{4}{N_C}]$, ($v_{\rm EW}=246$ GeV), 
which 
implies  
$F_\phi\simeq 5 \,v_{\rm EW} $ for  
$M_\phi \simeq
125\, {\rm GeV}\simeq \frac{1}{2} v_{\rm EW}$ in the 
one-family model ($N_C=4, N_F=8$), 
in good agreement with the current LHC Higgs data. 
The result reflects a generic scaling $ M_\phi^2/v_{\rm EW}^2\sim M_\phi^2/F_\phi^2  \sim m_F^2 /F_\phi^2   
 \sim 1/(N_F N_C) \rightarrow 0 $ as a vanishing trace anomaly, namely  the TD has a mass vanishing 
 in the anti-Veneziano limit, similarly to $\eta^\prime$ meson as a pseudo-NG boson  of the ordinary QCD with vanishing $U(1)_A$ anomaly in the Veneziano limit ($N_F/N_C \ll 1$). 
\end{abstract}
\maketitle

\section{Introduction}
The Higgs boson with mass nearly 125 GeV has been found at LHC. Still there remains a mystery about the electroweak symmetry breaking, or the
dynamical origin of the Higgs, which would be understood by physics beyond the Standard Model (SM).
An attractive idea for the origin of mass  beyond the SM is the {\it dynamical symmetry breaking traced back to Nambu} \cite{Nambu:1961tp}, the birthplace of all the variants of the concept of  the spontaneous symmetry breaking (SSB).  Together with the
Nambu's dynamical symmetry breaking producing composite Nambu-Goldstone (NG) bosons, we should mention the composite approach 
by Shoichi Sakata, who proposed the Sakata model \cite{Sakata:1956hs},
a composite model for the hadrons, which paved a way to the quark model and
eventually to the Standard Model.  We are inspired by his never-ending enthusiasm seeking the deeper level of matter.

In contrast to the SM Higgs boson which has a mass given ad hoc without explanation,
the origin of mass $M$ in the Nambu's  dynamical symmetry breaking resides in the {\it criticality} 
with  the {\it nonzero critical coupling}  $g_{\rm cr}\ne 0$, such that  the value $M$ in the Nambu-Jona-Lasinio model (NJL) is generated from nothing as 
$M \sim \Lambda \, (1/g_{\rm cr} -1/g)^{1/2}$ for {\it strong coupling} $g>g_{\rm cr}$, where $g$ and $\Lambda$ are the dimensionless coupling and  an intrinsic scale carried by the four-fermion coupling, respectively, as $G\sim g /\Lambda^2$. We all know now that the Nambu's great idea is essentially realized in the reality, the QCD, where the strong gauge coupling in the infrared scale $\Lambda_{\rm QCD}$ gives rise to the hadron mass on that scale.
In the case of Higgs, 
the top quark condensate model (Top-mode standard model) \cite{Miransky:1988xi,Nambu:1989jt,Bardeen:1989ds}  is  a straightforward application of the NJL dynamics, with only the top coupling set to be above the criticality and others below it \cite{Miransky:1988xi,Bardeen:1989ds}
\footnote{Note that the NJL dynamics with $g_{\rm cr}\ne 0$  is {\it in sharp contrast to the weakly-coupled BCS theory which has $g_{\rm cr}=0$} due to the ``dimensional reduction'' by the presence of the Fermi surface. The NJL criticality $g^{top}>g_{\rm cr}
>g^{others} $ is an essence of the top quark condensate model of Ref.\cite{Miransky:1988xi,Bardeen:1989ds} to ensure that only the top quark gets condensed to produce only three NG bosons to be absorbed into the weak bosons.
}:
The origin of the intrinsic scale $\Lambda$ could be 
the quantum mechanical origin
as the trace anomaly like $\Lambda_{\rm QCD}$ in the classically scale-invariant theory e.g, gauge theory,  or the explicit one such as the given four-fermion coupling in the NJL  with $\Lambda$ to be regarded as the Landau pole or the compositeness scale, or the intrinsic scale of certain underlying gauge theory at deeper level.
Note that   the existence of the scale $\Lambda$ does not necessarily implies the existence of the mass $M$:  The weak coupling  $g  <g_{\rm cr}$ does not produce the mass $M$, while
 the strong coupling does  create it, 
 picking up  the intrinsic scale $\Lambda$ a la dimensional transmutation, generically in the form $M\sim \Lambda f(g(\Lambda))=\mu f(g(\mu))$, with $f( g(\mu)) \rightarrow 0$ as $g(\mu) \rightarrow 
g_{\rm cr}$.

 One of the candidates for such a dynamical symmetry breaking theory beyond the SM is the walking technicolor (WTC)  
 \cite{Yamawaki:1985zg,Bando:1986bg,Yamawaki:1996vr}, having a {\it large anomalous dimension} $\gamma_m=1$
 \footnote{
 It was further shown~\cite{Miransky:1988gk} that the NJL model coupled to the walking gauge theories (``gauged NJL model'') has an even larger anomalous dimension $1<\gamma_m <2$, along the critical line \cite{Kondo:1988qd,Yamawaki:1988na} with strong four-fermion coupling. It was 
 further shown \cite{Kondo:1991yk} that such a theory is renormalizable without Landau pole, i.e.,  nontriviality theory having a finite nontrivial (non-Gaussian) ultraviolet fixed point, in contrast the pure NJL model which is a trivial theory. See later discussions.
 }
  to solve the 
 Flavor-Changing Neutral Currents (FCNC) problem \footnote{Solving FCNC problem by a large anomalous dimension was proposed earlier \cite{Holdom:1981rm},  based on a pure assumption of the existence of a gauge theory having the nontrivial UV fixed point
at large coupling, where a large anomalous dimension $\gamma_m >1$ was postulated.  See also \cite{Georgi:1981xw,Yamawaki:1982tg}. 
} of the original technicolor (TC) ~\cite{Weinberg:1975gm,Farhi:1980xs} and  a {\it technidilaton (TD)}, a pseudo NG boson of the approximate scale symmetry, as a composite Higgs ("Conformal Higgs" \cite{Yamawaki:2010ms})
\footnote{
Similar works on the FCNC solution \cite{Holdom:1984sk} were done without notion of the anomalous dimension, the scale invariance, and the technidilaton.
 }. 
It has recently been shown that the TD properties are consistent with the current data of LHC for the 125 GeV Higgs and hence TD can be identified with the 125 GeV
Higgs at LHC
\cite{Matsuzaki:2012gd,Matsuzaki:2012vc,Matsuzaki:2012mk,Matsuzaki:2012xx}.

The above results  \cite{Yamawaki:1985zg,Bando:1986bg} were originally obtained based on the  ladder Schwinger-Dyson (SD) gap equation for the fermion propagator, with the nonrunning gauge coupling constant   $\alpha(\mu^2) = \alpha$ for
$0<\mu^2<\Lambda^2$ as an input coupling: The theory is scale-invariant  (infrared conformality) in the infrared region below 
the cutoff  $\Lambda$ 
to be identified with the intrinsic scale of the theory, $\Lambda_{\rm TC}$, 
like $\Lambda_{\rm QCD}$ of ordinary QCD, which is quantum mechanically induced by the regularization as the trace anomaly. When the coupling exceeds the critical coupling $\alpha>\alpha_{\rm cr} \ne 0$, the chiral and scale symmetries simultaneously get SSB due to the generation of the technifermion dynamical mass $m_F$ in such a way that $m_F$ is much smaller than the intrinsic scale  $m_F \sim  \Lambda_{\rm TC} \,f(\alpha) \ll \Lambda_{\rm TC}$ by the Miransky scaling \cite{Miransky:1984ef} (similar to Berezinsky-Kosterlitz-Thouless (BKT) scaling),  with $f(\alpha)\rightarrow 0 \,(\alpha\rightarrow \alpha_{\rm cr})$ in the essential-singularity form, thus retaining the approximate scale invariance $\alpha(\mu^2) \approx \alpha$ for the wide infrared region $m_F^2 < \mu^2<\Lambda_{\rm TC}^2$. The generation of the tiny $m_F$ in units of the intrinsic scale $\Lambda$
breaks the scale symmetry explicitly as well as spontaneously, so that the TD as a pseudo NG boson was expected to acquire a tiny mass to be estimated by the anomalous Ward-Takahashi (WT) identity for the approximate scale symmetry via Partially Conserved Dilatation Current (PCDC) relation \cite{Bando:1986bg}, in the same manner as the pion mass estimate by the Partially Conserved Axial Current (PCAC).

Since then, 
the WTC has confronted other challenges, namely, the  $S,T,U$ parameters\cite{Peskin:1990zt}
from the electroweak precision measurements
\footnote{There are several solutions to the $S$ parameter problem. See the discussions in the last section.
}, 
the large top quark mass 173 GeV
\footnote{Possible resolutions are discussed in the last section. 
}, 
and finally the most serious and urgent problem from the discovery of the Higgs at 125 GeV. This created 
a widely spread folklore against TC including WTC: e.g., ``More intuitively, the measured mass of the Higgs tells us that it is weakly 
coupled. Strong coupling solutions like Technicolor tend to lead to a strongly coupled Higgs'' \cite{Seiberg}.

This is totally a misconception based on the linear sigma model, whose $\lambda |\phi|^4$ coupling for the would-be QCD $\sigma$ meson with mass
$M_\sigma ^2 =2 \lambda v^2\simeq (6 v)^2 \simeq (500 \, {\rm MeV})^2$ would be obviously strong $\lambda \sim (6 v)^2/(2v^2) \simeq 18\gg 1$, in sharp contrast to 
the 125 GeV Higgs with $\lambda \simeq (125 \, {\rm GeV)^2/[2 (246\, GeV})^2]\simeq 1/8 \ll 1$.
Actually, the linear sigma model is  not the right effective theory of
QCD, rather  the nonlinear sigma model corresponding to $\lambda$  or $ M_\sigma^2$$ \rightarrow \infty$ is the correct one, the Chiral Perturbation Theory (ChPT).
The ChPT is not scale-invariant, which is  in accord with the QCD having no scale invariance.  
However, the WTC does have an approximate 
scale invariance and hence its effective field theory must be approximately 
scale-invariant. The light composite Higgs, the TD as the pseudo NG boson of the approximate scale symmetry, does make the  nonlinear sigma model 
(approximately) scale-invariant, 
in a way fully consistent with  the strongly coupled underlying theory, the WTC 
 (``scale-invariant ChPT'', sChPT for short)\cite{Matsuzaki:2012vc,Matsuzaki:2013eva}. It will be shown in Eq.(\ref{selfcouplings:0}) that the self-interactions of the 
 TD are  even weaker than the SM Higgs!

Note that  all the bound states (techni-hadrons) 
are in principle strongly coupled to each other  {\it within the WTC sector} 
just as hadrons in QCD are, whereas
the couplings of 125 GeV Higgs so far observed at LHC are {\it not those among the techni-hadrons} but only  the couplings of 
{\it a special technihadron, TD,  to the SM sector particles}, 
which must be weak, through either the (weak) $SU(2)\times U(1)$ gauge couplings or the (weak) effective
Yukawa couplings (loop-suppressed and extended TC (ETC)-scale suppressed via ETC-like couplings), see Eq.(\ref{efffectiveyukawa}), all related to outside of the strongly coupled WTC sector. 
Moreover, it was shown that the TD couplings themselves characterized by  $1/F_\phi\ll 1/v_{\rm EW}$, 
are {\it even weaker} than those of the
SM Higgs~\cite{Matsuzaki:2012mk,Matsuzaki:2012xx}. 
See Eqs.(\ref{selfcouplings:0}) and (\ref{scaling}).

Another widely spread misconception comes from the ChPT, the opposite to the linear sigma model view.
It says that there is no light composite scalar meson with mass much lighter than the scale of the naive dimensional analysis (NDA), $4\pi F_\pi$,
which is based on the estimated breakdown scale of the conventional ChPT valid in the ordinary QCD.
The crucial assumption of NDA is that  no light spectrum other than the pions exist below $4\pi F_\pi$, which however 
is already in contradiction with the reality even in the
conventional QCD: $M_{f_0}\simeq 500\, {\rm MeV}$ and $M_\rho \simeq 770 \, {\rm MeV}$,  well below the NDA $4\pi F_\pi \simeq 1.2\, {\rm GeV}$.
Actually, the statement should be reversed:  
If there exists a light spectrum
lighter than  $4\pi F_\pi$, then the conventional ChPT should be modified so as to include the light spectrum in such a way that the effective theory must respect the symmetry of the underlying theory. In the case at hand, it is the sChPT \cite{Matsuzaki:2012vc,Matsuzaki:2013eva}.

There have been much progress of the WTC particularly on the light TD, 
not just in the ladder SD equation, but also in a  variety of approaches such as the ladder Bethe-Salpeter (BS) equation 
combined with the ladder SD equation~\cite{Harada:2003dc,Kurachi:2006ej}, the effective theory based on the sChPT \cite{Matsuzaki:2012vc,Matsuzaki:2013eva}, with possible extension including vector mesons via HLS in a scale-invariant manner~\cite{Kurachi:2014qma},
holographic method~\cite{Haba:2010hu,Matsuzaki:2012xx,Kurachi:2014xla}, and eventually, the first-principle calculation of the flavor-singlet scalar meson in the large $N_F$ QCD on the lattice \cite{Aoki:2014oha,Aoki:2013zsa,Fodor:2014pqa, Brower:2014ita}. 
In particular, it is remarkable that such a light flavor-singlet scalar meson as a candidate for the TD was observed in the lattice $N_F=8$ QCD \cite{Aoki:2014oha}, the theory shown to  have signatures of the lattice walking theory including the anomalous dimension $\gamma_m \simeq 1$ \cite{Aoki:2013xza, Appelquist:2014zsa,Hasenfratz:2014rna}. 
 Note that $N_F=8$ (four weak-doublets)  corresponds to the ``one-family model'' \cite{Dimopoulos:1979sp,Farhi:1980xs}  which is  
the  most straightforward  model building 
 of  the ETC~\cite{Dimopoulos:1979es} as a standard way to give mass to the quarks and leptons. The one-family model of the WTC with $N_C=4$ is in fact best fit to the 125 GeV Higgs data~\cite{Matsuzaki:2012gd,Matsuzaki:2012vc,Matsuzaki:2012mk,Matsuzaki:2012xx}, and is shown to be most natural for the ETC model building \cite{Kurachi:2015bva}.  
 
Among such many approaches,  the ladder SD equation is still a powerful and relatively handy tool to analyze the TD as a composite Higgs,
in spite of the fact that it is not a systematic approximation in the sense that high order corrections are not controllable (see below, however).
In fact it turned out to be more than a 
mnemonic of the physics guess: 
It well reproduced  numerically as well as the qualitatively the nonpertubative aspects of the ordinary QCD in the hadron physics, with additional ansatz simply replacing the nonrunning coupling 
by the one-loop running one as the input coupling of the SD equation \cite{Miranskybook}. 
Many ladder analyses on the dynamical symmetry breaking with large anomalous dimensions in the strongly coupled gauge theories and gauged NJL model gave many suggestive results in the applications for WTC, top quark condensate model, etc.. \cite{Yamawaki:1996vr,Miranskybook}.  
\\

\begin{figure}
  \begin{center}
    \includegraphics[scale=0.4]{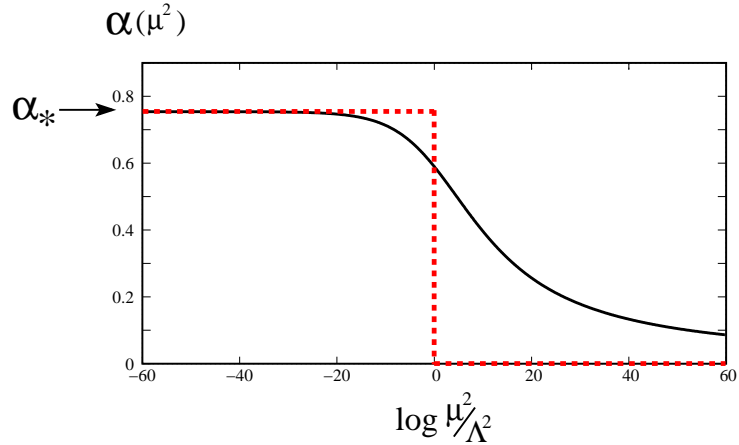}
  \end{center}
\caption{
Two-loop
running coupling (solid curve)  
in the case of 
$SU(3)$ gauge theory with $N_F=12$ massless fundamental fermions with intrinsic scale $\Lambda_{\rm TC}$, compared with the ladder coupling for $\Lambda=\Lambda_{\rm TC}$. The overall scale of $\alpha (\mu)$ shrinks like $1/N_C$ to zero in the large $N_C$ limit with $N_C \alpha ={\rm fixed}$ (i.e., $\Lambda=\Lambda_{\rm TC}$ is fixed) 
and $r=N_F/N_C ={\rm fixed} \gg 1$
(walking/anti-Veneziano limit). 
}
\label{fig:run}
\end{figure}

 In this paper, 
in the new light of the 125 GeV Higgs at LHC, we investigate full implications of the ladder SD gap equation for the WTC, in the context of  near conformal window of large $N_F$ QCD,  $SU(N_C)$ gauge theory with massless $N_F$ flavors \cite{Appelquist:1996dq,Miransky:1996pd}, in the particular walking limit,  ``anti-Veneziano limit'' (in distinction to the original Veneziano limit with $N_F/N_C\ll 1$):
\beq 
N_C \rightarrow \infty \quad {\rm and} \quad \lambda\equiv N_C \cdot \alpha = {\rm fixed},
 \quad  {\rm with} \quad r\equiv N_F/N_C ={\rm fixed} \,\, \gg 1,
 \label{antiVeneziano}
\eeq
(See \cite{Kurachi:2014xla, Kurachi:2014qma} for preliminary discussions).  Such a limit
realizes the ideal situation for the ladder SD equation,  where the input perturbative coupling becomes nonrunning (infrared conformality), $\alpha(\mu^2) \equiv \alpha_*$, thanks to the  
  perturbative infrared (IR) fixed point (Caswell-Banks-Zaks (CBZ) IR fixed point) \cite{Caswell:1974gg}, $\alpha(\mu^2) \simeq \alpha_*=\alpha(\mu^2=0)$   
 already near the anti-Veneziano limit  for $0<\mu^2<\Lambda_{\rm TC}^2$.   
 See  Fig.\ref{fig:run}. The present paper is an extension of Ref.\cite{Hashimoto:2010nw}, where a similar analysis was done without concept of the  anti-Veneziano limit.

 In the ladder SD equation, 
 the critical coupling $\alpha_{\rm cr}$ is given as 
 $\alpha_{\rm cr}=\frac{\pi}{3 C_2}$, with the quadratic Casimir $C_2=\frac{N_C^2-1}{2N_C}$. For   the strong coupling $\alpha=\alpha_*> 
 \alpha_{\rm cr}$, 
the technifermion acquires the dynamical mass $m_F$ in an essential-singularity (non-analytic) form 
a la Miransky-Berezinsky-Kosterlitz-Thouless \cite{Miransky:1984ef} of the conformal phase transition \cite{Miransky:1996pd}:
 \beq
 m_F \simeq 4  \Lambda\cdot  \exp \left(-\frac{\pi}{\sqrt{\frac{\alpha}{\alpha_{\rm cr}}-1}}\right)  
\ll \Lambda \quad  \left(0<\frac{\alpha}{\alpha_{\rm cr}} -1\ll 1,\quad \alpha_{\rm cr}=\frac{\pi}{3 C_2}=\frac{\pi}{3}\frac{2N_C}{N_C^2-1}\right),   
 \label{Miranskys} 
 \eeq 
which implies a large hierarchy $m_F \ll \Lambda$ near the criticality  $\alpha\simeq \alpha_{\rm cr}$, where the cutoff $\Lambda$
as a regulator may be regarded as the intrinsic scale $\Lambda=\Lambda_{\rm TC}$. 
 The would-be CBZ IR fixed point $\alpha_*$ is washed out by the mass, which however is a small mass  
  $m_F \ll \Lambda_{\rm TC}$, 
so that there still remains  an approximate scale symmetry  in a wide infrared region $m_F <\mu <\Lambda_{\rm TC}$. 
Note that in the walking/anti-Veneziano limit the ladder approximation becomes more trustable, since the coupling becomes ``weak'',
\beq
\alpha(\mu) \simeq \alpha_* \simeq \alpha_{\rm cr} \sim 1/N_C \rightarrow 0\,,
\eeq 
so that many non-ladder diagrams without $C_2$ factor multiplied on $\alpha$ are suppressed as in the usual 
$1/N_C$ expansion (also is the case in the NJL model where $(g, g_{\rm cr})  \sim 1/N_C$), in spite of the fact that 
the 't Hooft coupling $\lambda$ is really strong and the ``effective'' critical coupling to trigger the chiral condensate is strong,  $C_2 \alpha_{\rm cr}=\frac{\pi}{3}>1$ such that $\lambda > \lambda_{\rm cr} =(N_C/C_2) \pi/3\rightarrow 2\pi/3$. 

Eq.(\ref{Miranskys}) dictates that $\alpha$ is no longer constant due to $m_F\ne 0$ but does run depending on $\Lambda/m_F$ according to the nonperturbative beta function $\beta^{(NP)}(\alpha)$ \cite{Leung:1985sn}
(See Fig. 1 (a) of Ref.~\cite{Yamawaki:1985zg}):  
\begin{eqnarray}
\beta^{(NP)}(\alpha) &=&\Lambda \frac{\partial \alpha(\Lambda)}{\partial \Lambda} = - \frac{2\pi^2\alpha_{\rm cr}}{\ln^3 (\frac{4\Lambda}{m_F})} 
=  - \frac{2\alpha_{\rm cr}}{\pi} \left(\frac{\alpha}{\alpha_{\rm cr}}-1\right)^{\frac{3}{2}} 
\quad\quad  (\alpha> \alpha_{\rm cr})\,, 
\label{NPbeta}
\end{eqnarray}
as $\Lambda/m_F \rightarrow \infty$, and hence the coupling as a solution of 
$\frac{\partial \alpha}{\partial \ln \mu} =\beta^{(NP)}(\alpha)$ runs as renormalization point $\mu$:
\beq
\alpha(\mu) =
\alpha_{\rm cr} \left[ 1+ \frac{\pi^2}
{\ln^2\left(\frac{\mu}{\mu_{\rm IR}} 
\right)
}  \right]  
 \quad\quad  (\alpha (\mu) >\alpha_{\rm cr})\,,
\label{NPrunning}
\eeq
with $\mu_{\rm IR} ={\cal O} (m_F)$, even when the perturbative coupling (input coupling)
is nonrunning,
$\beta(\alpha)|^{\rm perturbative}\equiv 0$. 
This is completely different from the two-loop beta function having the CBZ IR fixed point,  
which is no longer valid for $\alpha>\alpha_{\rm cr}$, 
where $\alpha (\mu)\searrow \alpha_{\rm cr}$ ($\mu \nearrow$): 
$\alpha_{\rm cr}$ 
is now regarded as the ultraviolet (UV) fixed point, as was emphasized in Ref.~\cite{Yamawaki:1985zg} in the context of the WTC. 
Then the would-be {\it IR fixed point $\alpha_* \simeq \alpha_{\rm cr}$ is also regarded as 
the UV fixed point} of the nonperturbative running (walking) coupling $\alpha(\mu) \approx \alpha_{\rm cr}$ 
in the wide IR region $m_F < \mu < \Lambda=\Lambda_{\rm TC}$ for the characteristic large hierarchy (``criticality hierarchy'') $m_F \ll \Lambda_{\rm TC}~$\cite{Yamawaki:2007zz,Hashimoto:2010nw}.  
(See also Ref.\cite{Kaplan:2009kr} for a similar observation.) 

The scale symmetry is broken also explicitly by $m_F$ which is generated by the
SSB of the the same scale symmetry, the typical order parameter being the decay constant of the TD, $F_\phi$,
defined as $\langle 0|D_\nu|\phi (q)\rangle =- i F_\phi q_\nu$
\footnote{
$F_\phi$ is also defined as $\langle 0| \theta_{\mu\nu}|\phi(q)\rangle= F_\phi (q_\mu q_\nu- q^2 g_{\mu\nu})/3$, 
which yields the identical $F_\phi$. 
}.
 Different from the SSB of the internal symmetry 
like chiral symmetry, there exists no exact point where the scale symmetry is spontaneously 
broken without explicit breaking. Nevertheless there exists a limit where the explicit breaking is much smaller than the SSB  scale of the scale symmetry, that is, the walking/anti-Veneziano  limit, Eq.(\ref{antiVeneziano}).

Note that $m_F$ is an $N_F, N_C$-independent quantity  related to $\Lambda= \Lambda_{\rm TC}$ via Miransky scaling, Eq.(\ref{Miranskys}), with $\alpha/\alpha_{\rm cr}\simeq \alpha_*/\alpha_{\rm cr}$ being only dependent of the ratio $r=N_F/N_C$
in the anti-Veneziano limit. Since the dilatation current is a sum of all the $N_F$ and $N_C$ - technifermion species,  $D_\nu(x)  \sim N_FN_C$,  and the TD state is
normalized as $|\phi\rangle \sim 1/\sqrt{N_FN_C}$, we have  
$F_\phi \sim \sqrt{N_F N_C} \,m_F$ by definition of $F_\phi$,
so that the explicit breaking $m_F$ is much smaller than $F_\phi$,   
or the NDA associated with the TD loop $(4\pi F_\phi)^2/N_F$ in the walking/anti-Veneziano  limit,  in addition to the criticality hierarchy $m_F \ll \Lambda_{\rm TC}$ of direct relevance to the scale symmetry:
\beq
m_F^2\, \ll\, F_\phi^2 \, ,  
\qquad  
\frac{(4\pi F_\phi)^2}{N_F}\, \ll \Lambda_{\rm TC}^2. \,
\eeq

 Then the mass of the TD as a pseudo-NG boson   $M_\phi$ 
 can be evaluated, based on  the anomalous WT identity for the scale symmetry as the PCDC relation for the trace anomaly
\cite{Bando:1986bg} :
\begin{eqnarray}
M_\phi^2 F_\phi^2 &=& - F_\phi \langle 0 |\partial_\mu D^\mu |\phi \rangle^{(NP)} = - 4\langle 0 |\theta_\mu^\mu |0\rangle^{(NP)}
=-\frac{\beta^{(NP)}(\alpha(\mu))}{\alpha(\mu)} \langle G_{\nu\lambda}^2(\mu)\rangle^{(NP)} \nonumber \\
&=& 
N_F N_C\left(\frac{16 \xi^2  }{\pi^4} m_F^4\right),   \quad ( \xi \simeq 1.1),\,
\label{PCDCladder1}
\end{eqnarray}
where the last expression  was given through the ladder evaluation of the vacuum energy $E =\langle 0 |\theta_\mu^\mu |0\rangle^{(NP)}/4$ \cite{Gusynin:1987em}.  
Since $F_\phi^2 \sim N_F N_C m_F^2$ by definition of $F_\phi$, 
Eq.(\ref{PCDCladder1}) is also  in accord with the fact that  $M_\phi^2$ as well as $m_F^2$ has no
explicit  dependence on $N_F$ and $N_C$. 
Here all the quantities with $(\cdot)^{(NP)}$ to be defined later contain only the nonperturbative contributions arising from the dynamical mass $m_F\ne 0$ due to the SSB, 
and hence vanishes as $m_F\rightarrow 0$.

  We show  that  independent calculations of $\frac{\beta^{(NP)}(\alpha(\mu))}{4 \alpha(\mu)}$ and $\langle G_{\nu\lambda}^2(\mu)\rangle^{(NP)}$  in the ladder approximation yield the nonperturbative trace anomaly  as a product 
of them, which 
precisely agrees with the result  calculated independently from the vacuum energy of Ref.\cite{Gusynin:1987em}.  
The agreement is realized in a highly nontrivial manner,  fully consistent with the Renormalization-Group Equation (RGE)
point of view: 
Each of the $\frac{\beta^{(NP)}(\alpha(\mu))}{4 \alpha(\mu)}$ and $\langle G_{\nu\lambda}^2(\mu)\rangle^{(NP)}$ does depend on the renormalization point $\mu$:  See Eq.(\ref{NPbeta}) for 
$\frac{\beta^{(NP)}(\alpha(\mu))}{4 \alpha(\mu)} \sim 1/\ln^3\mu$, and explicit calculation reads $\langle G_{\nu\lambda}^2(\mu)\rangle^{(NP)} \sim  \ln^3\mu
$ for $m_F<\mu<\Lambda_{\rm TC}$.
Such a $\mu$-dependence is completely cancelled each other in the product to arrive at the $\mu$-independent   trace anomaly as  it should be. 
Similar cancellation of the $\mu$-dependence also takes place in the ordinary QCD, where $\frac{\beta(\alpha(\mu))}{4 \alpha(\mu)} \sim 1/(\ln \mu)$ and 
 $\langle G_{\nu\lambda}^2(\mu)\rangle \sim \ln \mu$, with the logarithm of $\ln \mu$ instead of $\ln^3 \mu$. 
 This result 
is the RGE view of the previous calculation \cite{Hashimoto:2010nw} based on the improved  ladder approximation
for the large $N_F$ $SU(N_C)$ gauge theories near conformal window.  
  
 The key observation of the present paper is that as far as the PCDC relation is satisfied as in the ladder approximation, 
the TD as a pseudo NG boson has a vanishing mass in the anti-Veneziano limit, quite independently of the numerical details  of the ladder calculation (See \cite{Kurachi:2014xla, Kurachi:2014qma} for preliminary discussions):
  Noting that $F_\phi^2={\cal O} ( N_F N_C m_F^2)$, Eq.(\ref{PCDCladder1}) naturally explains 
 \beq
 M_\phi = {\cal O} \left(\frac{4\xi m_F}{\pi^2}\right)={\cal O} \left(\frac{m_F}{2}\right) \, \ll \, F_\phi={\cal O} ( \sqrt{N_F N_C}\, m_F),\,
 \eeq
in the walking/anti-Veneziano limit, Eq.(\ref{antiVeneziano}), where we have $M_\phi^2/F_\phi^2 \sim 1/(N_FN_C) \rightarrow 0$.
The TD as the pseudo NG boson  has a vanishing mass limit, though not exact massless point, 
  in the anti-Veneziano limit,  
  where the nonperturbative trace anomaly vanishes in units of $F_\phi$ 
  as a measure of the SSB of the scale symmetry. 
  This is  similar to the $\eta^\prime$ meson in QCD, which is regarded as a pseudo NG boson whose mass, evaluated through the anomalous WT identity with the $U(1)_A$ anomaly,
   does vanish  in   the large $N_F$ and $N_C$ limit with $N_F/N_C$ fixed ($\ll 1$) (Veneziano limit): $M_{\eta^\prime}^2/F_\pi^2 
  \sim N_F/N_C^2 \rightarrow 0$, without the exact massless point.

Note that $m_F$ is related to the weak scale $v_{\rm EW}=246\, {\rm GeV}$ through  the Pagels-Stokar formula $F_\pi^2\simeq 
 (N_C\xi^2/2 \pi^2) \, m_F^2$ in the ladder approximation reads (Eq.(\ref{PSv})):
$ v_{\rm EW}^2=(246\,  {\rm GeV})^2 
= N_D F_\pi^2 \simeq  
\frac{N_F N_C\xi^2}{4\pi^2} \, m_F^2  
\simeq  m_F^2 \left[\frac{N_F}{8}\frac{N_C}{4}\right]$,
with $N_D (=N_F/2)$ being the number of the electroweak doublets. Then a natural estimate of the TD mass for the one-family model $N_F=8$ with $N_C=4$ is that 
$M_\phi = {\cal O} (m_F/2) ={\cal O} ( v_{\rm EW}/2) ={\cal O} (125 \, {\rm GeV})$, in agreement with the LHC Higgs as the TD.
More precisely, Eq.(\ref{PCDCladder1}) can be rewritten in terms of $v_{\rm EW}$: 
\beq
M_\phi^2 \simeq \left(\frac{v_{\rm EW}}{2}\right)^2 \cdot \left(  \frac{5\, v_{\rm EW}}{F_\phi} \right)^2 \cdot \left[\frac{8}{N_F}\frac{4}{N_C}\right]. 
\eeq 
It was first pointed out in Ref. \cite{Matsuzaki:2012gd} that this ladder PCDC result accommodates the $125\, {\rm GeV}$ Higgs
with $F_\phi = {\cal O}\, ( {\rm TeV})$ for the one-family model with $N_F=8$ and 
was shown to be the best fit to the current LHC data:
\beq
F_\phi \simeq 5\, v_{\rm EW} \simeq 1.25 \, {\rm TeV}\quad {\rm for}\, \quad M_\phi=125 \, {\rm GeV}\quad (N_F=8,\, N_C=4)
\eeq
\cite{Matsuzaki:2012gd,Matsuzaki:2012vc,Matsuzaki:2012mk,Matsuzaki:2012xx} (See also the later discussions). With
the fact that $v_{\rm EW}^2 \propto N_FN_C m_F^2 \sim F_\phi^2$, the result reflects
the generic scaling: 
\beq
\frac{M_\phi}{v_{\rm EW}} \sim \frac{M_\phi}{F_\phi} \sim \frac{m_F}{F_\phi} \sim \frac{1}{\sqrt{N_F N_C}} \rightarrow 0, 
\eeq
in the anti-Veneziano limit~\footnote{
In early days, ladder-like calculations \cite{Leung:1985sn,
Holdom:1986ub,Gusynin:1987em,Nonoyama:1989dq,Shuto:1989te}
both in pure gauge theories and
gauged NJL models
showed $M_\phi ={\cal O} (m_F)$, however without paying attention to the $N_F, N_C$ dependence, particularly
to the fact  that $ \frac{M_\phi}{v_{\rm EW}} \sim \frac{m_F}{v_{\rm EW}} \,\ll 1$ in the anti-Veneziano limit, which is actually  of the most phenomenological
relevance with respect to the 125 GeV Higgs such as in the one-family model with $N_F=8,\, N_C=4$. 
}. 
 
 On the other hand, all the non-NG boson technihadrons, such as the techni-rho, techni-$a_1$, technibaryon, etc., have {\it no constraints from the 
 PCDC as the explicit breaking} of the scale symmetry but do have {\it constraints from the SSB} of the scale symmetry, so that they
  should have masses on the scale of  SSB  of the scale symmetry, characterized by $F_\phi$ much larger than $2 m_F$ of the naive nonrelativistic quark model picture:
   \beq
 M_\rho, M_{a_1}, M_N, \cdots = {\cal O} ({\rm TeV's}) > {\cal O} (F_\phi) \gg 2 m_F \gg M_\phi.
 \eeq
 In fact, the IR conformal physics of the WTC should  be described by the low-lying composite fields as effective fields, 
 in a way to realize all the symmetry structure of the underlying theory.
 
  Such an effective theory of WTC as a straightforward extension of sChPT \cite{Matsuzaki:2012vc,Matsuzaki:2013eva}  is already constructed, i.e, the scale-invariant version \cite{Kurachi:2014qma} of the Hidden Local Symmetry (HLS) model \cite{Bando:1984ej,Harada:2003jx},  (the ``sHLS model''),
  where  the technirho mass terms have the {\it scale-invariance nonlinearly realized}  by the TD field $\chi=e^{\phi/F_\phi}$, with the SSB of the scale invariance  characterized by the scale of $F_\phi$,
 while the Higgs (TD) mass term in the TD potential,  on the  order of $m_F (\ll F_\phi)$,  is the only source of the explicit breaking of the scale symmetry related (via PCDC) to the nonperturbative trace anomaly 
 of the underlying theory. 
 
One interesting  candidate for such technihadrons may be a resonance behind the diboson excess recently observed at LHC at 2 TeV~\cite{Aad:2015owa, Khachatryan:2014hpa}, which can be identified with the walking technirho \cite{Fukano:2015hga}. A smoking gun  of the walking techni-rho is the absence of the decay to the 125 GeV Higgs (TD), which is forbidden by the scale symmetry explicitly
broken only 
by the Higgs (TD) mass term 
(corresponding to the nonperturbative trace anomaly in the underlying WTC)
\cite{Fukano:2015uga}.  Actually, the salient feature of the scale symmetry of the generic effective theory not just the sHLS model, containing the SM
 gauge bosons and the Higgs plus new vector bosons (any other massive particles as well), 
 is the {\it absence of the decay of the new vector bosons}  such as the technirho (and also other higher resonances) {\it  into  the 125 GeV Higgs plus the SM gauge bosons} \cite{Fukano:2015uga}. If such decays of new particles are not found at LHC Run II, then the 125 GeV Higgs is nothing but the dilaton (TD in the case of the WTC) responsible for the nonlinearly realized scale symmetry, i.e., the  SSB of the scale symmetry, no matter what underlying theory may be beyond the SM. 
This should be tested in the  ongoing LHC Run-II.

The paper is organized as follows: In section \ref{SDeqsol} we review the solutions of the  ladder SD equation in some details 
and the conformal phase transition a la Miransky-BKT in the context of CBZ IR fixed point of the large $N_F$ QCD in the anti-Veneziano limit.  Nonpertubative beta function and the corresponding running of the coupling is discussed. Large anomalous dimension $\gamma_m=1$
and its phenomenological implications are reviewed. 
In section \ref{NPtraceanomaly}  we explicitly
show the RG invariance of the nonperturbative trace anomaly in the broken phase of the ladder SD equation, in such a way that three independent calculations of $\frac{\beta^{(NP)}(\alpha)}{4 \alpha}$, $\langle G_{\mu\nu}^2\rangle^{(NP)}$ and $\langle(\theta_\mu^\mu)\rangle^{(NP)} $ yield precisely a correct trace anomaly relation.
We further check explicitly that the ladder calculation satisfies
the anomalous WT identity in the case of nonzero
fermion mass $m_0\ne 0$. This is to establish the consistency of the ladder calculation with the 
sChPT proposed in Ref.\cite{Matsuzaki:2013eva} for determining the mass $M_\phi$ and the decay constant $F_\phi$ of the TD on the lattice. In section \ref{TDmass} we give a 
mass and decay constant of the TD through the PCDC relation as an anomalous WT identity for the scale symmetry based on the nonperturbative trace anomaly.  We discuss that the
TD becomes a parametrical NG boson in the anti-Veneziano limit in accord with the walking regime of large $N_F$ QCD, in a  sense similar to the $\eta^\prime$ meson  in the ordinary QCD a la Witten-Veneziano.
In section \ref{LHCpheno} we show that the light TD is consistent with the current LHC data on the 125 GeV Higgs, as an update of the Ref.~\cite{Matsuzaki:2012mk}.
Section \ref{BeyondTD}  is for the technihadrons other than TD.
Section \ref{Summary} 
is devoted to summary and discussions. 
Appendix A is for the basic formulas of the ladder SD equation. 
In Appendix B we give a ladder result for the Pagels-Stokar formula for $F_\pi^2$. 
 Appendix C is for the details  about the contamination of the 125 GeV Higgs poduction between 
the gluon fusion production and the vector boson fusion production at the present LHC data.

  \section{Solution of the Ladder SD equation and Conformal Phase Transition}
\label{SDeqsol}
 \subsection{Ladder coupling as the CBZ IR fixed point in the anti-Veneziano limit}
 
  Let us first recapitulate the results in Ref.\cite{Yamawaki:1985zg} based on the 
  the ladder SD gap equation for the technifermion mass function $\Sigma(-p^2)$ ($p^2<0$) with the nonrunning coupling
  as an idea limit of the CBZ  IR fixed point of large $N_F$ $SU(N_C)$ gauge theories, which can be well described by 
the  improved ladder approximation with the running coupling $g^2(-p^2)$ \cite{Miransky:1983vj}: 
\beq
  S_F^{-1}(p) \ =\ S^{-1}(p) + \int \frac{d^4 k}{(2 \pi)^4}\ 
  C_2\, 
  g^2((p-k)^2)\ D_{\mu\nu} (p-k)
  \gamma^\mu \ S_F(k) \ \gamma^\nu,
\label{eq:improvedSDeq}
\eeq
where  $i S_F^{-1}(p)=Z^{-1} (-p^2) (\xbar{p} - \Sigma(-p^2))$ and $i S^{-1}(\xbar{p} - m_0)$ are the full and bare technifermion propagators, respectively, and
$i D_{\mu\nu}(p)$ the bare technigluon  propagator in the Landau gauge, with an ansatz $g^2((p-k)^2) \Rightarrow g^2(max\{-p^2,-k^2\})$.    $C_2$ is  the quadratic Casimir of the technifermion  of the gauge theory, with $C_2= (N_C^2 -1)/(2N_C)$ for 
  the fundamental  representation in $SU(N_C)$.   
After the angular integration,  the improved ladder SD equation in Landau gauge for $\Sigma (x\equiv -p^2)$ 
reads:
 \begin{equation}
 \Sigma(x) = m_0 +  
 \frac{3 C_2}{4\pi}  \int
 dy\,   \left[\frac{\alpha(x)}{x}\theta(x-y) + \frac{\alpha(y)}{y}\theta(y-x)\right] \frac{y \Sigma(y)}{ y +\Sigma^2(y)},   \quad  (Z^{-1}(x) \equiv 1).  
 \label{SDeq}
 \end{equation}

The original ladder SD gap equation as the basis for the WTC \cite{Yamawaki:1985zg,Bando:1986bg}  is a scale-invariant dynamics, having an input
coupling as {\it nonrunning}: 
 \beq
\alpha (x) =\frac{g^2(x)}{4\pi}\equiv \alpha,  
\label{laddercoupling}
 \quad \beta(\alpha) \equiv 0, \quad {\rm for}\,\,  0<x <\Lambda^2
\,. 
\eeq
The  cutoff $\Lambda$ breaks explicitly the scale symmetry, as does the 
the intrinsic scale $\Lambda_{\rm TC}$ analogous to the $\Lambda_{\rm QCD}$.
Such a scale-invariant coupling is indeed an idealization of the CBZ IR fixed point \cite{Caswell:1974gg} 
$\alpha=\alpha_*$, such that $\beta^{(2-loop)}(\alpha_*) =0$ and $\alpha(\mu^2) \approx \alpha_*$ ($\mu^2\ll\Lambda_{\rm TC}^2$) in the large $N_F$ QCD \cite{Appelquist:1996dq,Miransky:1996pd}, where the two-loop coupling is almost nonrunning particularly  in the walking/anti-Veneziano limit 
Eq.(\ref{antiVeneziano}), while
it is rapidly decreasing in the one-loop dominated asymptotically free UV region $\mu>\Lambda_{\rm TC}$, as in the ordinary QCD (See Fig.\ref{fig:run}):
\begin{eqnarray}
\mu \frac{\partial}{\partial \mu} \alpha = \beta^{(2-loop)}(\alpha) &=& -b_0 \alpha^2 -b_1 \alpha^3,\nonumber\\
b_0&=&\frac{1}{6\pi} (11N_C- 2N_F), \quad b_1= \frac{1}{24\pi^2}\left(34N_C^2-10N_C N_F -3 \frac{N_C^2-1}{N_C} N_F\right),\nonumber\\
 \alpha_*&=& -\frac{b_0}{b_1} \quad \longrightarrow \frac{4\pi}{N_C} \frac{11- 2 r}{13 r - 34} \quad \left(N_C\rightarrow \infty, \,\,\frac{34}{13} < r\equiv \frac{N_F}{N_C}= {\rm const}<\frac{11}{2}\right)\,.
 \label{2loop}
\end{eqnarray}
The analytic form of $\alpha$ is given as
\beq
\alpha(\mu^2) = \frac{\alpha_*}{ 1+ W(z(\mu))},\quad z(\mu)\equiv \frac{1}{e} \left(\frac{\mu}{\Lambda_{\rm TC}}\right)^{b_0\alpha_*},\quad {\rm with} \quad b_0\alpha_*
\longrightarrow \frac{2}{3}\frac{(11-2 r)^2}{34-13 r} ,
\eeq
 where $W(z)$ is the Lambert W function and $\Lambda= \Lambda_{\rm TC}$ is the intrinsic scale 
defined as  
\beq
\Lambda_{\rm TC}=\mu \cdot \exp  \left( -\int^{\alpha(\mu^2)} \frac{d \alpha}{\beta^{(2-loop)}(\alpha)}\right)\,,
\label{perturbativeDT}
\eeq
conventionally taken as $\alpha(\mu^2=\Lambda_{\rm TC}^2)=\alpha_*/[1+ W (e^{-1})] \simeq 0.78 \alpha_*$.
The UV and IR behaviors of $\alpha(\mu^2)$ are given by
\begin{eqnarray}
\alpha(\mu^2) &\sim& \frac{1}{b_0 \ln\frac{\mu}{\Lambda_{\rm TC}}}\quad \quad \left(\mu^2\gg \Lambda_{\rm TC}^2\right),
\nonumber \\
&\sim& \frac{\alpha_*}{1+ \frac{1}{e}\left(\frac{\mu}{\Lambda_{\rm TC}}    \right)^{b_0\alpha_*}}\quad \quad  \left(
\mu^2\ll \Lambda_{\rm TC}^2
\right)
\end{eqnarray}
The intrinsic  scale $\Lambda_{\rm TC}$ is generated by the regularization in the form of the perturbative trace anomaly:
\beq
\langle \theta_\mu^\mu\rangle^{(perturbative)} =\frac{\beta^{(2-loop)}(\alpha)}{4\alpha} \langle G_{\mu\nu}^2 \rangle 
\sim - (N_C \alpha)\langle G_{\mu\nu}^2 \rangle  \sim - N_F N_C \Lambda_{\rm TC}^4\,,
\label{pertanomaly}
\eeq  
where $G_{\mu\nu}$ is the technigluon field strength~\footnote{
Usual large $N_C (\gg N_F) $ counting would imply  $\langle\theta_\nu^\nu\rangle^{(perturbative)}  \sim - (N_C \alpha)  \langle G_{\nu\lambda}^2\rangle^{(perturbative)}
=  - {\cal O} (N_C^2\Lambda_{\rm TC}^4)$ from the gluon loop,  which 
would dominate the fermion-loop of order 
 $ - {\cal O} (N_C N_F\Lambda_{\rm TC}^4)$. In the case at hand with $N_F\gg N_C$, however, the fermion-loop dominates instead. See later discussions.
 }.   
Eq.(\ref{pertanomaly})  is of course RG invariant:  The $\mu$-dependence of $\frac{\beta^{(2-loop)}(\alpha)}{4\alpha}\sim - 1/\ln (\mu^2/\Lambda_{\rm TC}^2)$
is precisely cancelled by that of   $\langle G_{\mu\nu}^2 \rangle\sim \Lambda_{\rm TC}^4 \,\ln (\mu^2/\Lambda_{\rm TC}^2)$ 
in the UV region $\mu^2>\Lambda_{\rm TC}^2$ as in the ordinary QCD.

\begin{figure}[t]
\begin{center}
   \includegraphics[width=8.5cm]{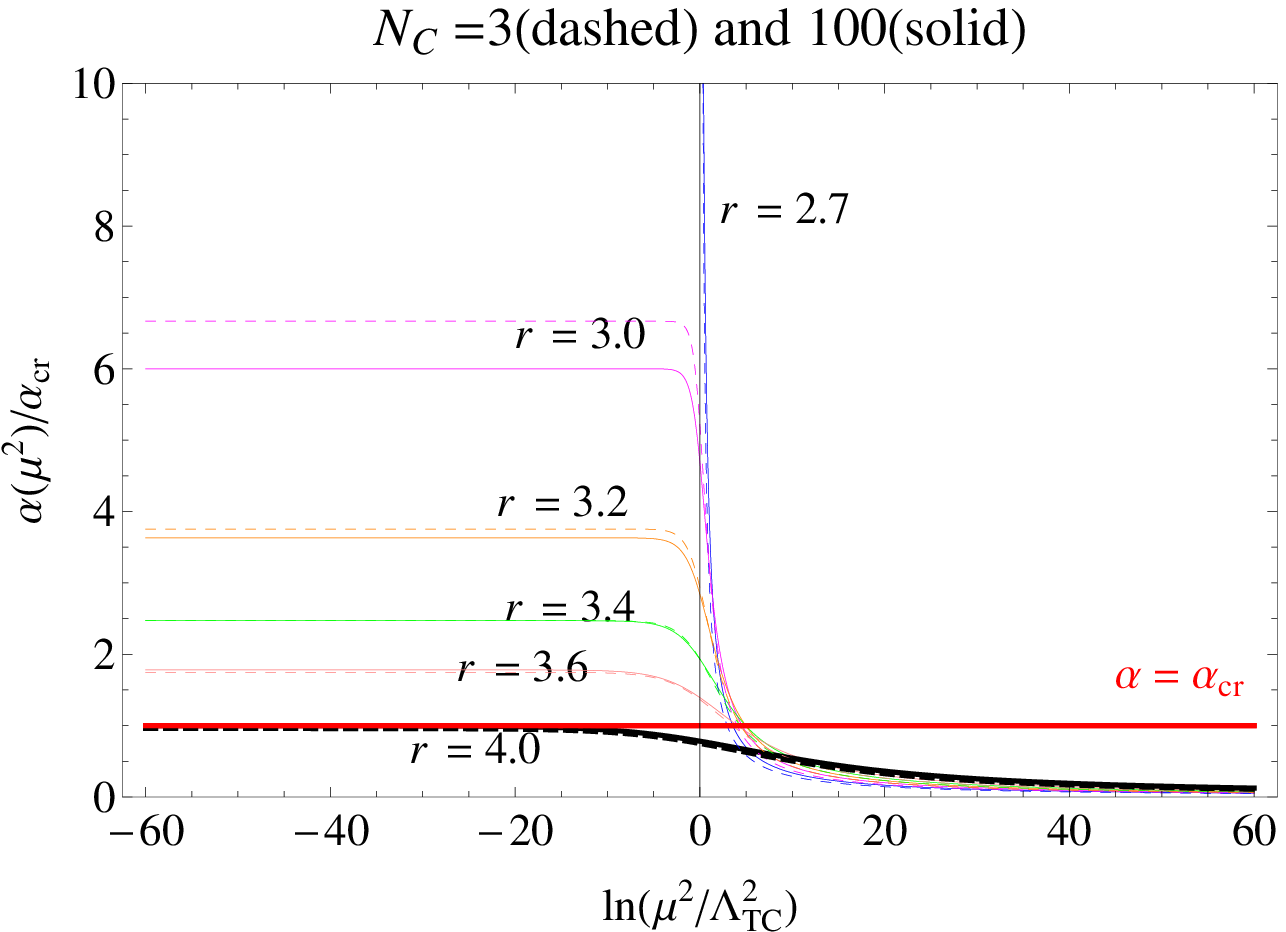}\ \ \ 
   \includegraphics[width=7.5cm]{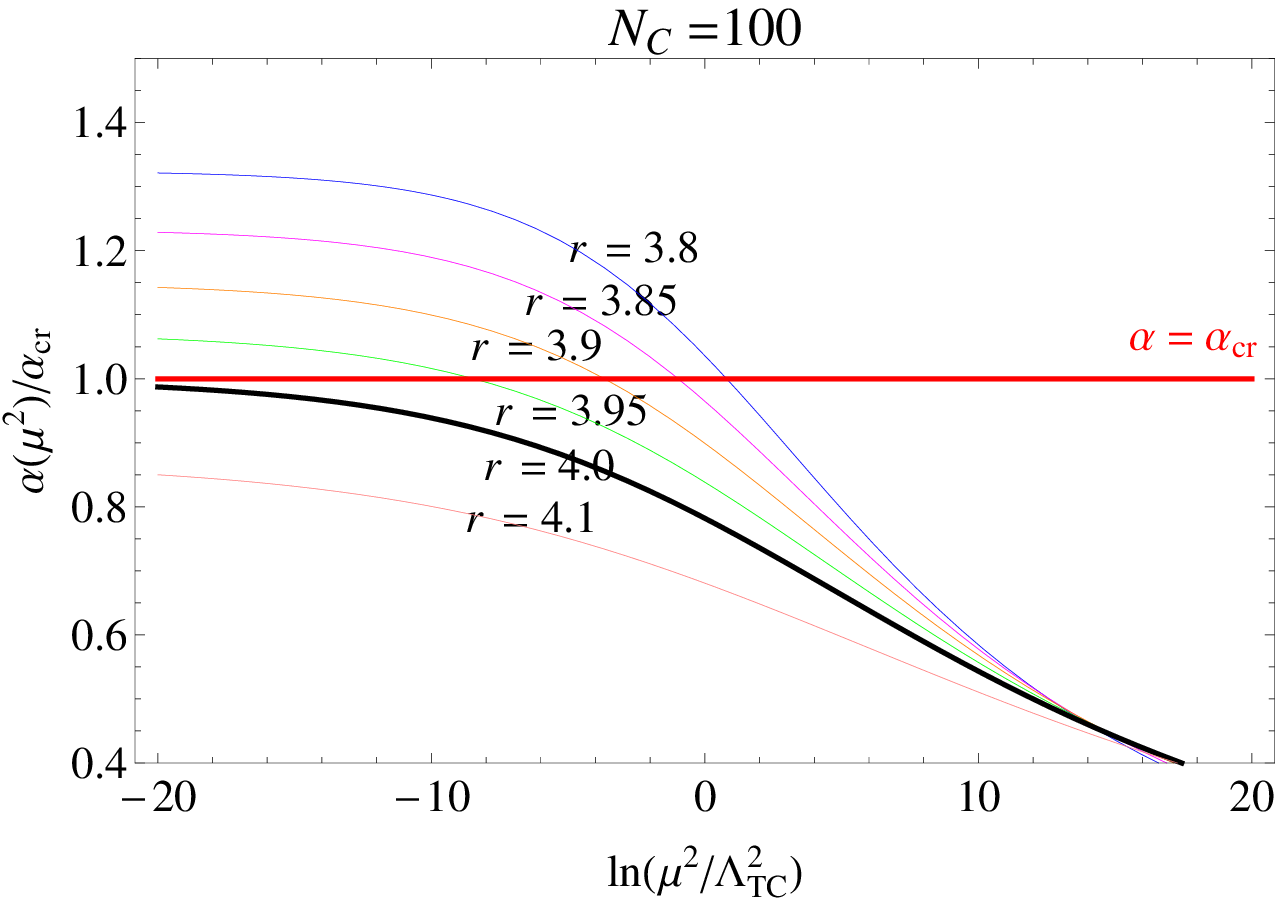}
\caption{ Two-loop coupling  constant $\alpha(\mu^2)$ normalized to the ladder critical coupling $\alpha_{\rm cr}=\pi/(3 C_2)$ 
at increasing $r=N_F/N_C$ in the anti-Veneziano limit (left). 
 The closed up view near $\alpha=\alpha_{\rm cr}$ (right).   
 } 
\label{alpha-beta-a-V}
\end{center} 
 \end{figure}

 The physics behind the walking/anti-Veneziano limit is very simple: The scale of  $m_F$ is determined by the criticality $\alpha(\mu^2=m_F^2) \sim \alpha_{\rm cr}$. Let us start with the QCD-like theory
 with $r_0\equiv N_F/N_C \sim 1$ where $\Lambda_{\rm TC}$ is specified as $\alpha(\mu^2=\Lambda_{\rm TC}^2) ={\cal O} (\alpha_{\rm cr}) ={\cal O} (1/N_C)$,  so that we have $m_F ={\cal O} (\Lambda_{\rm TC})$
 as in the usual QCD.  We then increase $r=r_1 >r_0$, which decreases the coupling mainly in the infrared region $\mu^2<\Lambda_{\rm TC}^2$  (biasing infrared-free against asymptotic-free) as a consequence of the increased screening effects of the fermion loop: $\alpha_1(\mu^2) <\alpha_0(\mu^2)$ for $\mu^2<\Lambda_{\rm TC}^2$.  The criticality $\alpha_1(\mu^2=m_F^2)={\cal O} (\alpha_{\rm cr})$ for the
 infrared-weakened coupling determines
 the new scale of $(m_F)_{r_1} <(m_F)_{r_0}$.  As we continue increasing $N_F$, we get smaller $m_F$ accordingly,  eventually $m_F=0$ at certain critical $r=r_{\rm cr}=N_F^{\rm cr}/N_C$, and
 the large hirerarchy $m_F \ll \Lambda_{\rm TC}$ is realized near $r_{\rm cr}$.
 Beyond that point $r_{\rm cr} < r <11/2$, called conformal window, the chiral symmetry is not spontaneously broken, $m_F\equiv 0$. 
 This is depicted in Fig. \ref{alpha-beta-a-V}. 
 Then the ladder coupling is regarded as the CBZ IR fixed point in the anti-Veneziano limit:  
 \beq
 \alpha(x) =\alpha_* \theta (\Lambda_{\rm TC}^2 -x)\,.
 \label{ladderCBZ}  
 \eeq

\subsection{Solution of the ladder SD equation}
 
  Eq.(\ref{SDeq}) with the ladder coupling Eq.(\ref{laddercoupling}) is converted into a differential equation plus IR and UV boundary conditions \cite{Fukuda:1976zb}:
\beq
\left( x \Sigma(x) \right)'' 
+ \alpha \frac{3 C_2}{4 \pi} \frac{\Sigma(x)}{x + \Sigma^2(x)} &=& 0, 
\label{eq:diffSDo}\\
\lim_{x\rightarrow 0} x^2 \Sigma'(x) &=& 0,
\label{eq:IRBC}\\
\left. \left( x \Sigma(x) \right)'\right|_{x=\Lambda^2} &=& m_0.
\label{eq:UVBC}
\eeq
Since Eq.(\ref{eq:diffSDo}) is a nonlinear equation, the absolute value of the $\Sigma(x)$ is determined by the equation itself.
In order to have analytical insights,  however, we 
may linearlize  
Eq.~(\ref{eq:diffSDo}) by replacing 
$\Sigma(x)$ in the denominator of the second term in the left-hand side  
by a constant, $m_P$. 
Then the linearlized SD equation reads~\cite{Fomin:1984tv} 
\beq
\left( x \Sigma(x) \right)'' 
+ \alpha \frac{3 C_2}{4 \pi} \frac{\Sigma(x)}{x + m_P 
^2} &=& 0\,, 
\label{eq:diffSD}
\eeq 
where the absolute value of $\Sigma(x)$  is determined 
custormarily by 
\beq
m_P \equiv \Sigma(x=m_P^2)\, .
\label{eq:m_P}
\eeq

A solution of Eq.~(\ref{eq:diffSD}) which satisfies boundary 
condition Eq.~(\ref{eq:IRBC}) can then be expressed in terms of 
the hypergeometric function as~\cite{Fomin:1984tv} 
\beq
\Sigma(x) \, = \,\left( \xi\, m_P\right) \cdot {}_2 F_1\left( \frac{1+\omega}{2}, 
\frac{1-\omega}{2}, 2, -\frac{x}{m_P^2} \right), 
\label{hypergeometric}
\eeq
where
\beq
\omega \equiv \sqrt{1-\frac{\alpha}{\alpha_{\rm cr}}} \left(\alpha<\alpha_{\rm cr}= \frac{\pi}{3 C_2} \right),\quad  i\sqrt{\frac{\alpha}{\alpha_{\rm cr}}-1} =i\tilde \omega \quad \left(\alpha>\alpha_{\rm cr}\right)\,,
\label{omega}
\eeq
and $\xi$ is a numerical coefficient which is determined from the 
definition of $m_P$ in Eq.(\ref{eq:m_P}):
\beq
\xi =  
{}_2 F_1 \left( \frac{1+\omega}{2}, \frac{1-\omega}{2}, 2, -1 \right)^{-1} \simeq 1.1\,\, (\omega \simeq 0)\quad 1.0 \,(\omega \simeq 1).
\eeq 
In the limit of $x \gg m_P^2$, the solution can be expanded as
\beq
\Sigma(x) \, \simeq \, \xi\, m_P\, 
\left[ 
\ 
\frac{\Gamma(\omega)}{\Gamma(\frac{\omega+1}{2})\,\Gamma(\frac{\omega+3}{2}) } 
\left( \frac{x}{m_P^2} \right)^{\frac{\omega-1}{2}}
\ +\ \ 
(\omega \leftrightarrow -\omega)
\ 
\right].
\label{eq:Sigma}
\eeq

The bare chiral condensate of the technifermion, $\langle \bar F F\rangle_0\equiv \langle \bar F_i F_i\rangle_0$  (for a single flavor $i$ with no sum over $i$), is written in terms of the mass function $\Sigma(x)$ as
 \begin{eqnarray}
     \langle \bar F F\rangle_0 &=&- \frac{N_C}{4\pi^2}\int_0^{\Lambda^2} dy  \frac{y \Sigma(y)}{ y +\Sigma^2(y)}
 \label{condensate1}
   \end{eqnarray}
From Eq.(\ref{SDeq})  we have 
 \beq
  \Sigma(\Lambda^2) &=&m_0 + \frac{3 C_2 \alpha(\Lambda^2)}{4\pi} \frac{1}{\Lambda^2} \int_0^{\Lambda^2} dy  \frac{y \Sigma(y)}{ y +\Sigma^2(y)},\,
\eeq
 which yields a formula for the technifermion condensate in terms of the mass function at the cutoff $\Sigma(x=\Lambda^2)$~\cite{Bando:1987we}: 
  \beq     
  \langle \bar F F\rangle_0&=& \frac{N_C}{ 3C_2 \pi \alpha(\Lambda^2)} \,\Lambda^2 \left(m_0- \Sigma( \Lambda^2)\right)=\frac{N_C}{\pi^2} \left[ \frac{\alpha_{\rm cr}}{\alpha(\Lambda^2)}\right] \,\left[ \Lambda^2 \left( m_0- \Sigma( \Lambda^2) \right)\right] 
  = -\frac{N_C \alpha_{\rm cr}}{\pi^2} \frac{\Sigma'(x)}{\left(\frac{\alpha(x)}{x}\right)^\prime}\Bigg|_{x=\Lambda^2}\,. 
\label{condensate2}
 \eeq
 For the nonrunning coupling, the chiral condensate Eq.(\ref{condensate2}) reads:
  \beq     
  \langle \bar F F\rangle_0= \frac{N_C}{\pi^2} \left[ \frac{\alpha_{\rm cr}}{\alpha}\right] \,\left[ \Lambda^4\cdot \Sigma^\prime (x)|_{x=\Lambda^2} \right] .
  \label{condensate3}
 \eeq

 \subsection{The conformal phase $\alpha \leq \alpha_{\rm cr}$}
 \label{weak coupling}
 \subsubsection{The exact massless case: $m_0\equiv 0$}
 
Let us start with the weak coupling case
when the coupling is smaller than the critical coupling, $\alpha=\alpha_* <\alpha_{\rm cr}=\frac{\pi}{3 C_2}$. In the chiral limit
 $m_0\equiv 0$, the power-damping  solution with Eq.(\ref{eq:Sigma}) can satisfy  the UV boundary condition Eq.(\ref{eq:UVBC}) only by the trivial solution:
\beq
\Sigma (p) \equiv 0,\quad   \langle \bar F F\rangle_0=0, \quad (\alpha <\alpha_{\rm cr}, \quad   m_0\equiv 0).\,
\eeq
The chiral symmetry is not spontaneously broken, $ \langle \bar F F\rangle_0=0$, producing no mass parameter nor bound states (unparticle phase), in the chiral symmetry
limit, even though the scale symmetry is explicitly broken by the intrinsic scale $\Lambda$.
 In this case conformality persists within the ladder approximation, producing no bound states,  
 the situation characteristic to the ``conformal phase transition'' \cite{Miransky:1996pd}. This is the explicit example that the theory having intrinsic scale $\Lambda$ breaking
 the scale symmetry but has no mass. The same happens e.g., in the  
 NJL model,  
 where the scale symmetry is badly broken by the coupling characterized by the intrinsic scale $G\sim g/\Lambda^{D-2}$ but has no mass in the weak coupling $g<g_{\rm cr}$.

Although the coupling does not run $\alpha(\mu) \equiv \alpha$   ($\beta(\alpha) \equiv 0$) for $\mu <\Lambda$, 
there exists the explicit breaking of the scale symmetry due to $\Lambda$ 
corresponding to the intrinsic scale $\Lambda_{\rm TC}$ 
which is induced quantum mechanically by the regularization. So the scale symmetry is operative only for the energy region $\mu<\Lambda$ (IR conformal).
Such an explicit scale-symmetry breaking induced by the regularization manifests itself as the 
trace anomaly relevant even in the perturbation, see Eq.(\ref{pertanomaly}): $\langle \theta_\mu^\mu\rangle^{(perturbative)}=\frac{\beta(\alpha)}{4 \alpha(\mu^2)}\langle G_{\nu\lambda}^2(\mu^2)\rangle
= -{\cal O}(\Lambda^4)$.
Accordingly,  there exists no extra scale  
and so does no nonperturbative trace anomaly: 
\beq
\langle \theta_\mu^\mu\rangle^{(NP)} \equiv \langle \theta_\mu^\mu\rangle^{(full)} -\langle \theta_\mu^\mu\rangle^{(perturbative)} =0 \quad (\alpha<\alpha_c)\,.
\eeq

\subsubsection{Small explicit breaking: $m_0 (\ne 0) \ll \Lambda$} 
 If we introduce the explicit fermion mass $m_0 =m_0(\Lambda^2) \ne 0$ which is another source of the explicit breaking of the  scale symmetry in addition to
 $\Lambda (\gg m_0) $, then the exact IR confomality is gone and 
 physical states  including the bound states can appear,  with the  generic mass parameter $M$ solely due to $m_0\ne 0$, all of which are obeying  
 the typical hyperscaling relation \cite{DelDebbio:2010ze}, 
\beq
M \sim \Lambda 
\left(
\frac{m_0}{\Lambda}
\right)^{\frac{1}{1+\gamma_m}},\quad {\rm or} \quad m_0\sim M \left(\frac{M}{\Lambda}\right)^{\gamma_m} ,\,
\label{hyperscaling}
\eeq
where the $\gamma_m$ is the mass anomalous dimension.   
If $M$ is the renormalized mass of the fermion $m_R$, Eq.(\ref{hyperscaling}) takes the conventional form: $m_0=Z_m m_R$, with
the renormalization constant $Z_m =(m_R/\Lambda)^{\gamma_m}$. 

In fact 
 a nontrivial solution of the ladder SD equation, Eq.(\ref{eq:Sigma}),   
satisfies the UV boundary condition Eq.(\ref{eq:UVBC}):  
\beq
m_0 \, = \, \xi\, m_P\, 
\left[ 
\ 
\frac{\Gamma(\omega)}{\Gamma(\frac{\omega+1}{2})^2 } 
\left( \frac{\Lambda^2}{m_P^2} \right)^{\frac{\omega-1}{2}}
\ +\ \ 
(\omega \leftrightarrow -\omega)
\ 
\right],
\label{eq:m0_mP-Omega}
\eeq
 where $m_P=m_R$ 
 is now  the renormalized mass (or current mass) due to this explicit scale breaking mass $m_0$, with~\cite{Miransky:1984ef} 
\beq
Z_m \equiv \frac{m_0}{m_R}
=\, \xi \,
\left[ 
\ 
\frac{\Gamma(\omega)}{\Gamma(\frac{\omega+1}{2})^2 } 
\left( \frac{\Lambda^2}{m_R^2} \right)^{\frac{\omega-1}{2}}
\ +\ \ 
(\omega \leftrightarrow -\omega)
\ 
\right]\, , 
\label{Z_m}
\eeq 
or we have \cite{Leung:1985sn} 
\beq
 \gamma_m 
 = 
 \lim_{m_R/\Lambda \rightarrow 0} \frac{\partial \log Z_m}{\partial \log (m_R/\Lambda)} =
 \ 1-\omega \ 
=1 - \sqrt{1-\frac{\alpha}{\alpha_{\rm cr}}} \,,
\label{gamma-}
 \quad (\alpha<\alpha_{\rm cr})\,. 
\eeq
 For $\alpha \ll 1\,\, (\omega \simeq 1)$ it coincides with the perturbative one
$\gamma_m \simeq \frac{\alpha}{2\alpha_{\rm cr}}=\frac{3 C_2 \alpha}{2\pi}\simeq A/\ln (\Lambda^2/m_R^2) $ 
and $Z_m \simeq \left(\ln (\Lambda/m_R)\right)^{-A/2}$, with $A\simeq 18 C_2/ (11N_C -2 N_F)$.
For $\alpha\rightarrow \alpha_{\rm cr} \, (\omega \rightarrow 0)$, on the other hand, 
we have $\gamma_m \rightarrow 1$ and $Z_m \rightarrow \frac{2\xi}{\pi} \frac{m_R}{\Lambda}$.

The asymptotic solution Eq.(\ref{eq:Sigma}) takes the form 
\beq
\Sigma(x) \sim m_R \left(\frac{x}{m_R^2} \right)^{-\gamma_m/2} \quad\quad \left( \alpha<\alpha_{\rm cr}\right)\,,
\eeq
 which
is  consistent with the Operator Product Expansion (OPE). Such a nonzero  running mass 
is a genuine effect of the nonperturbative dynamics of the ladder
SD equation having a set of particular all order diagrams in the
conformal phase $\alpha<\alpha_{\rm cr}$  without SSB of the chiral symmetry.
Accordingly, the beta function after including the $m_R\ne 0$ effects would  no longer be a constant, although the
ladder coupling as a input is treated as  a constant: $\beta^{\rm (ladder)}(\alpha)=0$. 

Note that $m_0=m_0(\Lambda) \rightarrow 0$ as $\Lambda \rightarrow \infty$.
  Here we mention that the cutoff $\Lambda$ plays a crucial role to identify the solution of the SD equation~\cite{Maskawa:1974vs}, whether it is a  spontaneously broken solution or explicitly broken one:
The spontaneous chiral symmetry breaking solution with $\Sigma(x) \ne 0$ for $m_0\equiv 0$ exists only for the strong coupling $\alpha >\alpha_{\rm cr}=\pi/(3 C_2)$  in the presence of the cutoff $\Lambda<\infty$, while for weak coupling $\alpha <\alpha_{\rm cr}$ there exists only the explicit chiral symmetry breaking solution such that  
$\Sigma(x) \ne 0$ for $m_0\ne 0$ and $\Lambda<\infty$, with $m_0 \rightarrow 0 $  while the renormalized mass $m_R\ne 0$ for $\Lambda\rightarrow  \infty$. 
The explicit breaking solution would be confused with the spontaneous breaking,  if we took (erroneously)  $\Lambda \rightarrow \infty$ from the onset in the SD equation~\cite{Johnson:1964da}. See the discussions in Ref. \cite{Maskawa:1974vs}.

\subsection{The SSB phase $\alpha>\alpha_{\rm cr}$
}

Now we discuss the strong coupling phase, 
$\alpha> \alpha_{\rm cr}=\frac{\pi}{3 C_2}$ and $m_0\equiv 0$, where the nontrivial solution $\Sigma(x)\ne 0$ exists even at $m_0\equiv 0$, that is, the chiral symmetry is spontaneously broken, i.e.,  $\langle \bar F F\rangle_0\ne 0$. 
The SSB solution $\Sigma(x)$ in Eq.(\ref{hypergeometric}) with $\omega=i \tilde \omega$ in Eq.(\ref{omega}) takes the oscillating form \cite{Maskawa:1974vs,Fukuda:1976zb,Miransky:1984ef}
\begin{equation} 
 \Sigma(x) \simeq \xi \frac{m_F^2}{\sqrt{x}} \sqrt{ \frac{8\,{\rm cth}\frac{\pi \tilde \omega}{2}}{\pi \tilde \omega ({\tilde \omega}^2 + 1)}} 
\, \sin \left(\frac{ \tilde \omega}{2}  \ln \left( \frac{16 x}{m_F^2} \right)  -  \tilde \omega  \right)  \quad \quad\left(x\gg m_F^2 \right)\,,\quad \tilde \omega =\left(\frac{\alpha}{\alpha_{\rm cr}} -1\right)^{1/2} \,, \label{Sigma:sol}
\end{equation}
where we set the dynamical mass $m_F$ as $m_P=m_F$ such that $\Sigma(x=m_F^2)=m_F$, and $\xi = F(1/2,1/2,2:-1)^{-1} \simeq 1.1$. 
The oscillating solution can satisfy the
UV boundary condition Eq.(\ref{eq:UVBC}) for $m_0=0$:
\begin{eqnarray}
0= m_0 &\simeq& \left( x \Sigma (x) \right)^\prime|_{x=\Lambda^2}=\xi \frac{m_F^2}{\Lambda} \sqrt{
  \frac{8\, {\rm cth} \frac{\pi \tilde \omega}{2}}{\pi \tilde \omega}
}
\, \sin 
\left( \frac{\tilde \omega}{2}  \ln 
\frac{16 \Lambda^2}{m_F^2} -\tilde \omega + \tan^{-1} ( \tilde \omega) 
\right) 
\nonumber 
\\
&\simeq& \frac{ 4\xi}{\pi \tilde \omega} \frac{m_F^2}{\Lambda} 
\, \sin 
\left(
\frac{\tilde \omega}{2} \ln 
\frac{16\Lambda^2}{m_F^2}
\right)
\label{UVBCm0}
\end{eqnarray}
which is fulfilled by the vanishing phase, $\frac{\tilde \omega}{2}  \ln \left( \frac{16 \Lambda^2}{m_F^2}\right) \simeq n \pi \quad (n=1,2,3,\cdots)$, with $n=1$ being the ground state \cite{Miransky:1984ef}: 
 \beq
 m_F \simeq 4  \Lambda\cdot  \exp \left(-\frac{\pi}{\sqrt{\frac{\alpha}{\alpha_{\rm cr}}-1}}\right)  \quad  (\alpha \gtrsim \alpha_{\rm cr}=\frac{\pi}{3 C_2}=\frac{\pi}{3}\frac{2N_C}{N_C^2-1})\,.   
\label{Miranskyscaling} 
 \eeq
Then the technifermion acquires the dynamical mass $m_F$ in an essential-singularity (non-analytic) form (Miransky scaling, or the BKT transition) which implies a large hierarchy $m_F \ll \Lambda$ for $\alpha\simeq \alpha_{\rm cr}$, where the cutoff $\Lambda$
as a regulator may be regarded as the intrinsic scale $\Lambda_{\rm TC}$.

\subsubsection{Nonperturbative running (walking) coupling, with the IR fixed point as a UV fixed point}
\label{NPrunning:sec}

As we already mentioned in the Introduction, the Miransky-BKT scaling can create a large hierarchy, ``criticality hierarchy'', $m_F\ll \Lambda =\Lambda_{\rm TC}$ for $\alpha \simeq \alpha_*  \simeq \alpha_{\rm cr}$,  which dictates that the coupling no longer constant but
does depend on the $\Lambda/m_F$ as in Eq.(\ref{Miranskyscaling}), in such a way that the scale symmetry still remains approximately as the coupling is walking $\alpha(\mu^2) \simeq {\rm constant} $ for the wide region
$m_F^2 \ll \mu^2 <\Lambda_{\rm TC}^2$ as shown in  Eq.(\ref{NPbeta}): 
\begin{eqnarray}
\beta^{(NP)}(\alpha) &=&\Lambda \frac{\partial \alpha(\Lambda)}{\partial \Lambda}
 = - \frac{2\pi^2\alpha_{\rm cr}}{\ln^3 (\frac{4\Lambda}{m_F})} 
 \Rightarrow  - \frac{2\alpha 
 }{\pi} \left(\frac{\alpha}{\alpha_{\rm cr}}-1\right)^{\frac{3}{2}} \approx 
 0  \quad\quad  \left(< \,0 \right) 
  \nonumber \,,\\
\alpha(\mu) &=&
\alpha_{\rm cr}\left(1 + \frac{\pi^2}
{\ln^2\left(\frac{\mu}{\mu_{\rm IR}} 
\right)
}  \right) \approx 
\alpha_{\rm cr} \,, 
\label{NPbeta2}
\end{eqnarray} 
even when the perturbative coupling (input coupling)
is nonrunning,
$\beta(\alpha)|^{perturbative}\equiv 0$.  Here $\mu_{\rm IR} (\sim m_F/4)$ is given as 
$\ln (\mu/\mu_{\rm IR}) \simeq \ln (4\mu/m_F) [1+\pi^2/\ln^2(4\mu/m_F)]^{-1}$~
\footnote{
Solution of $\frac{\partial \alpha}{\partial \ln \mu}=\beta^{(NP)}(\alpha) $ is 
$\frac{1}{\pi} \ln \mu
= (\frac{\alpha}{\alpha_{\rm cr}}  -1)^{-1/2} + \tan^{-1}(\frac{\alpha}{\alpha_{\rm cr}}  -1)^{1/2} \simeq \frac{\alpha}{\alpha_{\rm cr}} (\frac{\alpha}{\alpha_{\rm cr}}  -1)^{-1/2}$. Or, $\alpha(\mu) \simeq \alpha_{\rm cr} \left(
1+ \frac{\pi^2}{\ln^2\mu} \left[1+\frac{\pi^2}{\ln^2\mu}\right]^2
\right)$.
}.

Note \cite{Yamawaki:2010ms,Hashimoto:2010nw} that the form of the beta function in Eq.(\ref{NPbeta2}) for $\alpha>\alpha_{\rm cr}$ has a multiple zero, 
and is in fact
not Taylor-expandable, with $\frac{d \beta(\alpha)}{d\alpha}|_{\alpha=\alpha_{\rm cr}} =0$ (without linear zero term), $|\frac{d^n \beta(\alpha)}{d \alpha^n}|_{\alpha=\alpha_{\rm cr}}|=\infty\,  (n\geq 2)$, reflecting the conformal phase transition of 
the Miransky-BKT  essential singularity scaling. This is  in sharp contrast to
the two-loop beta function Eq.(\ref{2loop}) having a Taylor expansion with the first term  of the linear zero at $\alpha=\alpha_*$: 
$\beta^{({perturbative})} \sim \alpha-\alpha_* +{\cal O} ((\alpha-\alpha_*)^2)$. Such a perturbative IR zero 
makes sense only for $\alpha(\mu) \ll \alpha_*\lesssim \alpha_{\rm cr}$ (deep conformal phase). 
Since the beta function should be continuous across
the critical point $\alpha_{\rm cr}$, it should be continued to the conformal phase $\alpha<\alpha_{\rm cr}$ with zero curvature.
 In the broken phase $\alpha_{\rm cr} <\alpha_*$  where $\alpha_*$ is washed out, the two-loop beta function is operative only for $\alpha(\mu) <\alpha_{\rm cr}$ in the far-ultraviolet region $\mu >\Lambda_{\rm TC}$, where the dynamics is irrelevant to the electroweak symmetry breaking.

  \begin{figure}[t]
\begin{center}
\includegraphics[width=6.5cm]{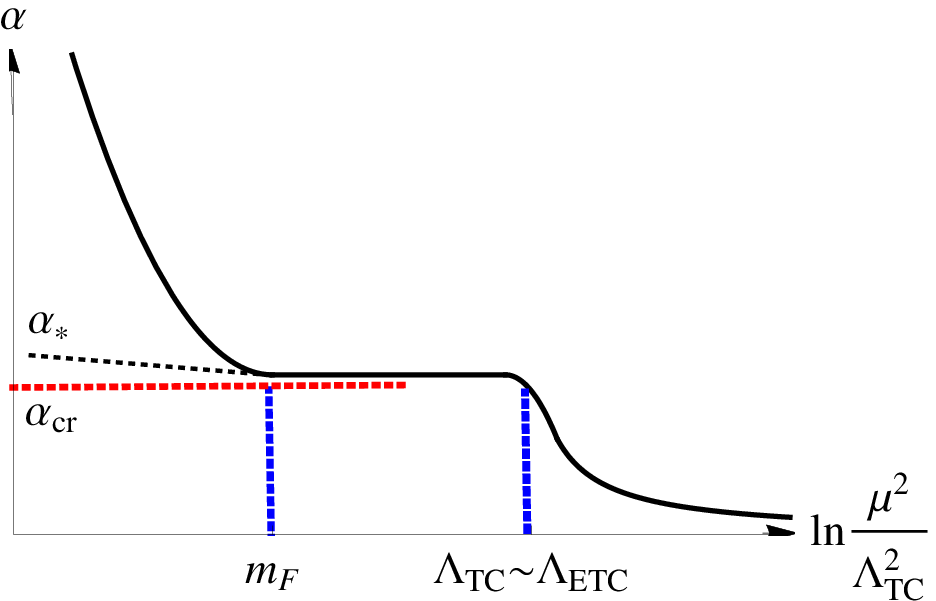} 
\hspace{15pt}
   \includegraphics[width=6.5cm]{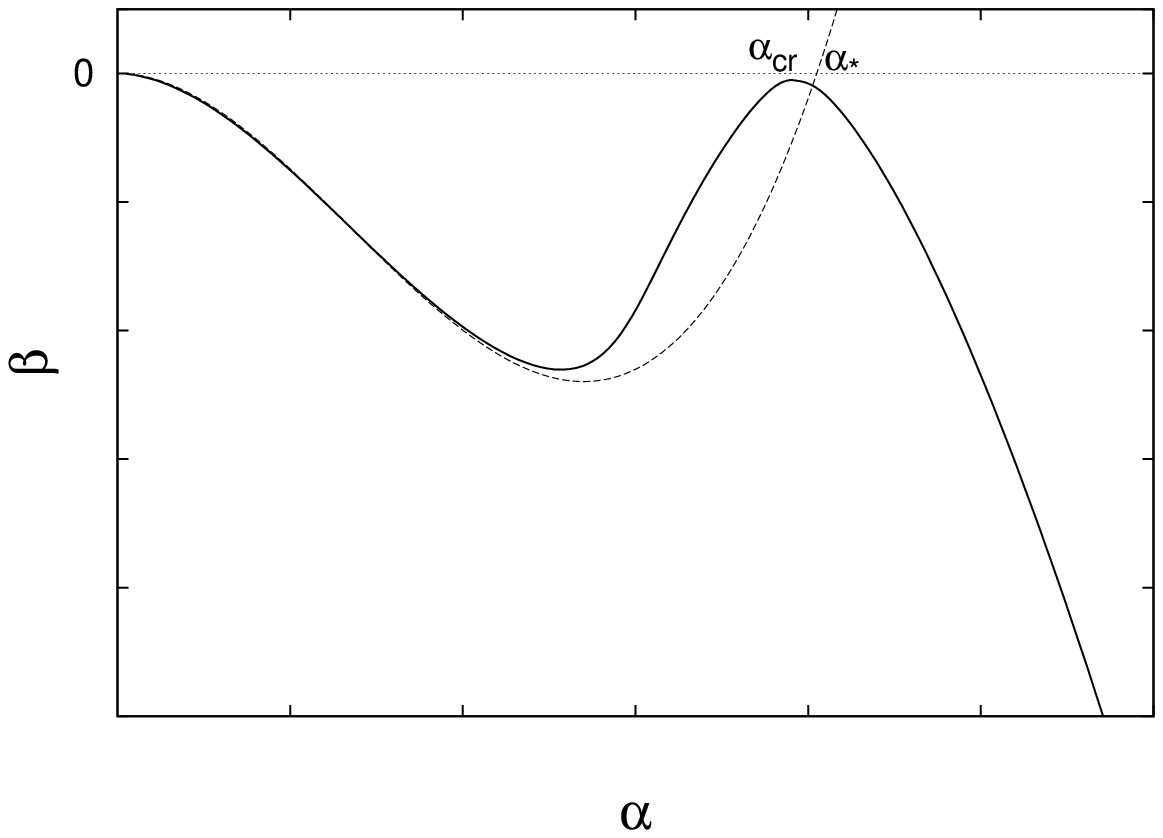} 
\vspace{15pt}
\caption{ Possible perturbative running coupling (left) and the beta function (right) in the region  $\alpha<\alpha_{\rm cr}$, in comparison with the nonperturbative region  $\alpha>\alpha_{\rm cr}$.
}
\label{beta:whole}
\end{center} 
 \end{figure}

 Since the critical coupling $\alpha_{\rm cr}$ behaves as the UV fixed point, the original ladder coupling as an ideal limit of the IR fixed point (viewed from the UV region $\mu^2>\Lambda_{\rm TC}^2$) in the anti-Veneziano limit may be identified with the UV fixed point viewed from the IR side $\mu^2<\Lambda_{\rm TC}^2$. Then  the effective coupling $N_C \alpha(\mu^2)$ keeps strong in IR region all the way up to
the intrinsic scale $\Lambda_{\rm TC}$ so that the anomalous dimension is very large in that  region. 
 Now 
the  would-be CBZ {\it IR fixed point $\alpha\simeq \alpha_* \simeq \alpha_c$ is  
regarded as 
the UV fixed point} of the nonperturbative running (walking) coupling $\alpha(\mu) \approx \alpha_c$ 
in the wide IR region $m_F < \mu < \Lambda=\Lambda_{\rm TC}$ for the characteristic large hierarchy $m_F \ll \Lambda_{\rm TC}$. See 
Fig.\ref{beta:whole}~\cite{Yamawaki:2007zz,Hashimoto:2010nw} 
(See also Ref.\cite{Kaplan:2009kr} for a similar observation.). 
This is the essence of the WTC.
The new scale $m_F$ (denoted as $\Lambda_{\rm TC}$ in Ref.~\cite{Yamawaki:1985zg}, which should not be confused with 
$\Lambda_{\rm TC}$ in this paper) is regarded as the second RG-independent quantity as, 
\beq
m_F =4 \mu\cdot  \exp \left(-\int^{\alpha(\mu)} \frac{d \alpha}{\beta^{(NP)}(\alpha)}\right) \simeq 4\Lambda_{\rm TC} \cdot
\exp \left(-\frac{\pi}{\sqrt{\frac{\alpha(\Lambda_{\rm TC})}{\alpha_c}-1}}\right)  \ll \Lambda_{\rm TC}\,, 
\eeq
 with $\beta^{(NP)}(\alpha)$ given in Eq.(\ref{NPbeta}).  Compare it with Eq.(\ref{perturbativeDT}).
 
On the other hand, in the UV region $\mu>\Lambda_{\rm TC}$ ($\alpha(\mu) <\alpha_c \simeq \alpha_*$), the coupling runs as the usual perturbative asymptotically free theory: 
$\alpha(\mu) \sim 1/\ln \mu$. See Fig.\ref{beta:whole}.  
Such a perturbative region $\alpha<\alpha_c$ is actually irrelevant to the physics of WTC, since the theory is expected to become only a part of more fundamental (unified) theory including the SM sector, say, the ETC \cite{Dimopoulos:1979es} or technicolored composite model \cite{Yamawaki:1982tg}.

Incidentally, the original setting  of the WTC~\cite{Yamawaki:1985zg} was an asymptotically 
non-free theory with small perturbative beta function $1\gg \beta(\alpha) >0$, as in the  technicolored preon  model \cite{Yamawaki:1982tg} where the technifermions as well as the quarks and leptons are composites on the same footing and the technicolor gauge at composite level is asymptotically non-free  in the perturbative sense due to the
formation of many composite technifermions (though the technicolor at the preon level is asymptotically free). This perturbative setting makes sense 
only in the weak coupling phase $\alpha<\alpha_c$.  
On the other hand,  in the strong coupling phase  $\alpha>\alpha_{\rm cr}$ ($\mu<\Lambda=\Lambda_{\rm TC} \sim \Lambda_{\rm ETC}$ (or $\Lambda_{\rm composite}$), both the asymptotically-free theories
with the CBZ IR fixed point and the asymptotically non-free theories yield the same nonperturbative beta function Eq.(\ref{NPbeta}), i.e., Eq.(5) and Fig.1(a) of Ref.~\cite{Yamawaki:1985zg}, having a UV fixed point at $\alpha=\alpha_c$, which is only the physical issue of the 
WTC.  In fact,
in the asymptotically non-free theory with  the perturbative coupling growing function of $\mu$ in units of the Landau pole $\Lambda=\Lambda_{\rm Landau}=\Lambda_{\rm Composite}$, the ladder SD equation tells us that the dynamical mass $m_F$ is generated as a scale when the coupling exceeds the critical coupling $\alpha(\mu=m_F) >\alpha_{\rm cr}$.
Then, in contrast to the infrared-free phase $\alpha<\alpha_{\rm cr}$ of the asymptotically-free theory (Coulomb phase), 
the physics in the strong coupling phase  is precisely the same as the WTC in the anti-Veneziano limit of the asymptotically free theory, with only exception being that 
 the $\Lambda$ in the Miransky scaling Eq.(\ref{Miranskys}) should now read the Landau pole scale $\Lambda_{\rm Landau}=\Lambda_{\rm Composite}$ 
(``compositeness condition''\cite{Bardeen:1989ds}, to be identified with the composite scale in the technicolored preon model
 \cite{Yamawaki:1982tg}, to generate the effective four-fermion interactions in Eq.(\ref{four-fermions})) instead of the intrinsic scale of the
 asymptotically-free theory.
 From the model building point of view, 
it does not make sense \cite{Bando:1987we} whether the WTC in isolation is asymptotically free or nonfree in the region, $\alpha<\alpha_{\rm cr}$ ($\mu>\Lambda=\Lambda_{\rm TC} \sim \Lambda_{\rm ETC}(\Lambda_{\rm Composite})$), where the theory is already
changed into a more fundamental unified theory, ETC or peon theory both being asymptotically-free anyway.

\subsubsection{Large anomalous dimension $\gamma_m =1$ and enhanced chiral condensate }
    \label{Anomalousdim}

    Eq.(\ref{Sigma:sol}) together with Eq.(\ref{UVBCm0}) yields the asymptotic form of $\Sigma(x)$ 
    at $m_F^2 \ll x \lesssim \Lambda^2= \frac{m_F^2}{16}\exp (\frac{2 \pi}{\tilde \omega})$~\cite{Yamawaki:1985zg}: 
\beq
    \Sigma(x) \simeq \xi \frac{m_F^2}{\sqrt{x}} \frac{4}{\pi \tilde \omega} \sin \left(\frac{\tilde \omega}{2} \ln (16x/m_F^2)  -\tilde \omega  \right)
    \simeq  \xi \frac{m_F^2}{\sqrt{x}} \frac{4}{\pi \tilde \omega} \sin \left( \pi -\tilde \omega  \right)\simeq \frac{4 \xi}{\pi} \frac{m_F^3}{x} \left(\frac{x}{m_F^2}\right)^{1/2}\,,
       \label{asym}
\eeq  
 where the logarithmic $x$-dependence is absent for the region
    $\frac{\tilde \omega}{2} \ln (16x/m_F^2) \sim \frac{\tilde \omega}{2} \ln (16\Lambda^2 /m_F^2) \simeq \pi$.    
In the proposal of the WTC \cite{Yamawaki:1985zg,Bando:1986bg}, this asymptotic form $\Sigma (x) \sim \frac{m_F^3}{x} \left(\frac{x}{m_F^2}\right)^{1/2}$   was identified with 
the OPE of $\Sigma(x)$ 
at $m_F^2\ll x\lesssim \Lambda^2$~\footnote{ The ladder SD solution  with respect to OPE was also discussed in \cite{Cohen:1988sq} in a way somewhat different  
  than Refs.~\cite{Yamawaki:1985zg,Bando:1986bg}, concerning the logarithmic dependence. The log peculiarity is just on the point $\alpha\equiv \alpha_{\rm cr}$ where
  no SSB takes place. Absence of log in the SSB phase  is consistently  seen in Eq.(\ref{barecondensate}) and Eq.(\ref{Z_mladder}).
  See also the OPE, Eq.(\ref{OPEfull}).
  }: 
 \beq
\Sigma(x) \sim \, \frac{m_F^3}{x} 
\left(\frac{x}{m_F^2}\right)^{\gamma_m/2}, 
\label{OPE}
 \eeq    
to conclude  a large anomalous dimension
in the SSB phase near criticality (UV fixed point)~\cite{Yamawaki:1985zg}:  
   \beq
  \gamma_m 
  = 1\,  \quad \quad \left( \tilde \omega =\sqrt{\frac{\alpha}{\alpha_{\rm cr}}-1} =\frac{\pi}{\ln \frac{4\Lambda}{m_F}} 
  \simeq 0\right)\,. 
  \label{anomdim}
  \eeq
The large anomalous dimension $\gamma_m=1$ in the SSB phase  was also compared with the anomalous dimension Eq.(\ref{gamma-}) in the conformal phase ($\alpha<\alpha_{\rm cr}$) at criticality: $\gamma_m=1-\sqrt{1-\alpha/\alpha_{\rm cr}} \rightarrow 1 $ ($\alpha\rightarrow \alpha_{\rm cr}$-0) (see Eqs.(6) and (7) and Fig. 1(b)  of Ref. \cite{Yamawaki:1985zg} ).

The ladder result, $\gamma_m =1$, in Eq.(\ref{anomdim}) 
is a direct consequence of the scale-symmetric strong dynamics 
 first found in the ladder SD equation in the proposal of WTC  \cite{Yamawaki:1985zg} as a solution of the
 FCNC problem of the original TC as a simple scale-up of the QCD \cite{Weinberg:1975gm}.   Before advent of the WTC, 
  a large anomalous dimension of the TC dynamics $\gamma_m\gtrsim 1$ was anticipated \cite{Holdom:1981rm} (see also \cite{Georgi:1981xw,Yamawaki:1982tg})  to enhance the
  bare condensate by the factor $Z_m^{-1}=(\Lambda/m_F)^{\gamma_m}$, as a solution of the  
 FCNC problem,   
 based on the pure assumption of the UV fixed point at strong coupling. 
 
 Masses of the quarks/leptons are generated through communication between quarks/leptons $\psi$ and the technifermions $F$ through extra dynamics such as the ETC \cite{Dimopoulos:1979es}, 
 or the technicolored peon  model \cite{Yamawaki:1982tg} (quarks, leptons and technifermions are composites on the same footing), 
 which generically  give effective four-fermion interactions: 
 \beq
G_a \left(\bar \psi \psi \right)^2\,, \, \frac{G_b}{N_C} \left(\bar F F\right)^2\,,  \,\frac{G_c}{N_C} \,\left(\bar \psi \psi\right)\left( \bar F F\right)\,,
\label{four-fermions}
\eeq
where the three types of  four-fermion couplings $G_{a,b,c}={\cal O}\left( \frac{a,b,c}{\Lambda^2}\right)$ are on the same order of magnitude 
characterized by the scale of the extra dynamics $\Lambda=\Lambda_{\rm ETC}\sim \Lambda_{\rm TC}$, except  for the numerical factors $a,b,c={\cal O} (1)$ depending on the explicit model, and the factor $1/N_C$ for $G_b, G_c$ is the effect of  the Fierz transformation from the current $\times$ current four-fermion coupling from the ETC gauge exchanges.
While $G_a$ yields FCNC, $G_c$ yields the quark/lepton mass: 
 \beq
 m_{q/l} =-\frac{G_c}{N_C} \, \langle \bar F F \rangle_0 \sim - \frac{c\, Z_m^{-1} }{ \Lambda^2} \frac{\langle \bar F F \rangle_R}{N_C} 
 \simeq c \frac{m_F^2}{\Lambda}\,,
  \quad 
 Z_m^{-1} 
 \sim 
 \frac{\Lambda}{m_F}
 \label{qlmass}
 \eeq
 where  $ \langle \bar F F \rangle_R/N_C =- {\cal O}(m_F^3)=- {\cal O} ({\rm \frac{1}{3} \,TeV})^3$ is the condensate renormalized at $\mu=m_F$, and 
$\Lambda_{\rm ETC} \sim 10^3 \, {\rm TeV}$, 
thus arriving at the typical order of quarks/leptons mass
 (except for the top quark) :
 $m_{q/l} \sim 0.1 \, {\rm GeV}$. 
 This is in sharp contrast to the ordinary QCD 
 with $Z_m^{-1} \sim \left(\ln (\Lambda/m_F)\right)^{A/2} ={\cal O} (1)$ and
 $\gamma_m \simeq \frac{3C_2 \alpha}{2\pi} \simeq A/\ln (\Lambda^2/m_F^2) \approx 0$ with $A= 18 C_2/(11 N_C- 2N_F) (<1)$:
 \beq
  m_{q/l} =G_c \, \langle \bar F F \rangle_0 \sim \frac{c }{\Lambda^2} \langle \bar F F \rangle_R \sim 0.1 \, {\rm MeV}\,,\quad  \langle \bar F F\rangle_0 
    ={\cal O}  \left(\langle \bar F F\rangle_R\right)\,.
 \eeq

In order to keep track of the concrete analytical expression of the ladder results (in a linearized version Eq.(\ref{eq:diffSD})),  we here list  results of the precise (linearized) ladder computation of the chiral condensate  $\langle \bar F F \rangle_0$,   using the explicit form of the SSB solution Eq.(\ref{Sigma:sol}), based on  Eq.(\ref{condensate2}) and/or (\ref{condensate3}).  (For details see Appendix A.)

The bare condensate and the mass renormalization constant $Z_m=m_0/m_R$ take the form in agreement  with Ref.~\cite{Miransky:1989qc}: 
\begin{eqnarray}
\langle \bar F F\rangle_0 &=&- \frac{\xi N_C}{\pi^2} \frac{\alpha_{\rm cr}}{\alpha(\Lambda^2)} \Lambda^2 \Sigma(\Lambda)
  \simeq - \frac{4\xi  N_C }{\pi^3} m_F^2\, \Lambda \,,
  \label{barecondensate}\\    
Z_m &=&\frac{m_0}{m_R}\simeq \frac{2\xi}{\pi} \frac{m_F}{\Lambda}\,,
\label{Z_mladder}\\
 \langle \bar F F\rangle_R&=& Z_m \langle \bar F F\rangle_0\simeq -\frac{8 \xi^2  N_C }{\pi^4}\, m_F^3\,.
 \label{Rcondensate}
\end{eqnarray}
Thus
the asymptotic form of $\Sigma(x)$ ($m_0\ne 0$) in Eq.(\ref{Sigma:sol}) with $m_P=\Sigma(x=m_P^2)\simeq m_F +m_R\, (m_R\ll m_F)$ is perfectly consistent with the OPE for $x$ such that $\tilde \omega \ln \left(\frac{16x}{m_F^2} \right) \simeq \pi$:
\begin{eqnarray}
\Sigma(x) &\sim& \frac{4\xi }{\pi}\frac{1}{ \tilde \omega} \frac{m_P^2} {\sqrt{x}} \sin \left( 
\frac{\tilde \omega}{2} \ln \left(\frac{16x}{m_P^2} \right) -\tilde \omega
\right) \simeq \frac{4\xi }{\pi}\frac{1}{ \tilde \omega} \frac{m_P^2} {\sqrt{x}} \sin \left( 
\frac{\tilde \omega}{2} \ln \left(\frac{16x}{m_F^2} \right) -\tilde \omega \ln (1-\frac{m_R}{m_F}) -\tilde \omega
\right)
\nonumber \\
 & \simeq&  \frac{4\xi }{\pi} \left[ \frac{m_F\, m_R} {\sqrt{x}} 
 +  \frac{m_F^2} {\sqrt{x}}  \right]
\nonumber \\
 & \sim& \frac{4\xi}{\pi}  m_R \left(\frac{x}{m_F^2}\right)^{- \gamma_m/2} 
  - \frac{\pi^3}{2\xi N_C} \frac{\langle \left(\bar F F\right)_R
  \rangle}{x} \left(\frac{x}{m_F^2}\right)^{\gamma_m/2 }\,.
  \label{OPEfull}
  \end{eqnarray} 
Combining Eq.(\ref{barecondensate})  with  the Pagels-Stokar formula Eq.({\ref{PSv}), $F_\pi^2 = \frac{N_C\xi^2}{2\pi^2} m_F^2$, or $m_F^2 \simeq \frac{4\pi^2}{\xi^2} \frac{1}{N_F N_C} \, v_{\rm EW}^2$,  we have 
for $\Lambda=\Lambda_{\rm TC} \sim \Lambda_{\rm ETC}$:
\beq
m_{q/l} = \frac{c}{N_C} \frac{\langle \bar F F\rangle_0}{\Lambda_{\rm ETC}^2}=
y^{\rm eff} \cdot v_{\rm EW},\quad  y^{\rm eff} =\frac{c}{N_FN_C} {\cal O} \left(\frac{4}{\xi \pi } \frac{4 v_{\rm EW} }{\Lambda_{\rm ETC}}\right) ={\cal O} \left( 10^{-3}\right)\,.
\label{efffectiveyukawa}
\eeq

\subsubsection{ Large Anomalous dimension and amplification of the  symmetry violation}

A striking feature of  the WTC having the large anomalous dimension $\gamma_m=1$ is  that the explicit symmetry breaking by a small Lagrangian parameter is 
enhanced by the strong dynamics near the criticality being persistent all the way up to the intrinsic scale $\Lambda_{\rm TC}$. The quark/lepton mass enhancement
already discussed is a typical such example: Such masses come from formally the small explicit breaking of the SM fermion chiral symmetry  by the small ETC gauge coupling, $g_{\rm ETC}$, leading to the
small  ETC-induced four-fermion coupling $G_c \sim g_{\rm ETC}^2 /M_{\rm ETC}^2\sim c/\Lambda_{\rm ETC}^2 (\ll 1/m_F^2) $, where $M_{\rm ETC}\sim g_{\rm ETC} \Lambda_{\rm ETC} $ is the ETC gauge boson mass generated by the SSB of the ETC gauge symmetry down to the WTC, with  the order parameter $v_{\rm ETC}$ of the SSB of the ETC gauge symmetry being $v_{\rm ETC} \sim \Lambda_{\rm ETC}$. Though the coupling is small, the resultant mass is amplified by the walking dynamics with $Z_m^{-1} \simeq \Lambda/m_F >10^3$, as we discussed in the above.

Here we briefly comment on yet another quantity subject to this enhancement effects due to the large anomalous dimension.
It is the technipions mass, another phenomenological issue of the generic WTC. The technipions are the left-over (pseudo) NG bosons besides the (fictitious) NG bosons absorbed into
SM gauge bosons. They exist in a large class of the WTC having large $N_F\, ( >2)$ and will be a smoking gun of this class of WTC in the future LHC. 

Technipion mass is all from explicit breaking outside of the WTC sector, i.e, SM gauge interactions and ETC gauge interactions ($G_b$ terms in Eq.(\ref{four-fermions})):
The estimation of the masses of 
the technipions in the WTC  is done, based on the first order perturbation of the explicit chiral symmetry breaking by the ``weak gauge couplings'' of SM gauge interactions and the ETC gauge interactions (Dashen's formula), 
\begin{eqnarray}
M_{\pi}^2({\rm SM}) &\sim& 
 \frac{C^{\rm SM}_2 \, \alpha_{\rm SM}}{4\pi F_\pi^2} 
\int d x  \left( \Pi_V(x)_V - \Pi_A(x) \right) \\
M_{\pi}^2({\rm ETC}) &\sim&  \frac{1}{F_\pi^2} \frac{\alpha_{\rm ETC}}{M_{\rm ETC}^2} \, \langle \frac{1}{N_C}
(\bar F F) 
\, (\bar F F) \rangle_0
\sim  \frac{1}{F_\pi^2} \frac{b}{\Lambda^2} \frac{1}{N_C}\, \left(\langle \bar F F \rangle_0\right)^2 
\end{eqnarray}
up to Clebsh-Gordan coefficient depending on the detailed model, where $\Pi_{V,A}(x)$ are current correlators of vector and axialvector currents.
This is the same strategy as the QCD estimate of the $\pi^+ - \pi^0$ mass difference, where the explicit chiral symmetry breaking is given by the QED lowest order coupling, while the full QCD nonperturbative contributions are estimated through the current correlators by various method like ladder, holography, lattice, etc..   
 
It is obvious that $M_{\pi}^2({\rm ETC})$ is enhanced through  the condensate by the anomalous dimension as $(Z_m^{-1})^2 \sim (\Lambda/m_F)^{2 \gamma_m}$, as  was noted before the advent of the WTC~\cite{Holdom:1981rm,Yamawaki:1982tg}, and was confirmed in the WTC with $\gamma_m=1$ based on the concrete scale-invariant dynamics, the ladder SD equation~\cite{Yamawaki:1985zg}.
  $M_{\pi}^2({\rm SM}) $ is also enhanced by the large anomalous dimension $\gamma_m=1$~\cite{Harada:2005ru},
since the high energy behavior is slower damping by the anomalous dimension $\Pi_V(x)- \Pi_A(x)|_{x>m_F^2}  \sim \alpha(x) \frac{\langle \bar F F\rangle_R^2}{x^2} \left(\frac{x}{m_F^2}\right)^{\gamma_m}
\sim \frac{N_C m_F^4}{x} $ (A similar observation was made without notion of the anomalous dimension \cite{Holdom:1987ed}). 
Then we have a large mass for the technipions~\cite{Harada:2005ru, Jia:2012kd,Kurachi:2014xla}: 
\beq
M_{\pi}^2({\rm ETC})\sim 2\pi^2 b\,  m_F^2={\cal O} \left( ({\rm TeV})^2    \right)\,, 
\quad \left[M_{\pi}^2({\rm SM})\right]_{x>m_F^2}\sim \left(C^{\rm SM}_2 \alpha_{\rm SM}\right) m_F^2\,
\ln\left(\Lambda^2/m_F^2\right) \lesssim 
({\rm TeV})^2
\,, 
\label{TCpionmass}
\eeq
where the Pagels-Stokar formula Eq.(\ref{PS}) is used~\footnote{
Note that $M_{\pi}^2({\rm SM})$ has also IR contributions from $x\lesssim m_F^2$, which is less than the UV contributions as far as the (techni-sector) $S$ parameter  is large $S>0.3$, thus
$M_\pi({\rm SM})^2=\left[M_{\pi}^2({\rm SM})\right]_{x>m_F^2} + \left[M_{\pi}^2({\rm SM})\right]_{x<m_F^2} < {\cal O} ((1.5 {\rm TeV})^2)$,  
 in somewhat tension with the present LHC limit for the colored technipions. (The techni-sector $S$ parameter can be cancelled by the ETC sector contribution 
 to be consistent with the $S$ parameter value constrained by the
 precision experiments.) 
 A possible way out besides the ETC cancellation would be the strong gluon condensate which has not been incorporated into the ladder
SD approach but has been shown in the holography to dramatically enhance the infrared part $\left[M_{\pi}^2({\rm SM})\right]_{x<m_F^2} $, in accord with the suppression of the 
techni-sector $S$ parameter  $S <0.1$~\cite{Kurachi:2014xla}. This gluonic effect enhances $M_{\pi}^2({\rm ETC})$.
}.

Striking fact is that 
although the explicit chiral symmetry breakings are formally
very small due to the ``weak gauge couplings'',  the nonperturbative contributions from the WTC sector lift all the technipions masses to the TeV region so that they all lose the nature of the ``pseudo NG bosons''.   
This is actually a universal feature of the dynamics with large anomalous dimension, ``amplification of the symmetry violation'' \cite{Yamawaki:1996vr},  as dramatically shown in the top quark condensate model \cite{Miransky:1988xi}, 
based  on the gauged NJL model with large anomalous dimension $\gamma_m \simeq 2$ \cite{Miransky:1988gk}. 

This amplification effect should not be confused with that of  the pseudo NG boson mass due to the technifermion bare mass effects, like the pion mass due to the current quark mass,
$F_\pi^2 m_\pi^2 = 2 m_0\langle \bar \psi \psi \rangle_0=2 m_R \langle \bar \psi \psi \rangle_R$, which are not amplified by the large anomalous dimension, 
since the bare mass operator as the explicit breaking is the RG invariant, $m_0 (\bar F F )_0= m_R (\bar F F)_R $,  
 and hence is ignorant about the anomalous dimension within the WTC sector. In the actual technicolor model, all the technifermions are set to be
 exactly massless and such a type of explicit breaking is not considered anyway.

Note that although the left-over light spectra are just three exact NG bosons absorbed into $W/Z$ bosons, 
our theory with $N_F \gg 2$ in the anti-Veneziano limit is completely  different from the model with massless flavors $N_f=2$ where the symmetry  breaking is  $SU(2)_L \times SU(2)_R/SU(2)_V$. In fact 
even though all the NG bosons, other than the three exact NG bosons to be absorbed into $W,Z$ bosons, are massive and decoupled from the low energy physics, 
they are composite of the linear combinations of all the $N_F$
technifermions not just 2 of them. 

In fact,  
the technifermions do not acquire  the explicit mass from these explicit breaking terms, and hence the walking behavior of the coupling of the large $N_F$ in the anti-Veneziano limit is not drastically changed.  
They actually get some effects on the dynamical masses, as a result of the vacuum alignment including the extra gauge interactions, which  are to be treated as the corrections to the ladder SD equation including 
not only the WTC gauge coupling but also the SM gauge interactions, with the modified criticality $C_2^{\rm WTC} \alpha^{\rm WTC} + C_2^{\rm SM} \alpha^{SM}>\frac{\pi}{3}$, and the ETC gauge interactions as corrections to the
WTC gauge interaction in the ladder kernel in a form of the four-fermion couplings $G_{b,c}$ in Eq.(\ref{four-fermions}). 

While $G_c$ is in general (except for the top quark) a small 
feedback of the SM fermion condensate 
to the technifermion condensate in the coupled SD equation, $G_b$ is potentially strong effects on the phase structure in
a
way that the critical coupling is replaced by the critical line (surface) of the two-dimensional (multi-dimensional) coupling space, $(\alpha, g)$, with $g= \frac{N_C}{4\pi^2}  \Lambda^2 G_b$ in the
 gauged NJL model \cite{Kondo:1988qd, Yamawaki:1988na}, as analyzed with the kernel having extra contributions of the SM (running) gauge couplings and
ETC-induced four-fermion interaction (strong ETC technicolor).
 In that case, the SSB solution of the SD equation exists even for the weak gauge coupling $\alpha<\alpha_{\rm cr}$ because of the additional strong NJL four-fermion coupling ($\alpha\rightarrow 0$ is the NJL limit)\footnote{
The NJL four-fermion coupling can be treated effectively as if a strong asymptotically nonfree gauge coupling  $\alpha(x) = g \frac{x}{\Lambda^2} \theta(\Lambda^2-x)$ in the improved ladder kernel in Eq.(\ref{SDeq}).    
}. 
 The result shows drastic effects with 
 the anomalous dimension even larger, $1< \gamma_m =1+\sqrt{1-\frac{\alpha}{\alpha_{\rm cr}}} <2$ \cite{Miransky:1988gk} at the critical line. This is particularly useful for reproducing the top quark mass which would need more enhancement than other quarks due to such a large anomalous dimension \cite{Miransky:1988gk,Matumoto:1989hf}. See the discussions in the last section.
 
In fact, 
the SSB solution   takes the form instead of the Miransky scaling:
  \beq
 m_F^{2\omega} = \Lambda^{2\omega} \left(
\frac{g-g_{\rm cr}^{(+)}}{g-g_{\rm cr}^{(-)}} 
 \right) \,,
 \quad g_{\rm cr}^{(\pm)}=\frac{1}{4} \left(1\pm \omega\right)^2\,, \quad \omega \equiv \sqrt{1-\frac{\alpha}{\alpha_{\rm cr}}}\quad \left(0<\alpha<\alpha_{\rm cr}\right)\,,
 \eeq
 and the anomalous dimension~\cite{Miransky:1988gk, Kondo:1991yk}: 
  \beq
  \gamma_m =2 g +\frac{\alpha}{2\alpha_{\rm cr}}\,, \quad
 \gamma_m^{(\pm)} =  \gamma_m\Bigg|_{g=g_{\rm cr}^{(\pm)}}=1\pm \omega= 1 \pm \sqrt{1-\frac{\alpha}{\alpha_{\rm cr}}} \,,
 \label{gaugedNJLgamma}
  \eeq
 where the critical line $g=g_{\rm cr}^{(+)}$ behaves as a UV fixed point, while the non-critical line $g=g_{\rm cr}^{(-)}$  an IR fixed point:
 \beq
 \beta^{NP)}(g)=\frac{\partial g}{\partial \ln \Lambda}\Bigg|_{\alpha, m_F}= - 2\left(g-g_{\rm cr}^{(+)}\right) \left(g-g_{\rm cr}^{(-)} \right)\,.
  \eeq
 The nonperturbative running coupling near the UV fixed point  $g=g_{\rm cr}^{(+)}$ is given as $g(\mu) = g_{\rm cr}^{(+)}(1 +\frac{m_F^{2\omega}}{\mu^{2\omega}-m_F^{2\omega}})$ for $g>g_{\rm cr}^{(+)}\, (\mu>m_F) $ and $g(\mu) = g_{\rm cr}^{(+)}(1 -\frac{m_F^{2\omega}}{\mu^{2\omega}+m_F^{2\omega}})$
 for $g<g_{\rm cr}^{(+)}  $.  At $\alpha\rightarrow \alpha_{\rm cr}$, the fusion of the UV and IR fixed points takes place: $g_{\rm cr}^{(+)}= g_{\rm cr}^{(-)} =1/4$, and hence
 $ \beta^{NP}(g)= - 2 (g-g_*)^2$ ($g_*=1/4$) ~\cite{Kondo:1991yk,Aoki:1999dv}. This beta function again has a multiple zero but not a simple zero at UV=IR fixed point, 
 with essential singularity scaling $m_F^2 =\Lambda^2 \exp(-1/(g-g_*))$,  similarly
 to conformal phase transition  at $\alpha_*=\alpha_{\rm cr}$ in the walking gauge theory without four-fermion coupling \cite{Yamawaki:2007zz,Hashimoto:2010nw,Kaplan:2009kr} (See the next subsection).

  The outstanding feature of the gauged NJL model with $\alpha\ne 0, \omega<1$ is the renormalizability
(in the sense of nontriviality, or no Landau pole)\cite{Kondo:1991yk}, when the gauge coupling is walking, $\alpha(\mu^2) \approx {\rm const.}$, with 
the four-fermion interaction having the full dimension $2< D = 2(3-\gamma_m) = 4-2\omega <4$ (relevant operator, or super renormalizable) , including $D\simeq 2(1+ A/\ln \mu^2)>2$
with a  moderately walking 
small coupling $\omega \simeq 1 -\frac{\alpha}{2\alpha_{\rm cr}} \simeq 1-\gamma_m \,\,(\gamma_m (\mu) \sim A/\ln \mu^2) $ with $A=18 C_2/(11 N_C -2 N_F) >1$, in sharp contrast to the pure (non-gauged) NJL model which is  a trivial theory having a Landau pole.

\subsection{Conformal Phase Transition}
\label{ConformalPT}

We now discuss a salient feature of the phase transition at $\alpha=\alpha_{\rm cr}$, what we call Conformal Phase Transition \cite{Miransky:1996pd}, which is characterized 
by the Miransky-BKT type non-analytic scaling. Let us first discuss the exact chiral limit $m_0\equiv 0$.

 In the conformal phase  $\alpha \leq \alpha_{\rm cr}$, there is no bound state (dubbed ``unparticle'' phase), although there is a UV scale 
$\Lambda$ which is identified with the intrinsic scale $\Lambda_{\rm TC}$ generated quantum mechanically (already by the perturbation) by the regularization as manifested in a form of the (perturbative) trace anomaly. It should be emphasized that just on the critical point $\alpha=\alpha_{\rm cr}$ 
the SSB does not take place in the same way as for $\alpha<\alpha_{\rm cr}$, and hence there are no bound states at all. In fact the solution of the ladder SD equation at $\alpha=\alpha_{\rm cr}$ takes the asymptotic form at $x \gg m_F^2$:
\begin{equation} 
  \Sigma(x) =\xi\cdot {}_2F_1 (1/2,1/2.2; -x) \sim  \frac{2\xi}{\pi} \frac{m_F^2}{\sqrt{x}} \left(\ln (\frac{16 x}{m_F^2}) -2 \right)\,  \quad\quad \left(\alpha=\alpha_{\rm cr}\right) 
  \,, \label{solcritical}
\end{equation}
which cannot satisfy the UV boundary condition Eq.(\ref{eq:UVBC}) for $m_0=0$, thus $\Sigma(x) \equiv 0$~\footnote{
When $m_0\ne 0$, this is an explicit breaking solution: $m_0=  (x\Sigma(x))^{\prime}|_{x=\Lambda^2} =  Z_m m_R$, with 
 $Z_m  
 = \frac{2\xi}{\pi} \frac{m_R}{\Lambda} \ln (\frac{4\Lambda}{m_R})$. This yields $\gamma_m (\mu) = 1-1/\ln (\frac{4\mu}{m_R})$
and the OPE: $\Sigma (x) \sim m_R e^{ -\int^t  d t^\prime \gamma_m (t^\prime)}\sim \frac{m_R^2}{\sqrt{x}} \ln (\frac{16x}{m_R^2} )$ , with $t \equiv \ln (4\sqrt{x}/m_R)$, in agreement with Eq.(\ref{solcritical}) up to trivial factors. Appearance of the 
log factor is peculiarity of the conformal phase at $\alpha=\alpha_{\rm cr}$ due to the collide/cancellation of the two terms, $ \omega$
and $-\omega$, at $ \omega=0$ in Eq.(\ref{eq:Sigma}). In the SSB phase satisfying  Eq.(\ref{Miranskyscaling}), no such a log factor exists when $\alpha\rightarrow \alpha_{\rm cr}+0$ ($m_F/\Lambda\rightarrow 0$ ), as already noted in sub-subsection \ref{Anomalousdim}.
}.

On the other hand, in the SSB phase $\alpha>\alpha_{\rm cr}$, bound states do appear with the mass on the order of the SSB scale ${\cal O} (m_F)\ll \Lambda$ up to factors depending on $N_F$ and $N_C$. Hence
the bound states spectra  change discontinuously  across the phase transition point, although the order parameter $m_F$ smoothly is changed as $m_F \rightarrow 0$ as $\alpha \searrow \alpha_{\rm cr}$ to the value $m_F\equiv 0$ for $\alpha\leq \alpha_{\rm cr}$ \cite{Miransky:1996pd}. 
For $\alpha>\alpha_{\rm cr}$ the massive bound states with masses proportional to $m_F$ approach to zero when $\alpha\searrow \alpha_{\rm cr}$ according to the
Miransky-BKT scaling, while the NG bosons of the chiral symmetry are exactly massless, all of which (including the NG bosons) suddenly dissappear when we reach the point $\alpha=\alpha_{\rm cr}$.   Hence it cannot be described by the 
Ginzburg-Landau effective theory (linear sigma model)\cite{Miransky:1996pd}.   This peculiarity is closely connected to the non-analytic form of the Miransky-BKT scaling in Eq.(\ref{Miranskyscaling}): The mass of the bound state A (other than the NG bosons of the chiral symmetry), $M_A$, has a {\it universal scaling function} $f\left(\frac{\alpha}{\alpha_{\rm cr}}\right)$ in the same form as the dynamical mass of the technifermions $m_F$ up to a constant $C_{\rm A}(r)$ depending on the each bound state~\cite{Harada:2003dc,Kurachi:2006ej}:  
 \beq
\frac{ M_{\rm A}}{\Lambda}  
  \simeq C_{\rm A} (r) \cdot f\left(\frac{\alpha}{\alpha_{\rm cr}}\right)\,,\quad 
 f\left(\frac{\alpha}{\alpha_{\rm cr}}\right) = \exp \left(-\frac{\pi}{\sqrt{\frac{\alpha}{\alpha_{\rm cr}}-1}}\right) \ll 1  \quad \quad (\alpha > \alpha_{\rm cr}
 )\,,   
\label{Miranskyscaling2}
 \eeq
where $M_{\rm A} /\Lambda \ll 1$ can be tuned only for $\alpha>\alpha_{\rm cr}$. 
This is an essential difference of the walking theory from the ordinary QCD, where
all the light bound states (except the NG boson pions) have masses on the same  order as the intrinsic scale $M_{\rm A}={\cal O} (m_F)={\cal O} (\Lambda_{\rm QCD})$: $M_{\rm A}/\Lambda_{\rm QCD} ={\cal O} (1)$ having 
no limit going to zero, 
in sharp contrast to  the walking theory. In the case at hand,
  all the light bound states have vanishing masses towards the criticality \cite{Chivukula:1996kg} in a universal way $f\left(\frac{\alpha}{\alpha_{\rm cr}}\right) \rightarrow 0$
  as $\alpha \rightarrow \alpha_{\rm cr}$
up to a constant $C_A(r)$ above as a consequence of the scale symmetry, and hence are a kind of ``dormant NG bosons'' of spontaneously broken scale (or chiral)  symmetry existing only in the broken phase (without the exact massless point): $M_A/M_B \rightarrow {\rm const.} \ne 0, \infty\,\, (\alpha\rightarrow \alpha_{\rm cr})$.

As we shall discuss later, the coefficient $C_A(r)$ for the spectra other than
the TD may depend on 
$r=N_F/N_C$ particularly in the anti-Veneziano limit, since only the TD has the mass subject to the explicit breaking of the scale symmetry
characterized by $m_F$, $M_\phi \sim m_F$, while others (except for technipions) reflect the SSB of the scale symmetry characterized by the dilaton decay constant 
$F_\phi \sim \sqrt{N_F N_C}\,
m_F$: (Technipions have masses $M_\pi \gg m_F$, see Eq.(\ref{TCpionmass}))
\beq
\frac{M_\phi}{M_A} \sim \frac{1}{C_A(r)} \ll 1\, \quad (r\rightarrow r_{\rm cr}).
\eeq
Thus the TD does tend to be massless (NG boson-like) faster than the others when approaching the criticality 
in a particular limit  
$N_C\rightarrow \infty,\, \lambda\equiv N_C \alpha={\rm fixed},  
 \, r \equiv N_F/N_C={\rm fixed} (\gg 1)$ (``anti-Veneziano limit'' in distinction to the original Veneziano limit
$r\ll 1$).  

Note that the  IR fixed point in the large $N_F$ QCD as a model of the ladder coupling  in the anti-Veneziano limit 
reads \cite{Shrock:2013pya} 
\beq
\frac{\alpha}{\alpha_{\rm cr}}=\frac{N_C \alpha_*}{N_C \alpha_{\rm cr}} \rightarrow \frac{66-12r}{13r -34}\,,
\quad f\left(\frac{\alpha}{\alpha_{\rm cr}}\right) \rightarrow \tilde f (r) \quad (N_C\rightarrow \infty,\, \,\frac{34}{13}<r \equiv \frac{N_F}{N_C}<\frac{11}{2})
\,, 
\eeq
which is an almost continuous parameter and hence $\frac{\alpha}{\alpha_{\rm cr}} (>1)$ can be tuned arbitrarily close to 1 by tuning
the ratio $r\equiv N_F/N_C \nearrow r_{\rm cr}=4$~\footnote{The value $N_F^{\rm cr} \simeq 4 N_C =12\,\, (N_C=3)$\cite{Appelquist:1996dq} should not be taken seriously, since it  is the result of two crude approximations: The IR fixed point value $\alpha_*$ of the two-loop 
beta function having a big error from higher loops in a scheme-dependent way \cite{Shrock:2013pya}  is not reliable near the lower end of $N_F/N_C$ where the loop expansion breaks down as its value $N_C \alpha_*$ is of ${\cal O}(1)$, and could trigger the spontaneous chiral symmetry breaking which washes out the IR fixed point
 (even though  $\alpha_* \sim 1/N_C \ll 1$, since the critical coupling also behaves as $\alpha_{\rm cr} \sim 1/C_2 \sim 1/N_C$). Also the critical value $\alpha_{\rm cr}$ of the ladder SD equation is subject to  20 $-$ 30 percent errors. Indeed lattice results suggest $N_F^{\rm cr} \simeq 8$ for $N_C=3$ \cite{Aoki:2013xza}.
 }.  
Thus the conformal phase transition as a continuous (non-analytic) phase transition
can also be realized in the large $N_F$ QCD  in the anti-Veneziano limit. 
Also note that the intrinsic scale $\Lambda_{\rm TC}$ as well as $m_F$ scales as $\sim N_F^0, N_C^0$ (fixed) in that limit, while the coupling scales like $\alpha_*\simeq \alpha_{\rm cr}={\cal O} (1/N_C) \, (\rightarrow 0)$ and hence the {\it spontaneous symmetry breaking is triggered by the weak coupling}, although the ``effective coupling''
$N_C \alpha_*\simeq N_C \alpha_{\rm cr}\simeq 2\pi/3$ is strong. Hence the ladder approximation is expected to give a
better result in the anti-Veneziano limit. This is somewhat analogous to the $1/N_C$ expansion of the NJL model with the
coupling $G\sim g/\Lambda^2$: Although the
effective critical coupling is strong, $g_{\rm cr}^{\rm eff} = N_C\cdot g_{\rm cr}=1$,   
the  coupling $g$ as well as $g_{\rm cr}$ scales like $1/N_C$, which justifies the NJL gap equation valid at
the leading order of $1/N_C$.

 This is the essence of the WTC where the explicit breaking of the 
scale symmetry is tiny compared with the intrinsic scale $\Lambda_{\rm TC}$: $m_F \ll \Lambda_{\rm TC}$, which is in contrast to the ordinary QCD where 
$m_F \sim \Lambda_{\rm QCD}$ with the scale symmetry violated completely. 
In fact    these properties are the universal features of the WTC not restricted to the ladder SD equation. We in fact find that the ladder SD results are useful for describing the 125 GeV Higgs as the TD, {\it not merely qualitatively but also quantitatively} in spite of the crude approximation.

\section{Nonperturbative Trace Anomaly}
\label{NPtraceanomaly}

\subsection{Nonperturbative Trace Anomaly and PCDC}

When the chiral symmetry is spontaneously broken, $\langle \bar F F\rangle \ne  0$,  
 {\it the scale symmetry is also spontaneously broken in the vacuum} with  the condensate of the chiral operator  $\bar F F$ transforming nontrivially under the scale transformation. 
 This leads to the TD as a NG boson of the scale symmetry.
The TD is actually not massless and thus is a pseudo NG boson, since the scale symmetry 
is {\it broken also explicitly by the same chiral condensate} as that breaks it spontaneously with 
a mass scale small compared with the intrinsic scale, $m_F\ll \Lambda= \Lambda_{\rm TC}$.
In fact such a  small mass generation in Eq.(\ref{Miranskys}) 
washes out the would-be  IR fixed point $\alpha \simeq \alpha_*$
in the deep IR region $\mu <m_F$, namely breaks the scale-invariance (nonrunning behavior or the perturbative IR fixed point) of the input coupling.

As we discussed in subsection \ref{NPrunning:sec}, the nonperturbative running of the coupling  is induced by the generation of $m_F$ through the regularization of the same chiral condensate as that breaks the scale symmetry spontaneously, where the intrinsic scale $\Lambda_{\rm TC}$ (already generated by the perturbative regularization as in Eqs.(\ref{perturbativeDT}) and (\ref{pertanomaly})) plays a role of regulator responsible for  the {\it nonperturbative trace anomaly} \cite{Miransky:1989qc} in distinction from the usual trace anomaly in the perturbation in Eq.(\ref{pertanomaly}):
\begin{eqnarray}
 \langle\partial_\mu D^\mu \rangle= \langle(\theta_\mu^\mu)\rangle^{(NP)}  &\equiv&  \langle\theta_\mu^\mu\rangle^{(full)} -  \langle\theta_\mu^\mu\rangle^{(perturbative)}  =
 \frac{\beta^{(NP)}(\alpha)}{4\alpha}  \langle G_{\mu\nu}^2\rangle^{(NP)} 
\,,\nonumber \\
 \langle G_{\mu\nu}^2\rangle^{(NP)}
 &\equiv& \langle G_{\mu\nu}^2\rangle^{(full)}
 -\langle G_{\mu\nu}^2\rangle^{(perturbative)}\,.
\label{NPanomaly}
\end{eqnarray}
The formal proof of this relation was given \cite{Miransky:1989qc} in terms of functional method for the Cornwall-Jackiw-Tomboulis effective potential $V[\Sigma(x)]$ whose stationary condition is the 
SD equation. The solution of the SD equation ${\bar \Sigma} (x)$ yields the vacuum energy  $E=V[{\bar \Sigma}(x)]$ and $\langle \theta_0^0\rangle= \langle\partial_\mu D^\mu \rangle= \langle(\theta_\mu^\mu)\rangle^{(NP)} =4 E$. 
 The IR conformality is manifested in the fact that the relevant mass scale $m_F$ is tiny, compared with that of the perturbative trace anomaly, $m_F \ll \Lambda_{\rm TC}$, 
$ \langle \theta_\mu^\mu\rangle^{(perturbative)}= - {\mathcal O} (N_F N_C \Lambda_{\rm TC}^4)$ in Eq.(\ref{pertanomaly}). This is in sharp contrast to the ordinary QCD, where $m_F\simeq \Lambda_{\rm QCD}$ and hence $\langle \theta_\mu^\mu \rangle \simeq  \langle \theta_\mu^\mu\rangle^{perturbative}$,  without the walking region and the IR conformality. 

Based on this  approximate scale symmetry in WTC,  
the light TD as a pseudo NG boson was predicted \cite{Yamawaki:1985zg, Bando:1986bg}  via the anomalous WT identity for the scale symmetry, so-called the PCDC  relation (Eqs.(6),(8) and (9) of  Ref.~\cite{Bando:1986bg}):
\begin{eqnarray}
M_\phi^2 F_\phi^2 = - F_\phi \langle 0 |\partial^\mu D_\mu |\phi \rangle= - d_\theta \langle 0 |\theta_\mu^\mu|0\rangle^{(NP)} &=&-\frac{\beta^{(NP)}(\alpha)}{\alpha} \langle G_{\mu\nu}^2\rangle^{(NP)} ={\cal O}
(N_F N_C m_F^4)\nonumber \\
&\ll& - d_\theta \langle \theta_\mu^\mu\rangle^{(perturbative)}={\cal O} (N_F N_C \Lambda_{\rm TC}^4) \,, 
\label{PCDC}
\end{eqnarray} 
where $D_\mu$ 
is the dilatation current and $F_\phi$ is the decay constant of $\phi$ defined as $\langle 0|D_\mu|\phi (q)\rangle =- i F_\phi q_\mu$, and $d_\theta$ $(=4)$ is the dimension of  $\theta_{\mu\nu}$. 
This is in sharp contrast to the ordinary QCD where $m_F \sim \Lambda_{\rm QCD}$ and hence $|\langle \theta_\mu^\mu \rangle| = {\mathcal O} (m_F^4) ={\mathcal O} (\Lambda_{\rm QCD}^4) =|\langle \theta_\mu^\mu\rangle^{({perturbative})}|$,  totally lacking the scale symmetry. 
Note that $\alpha \sim \alpha_*\sim \alpha_{\rm cr} \sim 1/N_C$ and  $\beta^{(NP)}(\alpha) \sim 1/N_C$ in the anti-Veneziano limit $N_C \rightarrow \infty$ with $r\equiv N_F/N_C=$ fixed $(\gg 1)$, so that we have $\frac{\beta^{(NP)}(\alpha)}{\alpha} \sim N_C^0$. 
(This is also the case for the perturbative beta function, see Eq.(\ref{2loop}).)

\subsection{RG invariance of the Nonperturbative trace anomaly}

Here we show  the RG invariance of the nonperturbative trace anomaly  $\frac{\beta^{(NP)}(\alpha)}{4 \alpha} \langle G_{\nu\lambda}^2\rangle^{(NP)}$  in the ladder approximation: 
\begin{eqnarray}
 \langle\partial^\mu D_\mu \rangle= \langle(\theta_\mu^\mu)\rangle^{(NP)}  = 
 \frac{\beta^{(NP)}(\alpha(\mu))}{4\alpha(\mu)}  \langle G_{\nu\lambda}^2\rangle^{(NP)}_{(\mu)} \,.
\label{NPanomaly2}
\end{eqnarray} 
Based on the result of Ref.\cite{Hashimoto:2010nw},  we  shall  show that the RG invariance is realized  in a nontrivial manner: The dependence of the renormalization point $\mu$ is precisely cancelled  among
$\beta^{(NP)}(\alpha(\mu))/(4\alpha(\mu)) \sim -(\frac{\alpha(\mu)}{\alpha_{\rm cr}}-1)^{3/2}\sim -1/\ln^3(\mu/m_F)$ and 
 $\langle G_{\nu\lambda}^2\rangle^{(NP)} _{(\mu)} \sim(\frac{\alpha(\mu)}{\alpha_{\rm cr}}-1)^{-3/2} \sim \ln^3 (\mu/m_F)$  for $m_F<\mu<\Lambda_{\rm TC}$, thereby yielding the same result as that  of the vacuum 
 energy calculation in Ref. \cite{Gusynin:1987em}. Comparing Eq.(\ref{NPanomaly2}) with Eq.(\ref{pertanomaly}), we see that the resultant trace anomaly of order ${\cal O} (m_F^4)$ is much smaller than the trace anomaly of order ${\cal O} (\Lambda_{\rm TC}^4)$ related to the 
 fundamental scale $\Lambda_{\rm TC}$ of the theory, and hence the use of the PCDC Eq. (\ref{PCDC})  is justified.

Let us first calculate the nonperturbative 
gluon condensate induced not from the gluon loop (already subtracted out) but from 
the fermion loop with the technifermion having dynamical mass $m_F$, which  is calculated at the two-loop level with the technifermion propagator as given in the ladder SD equation~\cite{Miransky:1989qc}: 
\begin{equation} 
 \langle G_{\mu\nu}^2 \rangle^{(NP)}   
 = - \frac{2 i g^2 N_F N_C}{(2\pi)^8} 
 \int d^4 k d^4 p \, {\rm tr}[S_F(p) \gamma_\mu S_F(k) \gamma_\nu] D^{\mu\nu}(p-k) 
\,.  \label{G2}
\end{equation}
By using the ladder SD equation for $S_F (p)$ in Eq.(\ref {eq:improvedSDeq}) 
with the nonrunning coupling, 
Eq.(\ref{G2}) can be rewritten into a simpler form 
\begin{eqnarray} 
  \langle G_{\mu\nu}^2 \rangle^{(NP)}   
&= &
 - \frac{2 i g^2 N_F N_C}{(2\pi)^4} 
 \int d^4 p \, {\rm tr}[1 - S_F(p)S^{-1}(p)] 
\nonumber \\ 
&=& 
\frac{N_F N_C}{2\pi^2} 
\int_0^{\Lambda^2} dx \frac{x \Sigma^2(x)}{x + \Sigma^2(x)}   = \frac{N_F N_C}{2\pi^2} \int^{\Lambda^2} dx\left[ \Sigma^2(x) -\frac{\Sigma^4(x)}{x+ \Sigma^2(x)}\right]\,,
 \label{G2:2}
\end{eqnarray} 
where the second term of the integral yields correction of order ${\cal O} (m_F^8/\Lambda^4)$, since 
$\Sigma(x) \sim m_F^2/\sqrt{x}$, and hence can be ignored.
For the high energy region where $x \gg m_F^2$, 
the mass function $\Sigma(x)$ takes the form given by Eq.(\ref{Sigma:sol}).
Using $\Sigma(x)$ in Eq.(\ref{Sigma:sol}), we find $\int^{\Lambda^2} dx \Sigma^2(x) \simeq  \xi^2 m_F^4 
 \frac{8\,{\rm cth}
\frac{\pi \tilde \omega}{2}}{\pi \tilde \omega ({\tilde \omega}^2 + 1)} 
 \ln \left( \frac{4 \Lambda}{m_F} \right) = \frac{16\xi^2}{\pi^2 {\tilde \omega}^2} m_F^4 
  \ln \left( \frac{4 \Lambda}{m_F} \right)\left[1 +{\cal O}(\tilde \omega^2)\right]$. 
From Eq.(\ref{G2:2}),  
we find  
\begin{eqnarray} 
    \langle G_{\mu\nu}^2 \rangle^{(NP)}   
&\simeq &     
    \frac{ N_F N_C }{2 \pi^2}  \cdot m_F^4 \frac{16\xi^2}{\pi^2 {\tilde \omega}^2}    
     \ln \left( \frac{4 \Lambda}{m_F} \right)  \simeq N_F N_C \frac{ 8\xi^2 }{\pi^3} m_F^4\cdot  \left(\frac{1}{\pi}\ln \left( \frac{4 \Lambda}{m_F}\right) \right)^3    \nonumber \\ 
&\simeq& 
N_F N_C \frac{8 \xi^2}{\pi^3} m_F^4 \cdot \left( \frac{\alpha}{\alpha_{\rm cr}}  -1 \right)^{-3/2}  \,,
  \label{G2:final} 
\end{eqnarray}
up to factor of $(1+ {\cal O} (\tilde \omega^2))$, where  we have used the Miransky scaling Eq.(\ref{Miranskyscaling}): $\tilde \omega = \sqrt{\alpha/\alpha_{\rm cr} -1} = \pi/\ln (4\Lambda/m_F)$.  Thus the gluon condensate is diverging as $\left(\ln \frac{\Lambda}{m_F}\right)^3$ much 
faster than the QCD-like theory with divergence $\ln \frac{\Lambda}{m_F}$ as noted in Ref.\cite{Hashimoto:2010nw}.

Note that the divergence of   $\langle G_{\mu\nu}^2 \rangle^{(NP)}$ is of the same origin as that for the amplification of the symmetry violation such as the technipion mass coming from the
UV contributions enhanced by the large anomalous dimension: $\Sigma (x) \sim \frac{m_F^3}{x} (\frac{x}{m_F^2})^{\gamma_m/2} \sim m_F^2/\sqrt{x}$ . Do not confuse it from the log divergence of the gluon condensate in the ordinary QCD, which comes from the gluon
loop in contrast to the present case coming from the fermion loop. We shall discuss it later.

Actually,  it was found \cite{Hashimoto:2010nw} that the divergence 
$\sim \left(\ln \left( \frac{4 \Lambda}{m_F}\right) \right)^3$ in Eq.(\ref{G2:final}) is precisely cancelled by the vanishing factor of the nonperturbative beta function in Eq.(\ref{NPbeta2}):
  \begin{equation} 
  \beta^{(NP)}(\alpha ) 
 = - \frac{2 \alpha_{\rm cr}}{\pi}
\left(
 \frac{1}{\pi} \ln (\frac{4\Lambda}{m_F})
 \right)^{-3} 
 = - \frac{2 \alpha}{\pi}  
 \left(\frac{\alpha}{\alpha_{\rm cr}} -1 \right)^{3/2}
  \left[1+\tilde \omega^2\right]^{-1}
  \,, \label{beta2}
 \end{equation}
such that 
the trace anomaly of the energy momentum tensor $\langle \theta_\mu^\mu \rangle$ is given by 
 \begin{equation} 
 \langle \theta_\mu^\mu \rangle^{(NP)} 
 = \frac{\beta^{(NP)}(\alpha)}{4\alpha} \langle G_{\mu\nu}^2 \rangle^{(NP)} 
 \simeq -  N_F N_C  \frac{4 \xi^2}{\pi^4} 
m_F^4 \left[1+{\cal O} (\tilde \omega^2)\right] 
\,.\label{theta}
\end{equation}
Thus the smallness of $\beta^{(NP)}(\alpha)$ as manifestation of the approximate scale symmetry  is in fact operative by canceling the otherwise amplified symmetry violation effects 
of the large anomalous dimension, and hence keeping the nonperturbative trace anomaly, the explicit breaking of the scale symmetry, on the order of $m_F^4$.

On the other hand, 
 the direct computation of $\langle \theta_\mu^\mu \rangle$ through the vacuum energy $\langle \theta_\mu^\mu \rangle=4  \langle \theta_0^0 \rangle$ is \cite{Gusynin:1987em}: 
\beq
 \langle \theta_\mu^\mu \rangle^{(NP)} = 4 V[\Sigma(x)] = - \frac{N_FN_C}{4\pi^2} \left[ \Lambda^4 \ln \left(1 +\frac{\Sigma(\Lambda^2)}{\Lambda^2}\right) \right] 
 \simeq  - \frac{N_FN_C}{4\pi^2} \Lambda^2 \Sigma^2(\Lambda^2)
\nonumber \\
  \simeq  - \frac{N_FN_C}{4\pi^2}   \xi^2  m_F^4
 \frac{8\,{\rm cth}
\frac{\pi \tilde \omega}{2}}{\pi \tilde \omega ({\tilde \omega}^2 + 1)} \sin \tilde \omega^2= -  N_F N_C  \frac{4 \xi^2}{\pi^4} 
m_F^4\left[1+{\cal O} (\tilde \omega^2)\right] \,.
\label{vacenergy}
\eeq 
Let us take $\Lambda \rightarrow \infty$ such that $\alpha(\Lambda) \rightarrow \alpha_{\rm cr}$ ($\tilde \omega \rightarrow 0$), then 
Eq.(\ref{theta}) precisely coincides with Eq.(\ref{vacenergy}). Thus the three independent calculations of different quantities are consistent 
with each other within the ladder approximation~\footnote{
   For idealized large $N_C$ in the anti-Veneziano limit, ladder calculation   
   becomes more reliable, as we demonstrated in Fig.2. 
The result of Eq.(\ref{vacenergy})  
   based on the ladder thus becomes more reliable in the anti-Veneziano  
   limit. As in the case of usual large $N_c$ arguments in QCD ($N_c=3 
   \to \infty$), the quantitative check of the validity of the anti-Veneziano  
   limit for the realistic value of $N_F$ and $N_C$ is of course subject to the  
   fully nonperturbative check by the lattice studies. 
}.

This is reformulated in terms of the nonperturbative running $\alpha (\mu) $ in the renormalization defined in Eq.(\ref{NPbeta2}) as 
\beq
 \langle \theta_\mu^\mu \rangle^{(NP)} =  \frac{\beta^{(NP)}(\alpha(\mu))}{4\alpha(\mu)} \langle G_{\nu\lambda}^2 \rangle^{(NP)}_{(\mu)}
 =-  N_F N_C  \frac{4 \xi^2}{\pi^4} m_F^4\,.
 \label{Traceanomalycontinuum}
 \eeq
 Then  the nonperturbative trace anomaly  
 $\langle \theta_\mu^\mu \rangle$ is written in the manifestly RG-independent way in the ladder approximation as it should be.

Such an RG invariance by cancellation is a well-known fact for the perturbative 
trace anomaly 
but  is explicitly recognized for the first time for the nonperturbative trace anomaly. 
It is in fact well-known that the perturbative trace anomaly is RG invariant, i.e., independent of the renomalization point $\mu$. 
In the chiral limit it reads $\langle \theta_\mu^\mu\rangle =\beta(\alpha)/(4\alpha) \langle G_{\mu\nu}^2\rangle$ which is RG invariant in such a way that  $\beta(\alpha)/(4\alpha) \sim  \alpha \sim 1/\ln(\mu/\Lambda_{\rm QCD})$ precisely  cancels the divergence in $\langle G_{\mu\nu}^2\rangle \sim \ln (\mu/\Lambda_{\rm QCD})$ as $\mu \rightarrow \infty$. This is also applied to the
WTC in the UV region $\mu>\Lambda_{\rm TC}$, where the perturbative trace anomaly in Eq.(\ref{pertanomaly}) is obviously RG invariant in the same way as in the ordinary QCD.  Note that in the  usual QCD the scale-invariance appear to exist  ``formally'' in the UV region $\mu \gg \Lambda_{\rm QCD}$ due to the vanishing $\beta(\alpha)/(4\alpha) \sim 1/\ln (\mu/\Lambda_{\rm QCD}) \rightarrow 0$ at the trivial UV fixed point $\alpha_*=0$, which is however compensated by the diverging gluon condensate  $\langle G_{\mu\nu}^2\rangle \sim \ln (\mu/\Lambda_{\rm QCD})$, and hence 
the scale invariance in QCD exists nowhere.

\subsection{Inclusion of  small bare mass of technifermions: Basis for the dilaton chiral perturbation theory}

For completeness we here show that with inclusion of the small bare mass $m_0$ or the renormalized mass (``current mass'' $m_R (\ll m_F)$) of the technifermion, the ladder calculations of various quantities 
yield a consistent  
trace anomaly for the anomalous WT identity, which is the basis of the sChPT~\cite{Matsuzaki:2013eva}. It is
vital for the lattice calculations of  the flavor-singlet scalar bound state as a candidate for the technidilaton, whose observed mass and decay constant 
should be extrapolated to the chiral limit.

Here we explicitly check that the ladder approximation is consistent with the anomalous WT identity for the SSB of the approximate scale symmetry 
 (for $\gamma_m=1$):
\begin{equation} 
 \langle \theta_\mu^\mu \rangle 
 = \frac{\beta^{(NP)}(\alpha)}{4\alpha} \langle G_{\mu\nu}^2 \rangle 
 + (1+\gamma_m)  N_F m_R \langle \bar{F} F \rangle_R = \frac{\beta^{(NP)}(\alpha)}{4\alpha} \langle G_{\mu\nu}^2 \rangle 
 + 2 N_F m_R \langle \bar{F } F \rangle_R  
 \,. \label{scale:WT:mc}  
\end{equation}
The formal proof of this relation was also given in Ref. \cite{Miransky:1989qc} in terms of functional method for the Cornwall-Jackiw-Tomboulis effective potential $V[\Sigma(x)]$. 
The relation is the basis of the sChPT~\cite{Matsuzaki:2013eva} for the TD mass in the presence of the technifermion explicit mass (current mass) $m_R$.
 The current mass $m_R$ and the associated-renormalized chiral condensate $\langle \bar{F} F \rangle_R$ 
are related to the bare mass $m_0$ in Eq.(\ref{eq:UVBC}) and the bare-chiral condensate $\langle \bar{F} F  \rangle_0$ 
involving the renormalization constant $Z_m$ in Eq.(\ref{Z_mladder}) as 
\begin{eqnarray} 
 m_R &=& Z_m^{-1} m_0 
\,. \\
 \langle \bar{F} F  \rangle_R &=& Z_m \langle \bar{F}F \rangle_0  
\,. \label{scaling:mc}
\end{eqnarray}
We then see that 
 Eq.(\ref{scale:WT:mc}) is nothing but the chiral expansion of $\langle \theta_\mu^\mu \rangle$ and/or the dilaton mass $m_\phi$ (sChPT~\cite{Matsuzaki:2013eva}): 
\begin{eqnarray} 
 \langle \theta_\mu^\mu \rangle &=& 
 \langle \theta_\mu^\mu \rangle\Bigg|_{m_R=0}   
 + 
\frac{\partial \langle \theta_\mu^\mu \rangle}{\partial m_0}\Bigg|_{m_R=0} \cdot m_R 
\,, \nonumber \\ 
 \langle \theta_\mu^\mu \rangle\Bigg|_{m_R=0}  
 &=&  
\frac{\beta_{\rm NP}(\alpha)}{4\alpha} \langle G_{\mu\nu}^2 \rangle \Bigg|_{m_R=0} 
 =- \frac{4\xi^2}{\pi^4} N_F N_C m_F^4\,, \nonumber \\ 
\frac{\partial \langle \theta_\mu^\mu \rangle}{\partial m_0}\Bigg|_{m_0=0} \cdot m_R 
&=& 
 2 N_F m_R \langle \bar{F}F \rangle_R  
  \,, \label{sWT}
\end{eqnarray} 
 where  $\langle \theta_\mu^\mu \rangle\Bigg|_{m_R=0} =- \frac{4\xi^2 }{\pi^4}N_F N_C m_F^4$ is given by Eq.(\ref{theta}).

We shall check Eq.(\ref{sWT}) by evaluating both sides with use of the ladder results. 
The bare-chiral condensate for $m_R=0$ is given as~\cite{Miransky:1989qc} 
\begin{equation} 
 \langle \bar{F} F \rangle_0
 \simeq - \frac{4 N_C}{\pi^3} \xi m_F^2 \Lambda   
 \,,  
\end{equation}
which is converted into the renormalized condensate through the scaling relation in Eq.(\ref{scaling:mc}) with the $Z_m=\frac{2\xi}{\pi} \frac{m_F}{\Lambda}$ 
in Eq.(\ref{Z_mladder}):  
\begin{equation} 
 \langle \bar{F} F \rangle_R 
= Z_m \cdot \langle \bar{F}F \rangle_0 
\simeq  
- \frac{8 N_C}{\pi^4} \xi^2 m_F^3 
\,. 
\end{equation} 
On the other hand, the  $ \langle \theta_\mu^\mu \rangle$, with the small current mass $m_R (\ll m_F)$ included into the full mass of the technifermion $m_P\simeq m_F+ m_R$, is given as~\cite{Miransky:1989qc}    
\begin{equation} 
  \langle \theta_\mu^\mu \rangle 
  \simeq - \xi^2 \frac{4 N_F N_C}{\pi^4} (m_R + m_F)^4 \simeq 
  \frac{\beta_{\rm NP}(\alpha)}{4\alpha} \langle G_{\mu\nu}^2 \rangle \Bigg|_{m_R=0}    +  \frac{\partial \langle \theta_\mu^\mu \rangle}{\partial m_R}\Bigg|_{m_R=0} \,m_R\,, \label{theta:full}
\end{equation} 
with 
\begin{equation}
 \frac{\partial \langle \theta_\mu^\mu \rangle}{\partial m_R}\Bigg|_{m_R=0} 
 \simeq 
 - \frac{16 N_F N_C}{\pi^4} \xi^2 m_F^3 
 = 2 N_F \langle \bar{F}F\rangle_{R} 
 \,,
\end{equation}
which reproduces Eq.(\ref{scale:WT:mc}).

It is straightforward to write down the effective theory to realize the relation, in Eq.(\ref{scale:WT:mc}) namely the sChPT~\cite{Matsuzaki:2013eva}:
\begin{eqnarray} 
{\cal L}&=& {\cal L}_{(2)}^{\rm inv}
+{\cal L}^{S}_{(2){\rm anomaly}} + {\cal L}^{S}_{(2){\rm mass}} +  {\cal L}_{(4)}\,, \nonumber  \\
 {\cal L}_{(2)}^{\rm inv}
 &=& \frac{F_\phi^2}{2} (\partial_\mu \chi)^2 
+ \frac{F_\pi^2}{4} \chi^2 {\rm tr}[\partial_\mu U^\dag \partial^\mu U] \,,\nonumber \\
{\cal L}^{S}_{(2){\rm anomaly}} 
&=&
-\frac{F_\phi^2}{4} m_\phi^2 \chi^4 \left( \log \frac{\chi}{S} - \frac{1}{4}\right)\,,\nonumber\\
{\cal L}^{S}_{(2){\rm mass}} 
 &=& 
\frac{F_\pi^2}{4} \left( \frac{\chi}{S} \right)^{3-\gamma_m} \cdot S^4 {\rm tr}[ {\cal M}^\dag U + U^\dag {\cal M}] 
- \frac{(3-\gamma_m) F_\pi^2}{8}  \chi^4 \cdot  \left( N_F{\rm tr}[\langle {\cal M}^\dag {\cal M} \rangle] \right)^{1/2}
\,,   
\label{sChPT}
\end{eqnarray} 
where $U(x) =e^{2i \pi(x)/F_\pi}$, $\chi(x)=e^{\phi(x)/F_\phi}$ ($\langle \chi \rangle=1, \langle \phi\rangle=0$) are nonlinear bases for the chiral and scale transformations, and ${\cal M}$ and $S(x)$ are chiral and scale spurion fields transforming in the same way as
$U(x)$ and $\chi(x)$, respectively, with $\langle{\cal M}\rangle=m_R$, $\langle S(x)\rangle =1$. $ {\cal L}_{(4)} $ contains the ${\cal O}(p^4)$ counter terms of the ChPT with $M_\phi^2 ={\cal O} (p^2)$
and explicitly given in Ref.~\cite{Matsuzaki:2013eva}. 
The ${\cal O}(p^2)$ terms in Eq.(\ref{sChPT}) lead to the TD mass formula~\cite{Matsuzaki:2013eva}:
 \begin{eqnarray} 
  M_\phi^2   
&=&  m_\phi^2 + 
\frac{(3-\gamma_m)(1+\gamma_m)}{4}  
\cdot \frac{2N_F F_\pi^2}{F_\phi^2}  m_\pi^2
\simeq m_\phi^2 +  \frac{2N_F F_\pi^2}{F_\phi^2} m_\pi^2 
, \label{mphi:p2}
 \end{eqnarray}
 where $m_\phi=M_\phi|_{m_R=0}$ is the TD mass in the chiral limit. 
 The same result is also derived directly from  the anomalous WT identity for the scale symmetry and
 chiral WT identity.
 The result is useful for determining the mass and decay constant of TD by the lattice simulations through chiral extrapolation. 

   Note that the nonperturbative trace anomaly is given by 
   $\langle \theta_\mu^\mu \rangle = \langle \partial^\mu D_\mu \rangle= \langle \delta_D {\cal L}^{S}_{(2){\rm anomaly}} \rangle =- m_\phi^2 F_\phi^2\langle\chi\rangle^4 /4 =- m_\phi^2 F_\phi^2 /4 $ for $S(x) =1$,  in accord with the PCDC relation, Eq.(\ref{PCDC}), where $\delta_D \chi =\chi +x^\mu\partial_\mu \chi$ is the dilatation transformation. 
 The form of  ${\cal L}^{S}_{(2){\rm anomaly}} $ is unique in the sense that it correctly reproduces the nonperturbative trace anomaly through the log factor and the factor $-1/4$ in the parenthesis is crucial both for eliminating the $\phi$ tad pole (linear term in $\phi$) so as to have the correct vacuum $\langle \chi\rangle=1$ ($\langle\phi\rangle=0$) and the correct vacuum energy $E=\langle \theta_0^0\rangle= \langle \theta_\mu^\mu\rangle/4 =-m_\phi^2F_\phi^2/16$, as well as the correct mass term of $\phi$, see later Eq.(\ref{dilatonpotential}). The form has a characteristic log form of reflecting the trace anomaly generated by the nonperturbative dynamics, similarly to the Coleman-Weinberg potential which is generated by the perturbative trace anomaly.

\section{Mass and Decay constant of the Technidilaton}
\label{TDmass}

From Eq.(\ref{Traceanomalycontinuum}) the PCDC relation in the  ladder approximation reads 
\beq
M_\phi^2 F_\phi^2 = - F_\phi \langle 0 |\partial_\mu D^\mu |\phi \rangle= - 4 \langle 0 |\theta_\mu^\mu|0\rangle^{(NP)} =-\frac{\beta^{(NP)}(\alpha(\mu))}{\alpha(\mu)} \langle G_{\mu\nu}^2\rangle^{(NP)}_{(\mu)} 
= N_F N_C \left( \frac{16 \xi^2}{\pi^4} m_F^4\right) \,.
 \label{PCDCladder}
\eeq  
Let us consider the saturation of the anomalous WT identity for the scale symmetry in the anti-Veneziano limit:
\beq
 F_\phi^2 M_\phi^2 = {\cal F.T.} \left\langle T\left(\partial^\mu D_\mu(x) \cdot \partial^\mu D_\mu(0)\right)\right\rangle  = {\cal F.T.}  
 \left\langle T   \left( \frac{\beta^{(NP)}(\alpha)}{4\alpha} G_{\mu\nu}^2 (x)^{(NP)}\cdot \frac{\beta^{(NP)}(\alpha)}{4\alpha} G_{\mu\nu}^2 (0)^{(NP)}\right)
 \right\rangle\,,
 \label{Vlimit}
    \eeq
which is dominated by the fermion loop in Fig.~\ref{GG-f-loop} 
and hence scales like $N_F\, N_C^3\, \alpha^2 \sim N_F\, N_C$, in accord with the explicit ladder computation in Eq.(\ref{PCDCladder}).

 \begin{figure}
  \begin{center}
    \includegraphics[scale=0.5,clip]{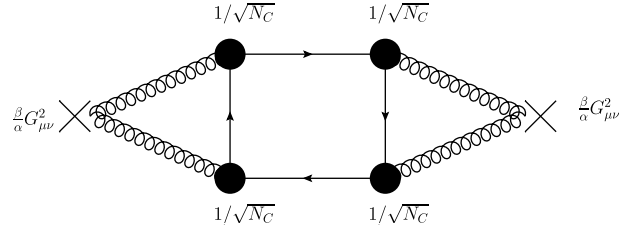}
  \end{center}
\caption{
The Feynman graph and the large $N_C$ and $N_F$ countings  
for the fermion loop contribution to the correlation function of gluon condensate operators reflecting the anti-Veneziano limit.  
}
\label{GG-f-loop}
\end{figure}

Instead of the notion of the nonperturbative running coupling,  
Eq.(\ref{NPbeta}), one may use
the ladder SD solution $\Sigma(x)$ in Eq.(\ref{Sigma:sol}) and the Miransky scaling Eq.(\ref{Miranskyscaling2}),
 in terms of the  parameter $N_F$, with $\Lambda=\Lambda_{\rm TC}$ fixed ($\gg m_F$), through the CBZ IR fixed point $\alpha_* (\gtrsim \alpha\gtrsim \alpha_{\rm cr})$:
 \cite{Appelquist:1996dq}
  \beq
m_F =4 \Lambda_{\rm TC}  \exp 
\left(-
\frac{\pi}
{
\sqrt{
\frac{
\alpha_*}{\alpha_{\rm cr}
}-1
}
}
\right)\,, \quad 0< \frac{\alpha}{\alpha_{\rm cr}}-1 \lesssim \frac{\alpha_*}{\alpha_{\rm cr}}-1 = \frac{\pi^2}{\ln^2 (\frac{4\Lambda_{\rm TC}}{m_F})}\propto N_F^{\rm cr}-N_F
\,\,\,\ll 1\,. 
\label{MscalingFP}
\eeq
 Then Eq.(\ref{G2:final}) and Eq.(\ref{vacenergy}) read for $\alpha_* \gtrsim \alpha \gtrsim \alpha_{\rm cr}$~\footnote{
 More precisely, $\frac{\alpha_*}{\alpha_{\rm cr}}-1 \propto r_{\rm cr}-r$ instead of $\propto N_F^{\rm cr}-N_F$ in the anti-Veneziano limit, where $r=N_F/N_C$.
The two-loop CBZ value of $\alpha_*$ plus ladder SD value of $\alpha_{\rm cr}$ implies  $r_{\rm cr}=4$ and $\frac{\alpha_*}{\alpha_{\rm cr}}-1
\simeq (25/18)(4-r)$ in that limit.
}: 
 \beq
 \langle G_{\mu\nu}^2\rangle^{(NP)} \sim N_C N_F m _F^4  \cdot (N_F^{\rm cr}-N_F)^{-3/2}\,, \quad 
  \langle \theta_\mu^\mu \rangle^{(NP)} 
 =-  N_F N_C  \frac{4 \xi^2}{\pi^4} m_F^4\,,
  \eeq
 from which for consistency with the trace anomaly we necessarily have the nonperturbative beta function in the broken phase $N_F<N_F^{\rm cr}$:  
  \beq
  \beta^{(NP)}(\alpha) \sim - (N_F^{\rm cr}-N_F)^{3/2}  \,\, (<0 \quad {\rm for} \quad N_F < N_F^{\rm cr} )\,. 
    \eeq
This agrees with $\beta^{(NP)}(\alpha) =- \frac{\partial\alpha}{\partial \ln m_F}$ along with Eq.(\ref{MscalingFP}), in contrast to  the arguments based on the 
two-loop beta function Eq.(\ref{2loop}), $\beta^{(2-loop)} (\alpha)\sim - (N_F^{\rm cr}-N_F )$
\cite{Dietrich:2005jn, Appelquist:2010gy}, which cannot cancel the divergence of $\ln^3 (\Lambda_{\rm TC}/m_F) \sim  (N_F^{\rm cr}-N_F)^{-3/2}$ . 
Hence our results arrive at  
the same for the TD mass as Eq.(\ref{PCDCladder}): 
\begin{eqnarray}
\beta^{(NP)}(\alpha) &\sim& - (N_F^{\rm cr}-N_F)^{3/2} \,, 
\quad\langle G_{\mu\nu}^2\rangle^{(NP)} \sim N_C N_F m _F^4  \cdot (N_F^{\rm cr}-N_F)^{-3/2}\,, \\
{\rm s. t.}  \quad M_\phi^2 F_\phi^2 &\simeq&  \frac{\beta^{(NP)}(\alpha)}{\alpha} \langle G_{\mu\nu}^2 \rangle^{(NP)}
 =  N_F N_C \left( \frac{16 \xi^2}{\pi^4} m_F^4\right)\, .
 \end{eqnarray}
Note that the two-loop beta function Eq.(\ref{2loop}) having the linear zero at the CBZ IR fixed point $\alpha_*$, $\beta^{(2-loop)} (\alpha)\sim - (N_F^{\rm cr}-N_F )$, is obviously  invalid in the broken phase $\alpha_*>\alpha_{\rm cr}$ ($N_F <N_F^{\rm cr}$), where tuning 
$m_F/\Lambda_{\rm TC} \ll 1$ should be made through the Miransky scaling Eq.(\ref{MscalingFP})  as $\alpha_*/\alpha_{\rm cr} \searrow 1 $ ($N_F\nearrow N_F^{\rm cr}$). See the discussions below Eq.(\ref{NPbeta2}). 

Thus as note in Ref. \cite{Hashimoto:2010nw}, the suppression effect by the small beta function $\beta^{(NP)}(\alpha)/(4\alpha)\ll 1$ as naively expected~\cite{Bando:1986bg,Dietrich:2005jn, Appelquist:2010gy} for the $M_\phi^2 F_\phi^2$ is actually compensated by the enhancement of $\langle G_{\mu\nu}^2\rangle^{(NP)} $ due to the large anomalous dimension $\gamma_m =1$, both being
the two sides of the same coin, characteristic to the approximate scale invariance 
for $\alpha\simeq \alpha_* \approx \alpha_{\rm cr}$ ($m_F\ll  \mu< \Lambda_{\rm TC}$). 
Actually, it is in a more sophisticated way that the smallness of the beta function or the approximate scale symmetry
 is responsible for the lightness of the TD : Lightness of the TD is guaranteed first by the hierarchy $m_F\ll \Lambda_{\rm TC}$ corresponding to the smallness of $\beta(\alpha)$, 
 which implies the nonpertubative trace anomaly of order of ${\mathcal O} (m_F^4)$ is much smaller than the perturbative trace anomaly on the order of ${\mathcal O} (\Lambda_{\rm TC}^4)$~\cite{Yamawaki:2010ms}. Additional small hierarchy $M_\phi  \ll F_\phi$  $(M_\phi \ll v_{\rm EW}$) comes from the $N_C, N_F$ scaling related with the same requirement $m_F \ll \Lambda_{\rm TC}$ via more concrete setting of the anti-Veneziano limit
$N_C \rightarrow \infty$ with tuning of $r\equiv N_F/N_C$ ($\gg 1$).

We now discuss the TD mass based on the PCDC relation, Eq.(\ref{PCDCladder}),  in the ladder approximation.
From Eq.(\ref{Traceanomalycontinuum}) with $m_F\ll \Lambda_{\rm TC}$, we in fact have a small  nonperturbative explicit breaking of the scale symmetry:
 $|\langle(\theta_\nu^\nu)^{(NP)} |\ll  |\langle\theta_\nu^\nu\rangle^{(perturbative)} |$, and hence a small TD mass compared with the intrinsic scale $\Lambda_{\rm TC}$. 
 Such a small pseudo NG boson mass can be estimated by the anomalous WT identity for the PCDC~\cite{Bando:1986bg} as in Eq.(\ref{PCDCladder}).

As already noted in the Introduction, $m_F$ is independent of $N_F, N_C$, since it is related to the $N_F, N_C$-independent quantity $\Lambda=\Lambda_{\rm TC}$ via the Miransky scaling in Eq.(\ref{Miranskyscaling2}), 
which is $N_F, N_C$ -independent, with $\alpha/\alpha_{\rm cr}$ and/or $\alpha_*/\alpha_{\rm cr}$ is independent of $N_F, N_C$.
Since the dilatation current $D_\mu (x)$ is sum of contributions from $N_F N_C$ fermions, and $|\phi\rangle$ is a
flavor/color singlet state normalized as $1/\sqrt{N_F N_C}$, 
it follows that $F_\phi={\cal O} (\sqrt{N_F N_C} m_F)$  by definition of $F_\phi$, $\langle 0| D_\mu|\phi\rangle =- i q_\mu F_\phi$.
\footnote{This can also be seen explicitly in the linear sigma model where TD can be  
a radial mode $\phi$ (Higgs in the SM) under certain condition  \cite{Fukano:2015uga}.  
In the polar decomposition of the chiral filed $ M= H U$, where $M\sim {\bar F}_R F_L$ is a $N_F\times N_F$ complex matrix  transforming 
as $M \rightarrow g_L M g_R^\dagger$ with $g_{L,R} \in SU(N_F)_L\times SU(N_F)_R$, and $H_{\alpha\beta}=(\phi +F_\pi) \delta_{\alpha\beta}$ and $U$ are hermitian and unitary 
matrix, respectively. The decay constant of $\phi$ in the linear sigma model (SM) is given by
$F_\phi^2 =(3-\gamma_m)^2 \frac{N_F}{2} F_\pi^2$~\cite{Miransky:1989qc}, where $\gamma_m$ is the anomalous dimension of the filed $M$ ($\gamma_m=1$ for the case that $M$ is a composite field ${\bar F}_R F_L$ in the WTC). Under the condition that  the linear sigma model is regarded as an effective theory of the
WTC \cite{Fukano:2015uga},
 this would yield $F_\phi^2 \simeq N_F N_C \frac{2\xi^2}{\pi^2}  m_F^2 ={\cal O} (N_F N_Cm_F^2) $, when combined with the Pagels-Stokar formula in the ladder. (When the nonlinear sigma model limit is taken, the relation of
  $F_\phi^2/[(3-\gamma_m)F_\pi]^2 =\frac{N_F}{2}$ would become arbitrary, in agreement with our PCDC relation for TD.)
  In passing, the linear sigma model result 
coincides with the holographic estimate of the $F_\phi$ \cite{Matsuzaki:2012xx}.
}
Hence Eq.(\ref{PCDCladder}) generically implies that  $M_\phi$ is independently of $N_F$ and $N_C$.
From the above rough estimate $F_\phi^2={\cal O} (N_F N_C m_F^2 ) $,   
Eq.(\ref{PCDCladder}) reads  
\beq
M_\phi =   {\cal O} \left(\frac{4}{\pi^2} \,m_F\right) ={\cal O} \left(\frac{m_F}{2}\right)   
\, .
\label{roughmphi}
\eeq
Furthermore the Pagels-Stokar formula for $F_\pi^2\simeq 
(N_C\xi^2/2 \pi^2) \, m_F^2$ in the ladder approximation (see Eq.(\ref{PSv})), 
\beq
 v_{\rm EW}^2=(246\,  {\rm GeV})^2 
= N_D F_\pi^2 \simeq 
N_F N_C\frac{\xi^2}{4\pi^2} \, m_F^2   
\simeq  m_F^2 \left[\frac{N_F}{8}\frac{N_C}{4}\right],
\label{PSformulaEW}
\eeq
with $N_D (=N_F/2)$ being the number of the electroweak doublets, which combined with Eq.(\ref{PCDCladder}), leads to 
\beq
F_\phi = {\cal O} \left(\frac{2\pi}{\xi}\,  v_{\rm EW}\right) 
= {\cal O} \left( 5\,  v_{\rm EW} \right) \,.
\label{roughfphi}
\eeq
Note that both Eqs.(\ref{roughmphi}) and (\ref{roughfphi}) are {\it independently of $N_F$ and $N_C$}, as far as the PCDC makes sense (as in WTC in the anti-Veneziano limit).

At this point, we should comment on a widely spread folklore claiming that the natural scale of the technicolor would be  ${\cal O} ({\rm TeV})$ and hence the Higgs mass 125 GeV cannot be 
obtained without fine tuning. This is totally unjustified statement tinted by the naive scale up of the QCD with $N_F=2,\, N_C=3$ where $m_F ={\cal O} (650\,\,  {\rm GeV})$ from  Eq.(\ref{PSformulaEW}),   in sharp contrast to
$m_F \simeq$  246 GeV in our walking theory with $N_F=8, N_C=4$ based on the same PS formula. Moreover,  the folklore presumes  the naive
non-relativistic estimate $M_\phi \sim 2 m_F$ which would give  $M_\phi ={\cal O}({\rm TeV})$ for $N_F=2$ $N_C=3$ in the QCD scale-up, where the PCDC does not make sense and no particular constraint on the flavor-singlet scalar bound state (no longer a dilaton-like object), since the ordinary QCD has no scale symmetry at all. In contrast, the approximate scale symmetry in the walking theory dictates the
PCDC relation, which yields $M_\phi \simeq 125$ GeV $\ll 2 m_F$, instead of the above naive non-relativistic guess.

The result in fact reflects a generic scaling law, 
\beq
\frac{M_\phi}{v_{\rm EW} }\sim \frac{M_\phi}{F_\phi} \sim \frac{1}{\sqrt{N_C N_F}} \rightarrow 0\,,
\label{TDscaling}
\eeq
 {\it independently of the ladder approximation},  
since it is a direct consequence of the the anti-Veneziano limit,  $N_F,N_C$ scaling of the PCDC relation $M_\phi^2 F_\phi^2 \propto N_F N_C m_F^4$ and  of 
$F_\phi^2 \propto v_{\rm EW}^2 \propto N_F N_C m_F^2$ coming from the definition of $F_\phi$ and $v_{\rm EW}$ in terms of the dynamical 
mass of the technifermions. 
Then in 
 the ``anti-Veneziano limit'' $N_C \rightarrow \infty$ with  $N_F/N_C = {\rm fixed}$ ($ \gg 1$, in accord with the IR conformality near the conformal window), 
 the TD parametrically has a 
vanishing mass compared with the spontaneous scale-symmetry breaking scale $F_\phi\, (\ll \Lambda_{\rm TC})$: $ M_\phi/F_\phi  \, (\gg M_\phi/\Lambda_{\rm TC}) \rightarrow 0$~\cite{Kurachi:2014xla,Kurachi:2014qma}. 
 
 Thus the light TD with the mass of 125 GeV can be regarded as a pseudo NG boson in the anti-Veneziano limit near the conformal window \cite{Kurachi:2014xla}:
Such a light TD  is in fact similar to the $\eta^\prime$ meson in the sense that $\eta^\prime$ is widely accepted to be  a pseudo-NG boson having a parametrically vanishing mass $M_{\eta^\prime}/F_\pi ={\cal O} (\sqrt{N_F}/N_C) <M_{\eta^\prime}/\Lambda_{\rm QCD}
={\cal O} (\sqrt{N_F/N_C}) \rightarrow 0$ 
in the large $N_C$ limit 
 with $N_F/N_C$ fixed ($\ll 1$) in the ordinary QCD (original Veneziano limit), a la Witten-Veneziano. 
 In fact the anomalous chiral WT identity for $ A_\mu^0(x) =\sum^{N_F}_{i=1} \bar q_i(x) \gamma_\mu \gamma_5 q_i (x)$ reads:
  \begin{eqnarray}
 N_F F_\pi^2 M_{\eta^\prime}^2 &=& {\cal F.T.} \left\langle T\left(\partial^\mu A^0_\mu(x) \cdot \partial^\mu A^0_\mu(0)\right)\right\rangle = 
{\cal F.T.} \left\langle T\left(N_F \frac{\alpha}{4\pi} G^{\mu\nu} {\tilde G}_{\mu\nu} (x)\cdot  N_F \frac{\alpha}{4\pi} G^{\mu\nu} {\tilde G}_{\mu\nu}(0)
\right)\right\rangle 
\nonumber\\
 &\sim& N_F^2\alpha^2 \times \left[
 N_C^2 \,\, ({\rm gluon\,\, loop\, , Fig.5})  +
 N_C^3 N_F \, \alpha^2 \,\ ({\rm fermion\,\,loop\,, Fig. 5 })
 \right]\,. 
 \end{eqnarray}
In the Veneziano limit $N_F/N_C \ll 1$ the gluon loop dominates the fermion loop, and hence we have 
\beq
 M_{\eta^\prime}^2 \sim \frac{N_F}{F_\pi^2} \Lambda_{\rm QCD}^4 \sim \frac{N_F}{N_C}\Lambda_{\rm QCD}^2 \ll \Lambda_{\rm QCD}^2\quad \frac{M_{\eta^\prime}^2}{F_\pi^2} \sim \frac{N_F}{N_C^2} \ll 1\,. 
 \eeq

 \begin{figure}
  \begin{center}
    \includegraphics[scale=0.7,clip]{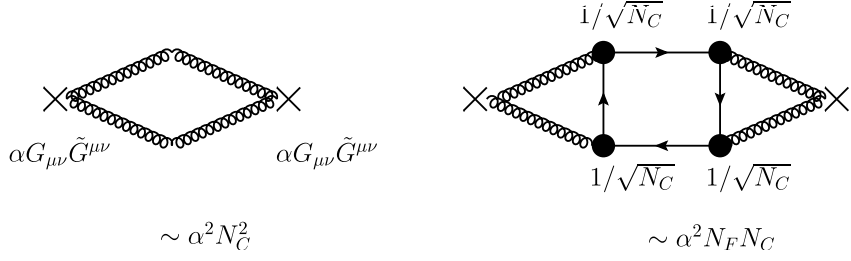}
  \end{center}
\caption{
 The loop diagrams contributing to the correlation function of $\alpha G_{\mu\nu} \tilde{G}^{\mu\nu}$ coming from 
 the gluon loop (left panel) and fermion loop (right panel). 
 The large $N_C$ and $N_F$ scalings have also been specified. 
}
\label{GGtil-loops}
\end{figure}

\noindent 
 Thus the TD in the anti-Veneziano limit and $\eta^\prime$ in the Veneziano limit are resemblant.

What about the $\eta^\prime$ in the anti-Veneziano limit, then? (No TD exists in the Veneziano limit, since it is not a walking theory.)
From Eq.(\ref{Vlimit}) and  Fig.~\ref{GGtil-loops}, we see the fermion loop dominates the gluon loop, contrary to the Veneziano limit.
Then we infer 
\beq
M_{\eta^\prime}^2 \sim \frac{N_F \alpha^2}{F_\pi^2} \left(N_C^3 N_F\alpha^2 m_F^2\right)
\sim \left(\frac{N_F}{N_C} m_F\right)^2 \gg m_F^2 
\,,
\label{etaprime}
\eeq 
 where we have again subtracted the perturbative contribution to the $U(1)_A$ anomaly.  
This could be tested on the lattice simulation \cite{Lattice2015}.
In the anti-Veneziano limit 
the $\eta^\prime$ mass does not go to zero and hence 
has no NG boson nature in contrast to the TD.  
In the walking case with $N_C/N_F \gg 1$ and $m_F\ll \Lambda_{\rm TC}$,  a simple scaling suggests that  
$M_\phi^2 ={\cal O} (m_F^2)$ and $ M_{\eta^\prime}^2 ={\cal O} (N_F^2/N_C^2) \, m_F^2 (\gg M_\phi^2)$. 

For the phenomenological studies, the PCDC in Eq.(\ref{PCDCladder}) together with the Pagels-Stokar formula in Eq.(\ref{PSformulaEW}) yields
 a more concrete result: 
\beq
M_\phi^2 \simeq \left(\frac{v_{\rm EW}}{2}\right)^2 \cdot \left(  \frac{5\, v_{\rm EW}}{F_\phi} \right)^2 \cdot \left[\frac{8}{N_F}\frac{4}{N_C}\right].
\label{PCDCrel}
\eeq
which is in accord with  \cite{Hashimoto:2010nw}  based on the improved ladder result (with the two-loop coupling as the input coupling).
It was first pointed out in Ref. \cite{Matsuzaki:2012gd} that this ladder PCDC result accommodates the $125\, {\rm GeV}$ Higgs
with $F_\phi = {\cal O}\, ( {\rm TeV})$ for the one-family model with $N_F=8$. 

Phenomenologically, the most interesting case is the one-family model ($N_F=8$) \cite{Dimopoulos:1979sp,Farhi:1980xs} with $N_C=4$, 
where we have $ m_F\simeq v_{\rm EW} =246 \,{\rm GeV}$ (Eq.(\ref{PSformulaEW})), and
Eq.(\ref{PCDCrel}) quite naturally accommodates the realistic point~\cite{Matsuzaki:2012vc,Matsuzaki:2012mk}: 
 \beq
 M_\phi \simeq \frac{v_{\rm EW}}{2}\simeq
 \frac{m_F}{2}\simeq 125\, {\rm GeV}, \quad
F_\phi
\simeq 5  \,v_{\rm EW} \simeq 1.25 \, {\rm TeV} \quad (N_C=4, N_F=8) \,,
\label{Fphi}
\eeq
which 
is in accord with the above rough estimate $F_\phi \sim \sqrt{N_C N_F} \, m_F =\sqrt{4 \times 8} \, v_{\rm EW}$.   
Amazingly, this value of $F_\phi$ turned out to be
consistent with 
the current LHC Higgs data \cite{Matsuzaki:2012mk}, as we shall discuss later.

In passing, the TD potential in ${\cal L}^{S}_{(2){\rm anomaly}} $ of 
Eq.(\ref{sChPT}) (with $m_\phi$ denoted as $M_\phi$ here) written in terms $\chi=e^{\phi/F_\phi}$ is rewritten in the TD field $\phi$ as~\cite{Matsuzaki:2012vc}
\begin{equation} 
V(\phi) =-{\cal L}^{S}_{(2){\rm anomaly}} 
= - \frac{M_\phi^2 F_\phi^2}{16} +\frac{1}{2}M_\phi^2\,\phi^2 +\frac{4}{3} \frac{M_\phi^2}{F_\phi} \,\phi^3 
+ 2 \frac{M_\phi^2}{F_\phi^2}\, \phi^4 
+ \cdots 
\,. 
\label{dilatonpotential}
\end{equation}
It is remarkable to notice that in the anti-Veneziano limit 
the TD self couplings (trilinear and quartic couplings) are highly suppressed:
\beq
\frac{4}{3} \frac{M_\phi^2}{F_\phi} \sim \frac{1}{\sqrt{N_F N_C}}\,,\quad 
2 \frac{M_\phi^2}{F_\phi^2} \sim \frac{1}{N_F N_C}
\eeq
by $M_\phi/F_\phi \sim 1/\sqrt{N_F N_C}$ and $M_\phi \sim N_F^0 N_C^0$. 
It is also interesting to numerically compare the TD self couplings for 
the one-family model ($N_F=8, N_C=4)$ having $v_{\rm EW}/F_\phi \simeq 1/5$ 
with 
the  self couplings of the SM Higgs with $m_h=M_\phi$, by making the ratios: 
\begin{eqnarray} 
\frac{g_{\phi^3}}{g_{h_{\rm SM}^3}}\Bigg|_{M_\phi=m_h} 
&=& \frac{\frac{4 M_\phi^2}{3 F_\phi}}{\frac{m_h^2}{2 v_{\rm EW}}} \Bigg|_{M_\phi=m_h}
\simeq \frac{8}{3} \left( \frac{v_{\rm EW}}{F_\phi}\right) \simeq 0.5 
\,, \nonumber \\ 
\frac{g_{\phi^4}}{g_{h_{\rm SM}^4}} \Bigg|_{M_\phi = m_h} 
&=& \frac{\frac{2 M_\phi^2}{F_\phi^2}}{\frac{m_h^2}{8 v_{\rm EW}^2}}\Bigg|_{M_\phi=m_h} 
= 16 \left( \frac{v_{\rm EW}}{F_\phi} \right)^2 \simeq 0.6 
\,. 
\label{selfcouplings:0}
\end{eqnarray}
This shows that the {\it TD self couplings, although generated by the strongly coupled interactions, are even smaller than those of the 
SM Higgs}, a salient feature of the approximate scale symmetry in the ant-Veneziano limit. 
This is in sharp contrast to 
the widely-believed folklore, ``Strong coupling solutions like Technicolor tend to lead to a strongly coupled Higgs''~\cite{Seiberg},  
as noted in the Introduction.

Finally, we should stress that the above estimated TD mass is stable against the feedback effects of the ETC ($G_c$ term) through particularly the top quark loop,
because of the large $F_\phi \simeq 5 v_{\rm EW}$. The loop corrections at the effective theory level including the SM sector and ETC effects were  estimated to
be~\cite{Matsuzaki:2012vc}
\beq
\frac{\delta M_\phi^2|_{\phi^4}}{M_\phi^2} \simeq  24 \frac{m_F^2}{(4\pi F_\phi)^2} \simeq 6\times 10^{-3}\,, \quad
\frac{\delta M_\phi^2|_{\rm ETC/Yukawa}}{M_\phi^2} \simeq - 12 (3-\gamma_m)^2 \frac{m_F^2}{(4\pi F_\phi)^2} \frac{m_t^2}{M_\phi^2} \simeq -(3-\gamma_m)^2 \frac{\delta M_\phi^2|_{\phi^4}}{M_\phi^2} 
\,, \eeq
which cancels each other as
 $\delta M_\phi^2/M_\phi^2 \approx 0$  for $\gamma_m=2$ (see the comments of the last section), and are within 1\% corrections to $M_\phi$ ($ \delta M_\phi^2/M_\phi^2 \simeq - 1.8 \times 10^{-2} $)  
 even for $\gamma_m=1$. Other loop effects are negligibly small.

 \section{The LHC Phenomenology of technidilaton}
 \label{LHCpheno}

Now that we have established ladder estimate of the mass and the decay constant of the TD, 
Eq.(\ref{PCDCrel}) and Eq.(\ref{Fphi}), we now discuss 
up-dating the previous analyses of the TD \cite{Matsuzaki:2012mk,Matsuzaki:2012xx} in view of the 
latest LHC data of the 125 GeV Higgs.

  One can obtain the TD couplings to the SM gauge bosons and the SM fermions just by 
scaling from the SM Higgs as $v_{\rm EW} \to F_\phi$~\cite{Matsuzaki:2012vc,Matsuzaki:2012mk}:  
\begin{eqnarray} 
  \frac{g_{\phi WW/ZZ}}{g_{ h_{\rm SM} WW/ZZ }}   
 &=&  \frac{g_{\phi ff}}{g_{h_{\rm SM} ff}}  \,
\quad ({\rm for} \quad f=t,b,\tau) 
\nonumber \\ 
  &=& \frac{v_{\rm EW}}{F_\phi} \,\quad \left[ \simeq \frac{1}{5} \ll 1\quad \left(N_F=8\,, N_C=4 \right)\right]
  \,.  \label{scaling}
\end{eqnarray} 
On the other hand, in the one-family model with $N_F=8$  
the couplings to digluon and diphoton 
include the colored/charged 
techni-fermion loop contributions along with a factor $N_C$~\cite{Matsuzaki:2012vc,Matsuzaki:2012mk},  
\begin{eqnarray} 
 {\cal L}_{\rm eff}^{\gamma\gamma,g g} =\frac{\phi}{F_\phi} \left\{ 
 \frac{\beta_F(g_s)}{2g_s} G_{\mu\nu}^2 + \frac{\beta_F(e)}{e} F_{\mu\nu}^2 
\right\}\,,\\
 \beta_F(g_s) = \frac{g_s^3}{(4\pi)^2} \frac{4}{3} N_C \,,\quad
\nonumber  
  \beta_F(e) = \frac{e^3}{(4\pi)^2} \frac{16}{9} N_C 
\,,   \label{betas}
\end{eqnarray} 
where the beta functions have been evaluated at the one-loop level. 
Thus one finds  
the scaling from the SM Higgs~\cite{Matsuzaki:2012mk} 
(Detailed formulae are given in the Appendix of Ref.\cite{Matsuzaki:2012vc})~\footnote{ 
One might think that the QCD interaction, which could be significant for the technidilaton, 
could spoil the walking picture based on the fixed point structure in the underlying one-family walking-technicolor dynamics. 
However, it is not the case since the QCD coupling is extremely small in magnitude 
for the energy region relevant to the walking technicolor dynamics, and 
the fixed point structure should be more sensitive to the higher-loops of the pure technicolor dynamics~\cite{Shrock:2013pya} 
rather than the QCD effects,  
if any.      
},  
 \begin{eqnarray} 
\frac{g_{\phi gg}}{g_{h_{\rm SM} gg}} 
&\simeq & 
\frac{v_{\rm EW}}{F_\phi} 
\cdot 
\left( 1 + 2 N_C \right)   \,,
\nonumber \\ 
\frac{g_{\phi \gamma\gamma}}{g_{h_{\rm SM} \gamma\gamma}} 
&\simeq & 
\frac{v_{\rm EW}}{F_\phi} 
\cdot 
 \left( \frac{63 -  16}{47} - \frac{32}{47} N_C \right)  
\,,  \label{g-dip-dig}
\end{eqnarray} 
where in estimating the SM contributions  
we have incorporated only the top (the terms of 1 and 16/47 for $gg$ and $\gamma\gamma$ rates, respectively) and the 
$W$ boson (the term of 63/47 for $\gamma\gamma$ rate) loop contributions.  
 In Table~\ref{tab:BR} the branching fractions for relevant decay channels of the TD at 125 GeV 
are listed in the case of 
 the one-family model with $N_C=4$.  Note that the total width $\Gamma_{\rm tot}=1.15\, {\rm MeV}$  is smaller than the SM Higgs, which 
 reflects the weaker couplings than the those of the latter, in contrast to the widely spread folklore mentioned in the Introduction. 
 
\begin{table} 
\begin{tabular}{|c|c|c|c|c|c|c|c|c|}
\hline 
\hspace{15pt}
BR[\%] \hspace{15pt}
 &\hspace{15pt}
 $gg$ \hspace{15pt}
& \hspace{15pt}
 $bb$ \hspace{15pt}
 & \hspace{15pt}
 $WW$ \hspace{15pt}
& \hspace{15pt}
$ZZ$ \hspace{15pt}
 & \hspace{15pt}
 $\tau\tau$ \hspace{15pt}
 & \hspace{15pt}
 $\gamma\gamma$ \hspace{15pt}
&\hspace{15pt}
 $Z \gamma$ \hspace{15pt}
& \hspace{15pt}
$\mu\mu$ \hspace{15pt}
 \\  
\hline \hline  
$\Gamma_{\rm tot}=1.15\, {\rm MeV}$ 
& 75.1 & 19.6 & 3.56 & 0.38 & 1.19 & 0.068 & 0.0048 & 0.0042 \\ 
\hline 
\end{tabular}
\caption{ The TD branching ratios at 125 GeV in the one-family model with $N_C=4$. The total width is also given.}
\label{tab:BR}
\end{table}

 Calculating the signal strengths for the LHC production categories (gluon gluon fusion (ggF), vector boson fusion (VBF), 
vector boson associate production (VH) and top associate production (ttH)), 
\begin{equation} 
 \mu^{i}_{X_1X_2} = \frac{\sigma^i_{\phi} \times {\rm BR}(\phi \to X_1X_2)}{\sigma^i_{\rm h_{\rm SM}} \times {\rm BR}(h \to X_1X_2)}
 \,, 
\end{equation} 
as a function of the overall coupling $v_{\rm EW}/F_\phi$ for given the number of $N_C$, 
we may fit the $\mu^i_{X_1X_2}$ to the latest data on the Higgs coupling measurements~\cite{ATLAS:2015:Higgs,ATLAS:2013oma,Aad:2013wqa,ATLAS:2013wla,ATLAS-tautau,TheATLAScollaboration:2013lia,CMS:ril,CMS:xwa,Chatrchyan:2013iaa,CMS:2013yea,Chatrchyan:2014nva,Chatrchyan:2013zna}.  
to determine the best-fit value of $v_{\rm EW}/F_\phi$. 
The result of the goodness of fit is shown in Table~\ref{TDfit}, which updates the analysis in Ref.~\cite{Matsuzaki:2012mk}.  
 The Table~\ref{TDfit} shows that the TD in the one-family model with $N_C=4$ is favored by the current LHC Higgs data 
 as much the same level as the SM Higgs. 
Remarkably, the best fit value $[v_{\rm EW}/F_\phi]_{\rm best}\simeq 0.2$, i.e. $F_\phi \simeq 5 v_{\rm EW}$ for $N_C=4$ 
is in excellent agreement with 
the ladder estimate of the TD mass $\simeq 125$ GeV in Eq.(\ref{Fphi})!

In Table~\ref{mu-vals} 
we also make a list of the predicted signal strengths for each production category 
for the best fit value of $v_{\rm EW}/F_\phi \simeq 0.23$ in the case with $N_C=4$, 
along with the latest result reported from the ATLAS and CMS experiments~\cite{ATLAS:2015:Higgs,ATLAS:2013oma,Aad:2013wqa,ATLAS:2013wla,ATLAS-tautau,TheATLAScollaboration:2013lia,CMS:ril,CMS:xwa,Chatrchyan:2013iaa,CMS:2013yea,Chatrchyan:2014nva,Chatrchyan:2013zna}. 
Note the TD signal strengths in the dijet category (VBF), which involves the contamination by about 30\% from the ggF + gluon jets, 
$gg \to \phi + gg$. The contribution from the ggF is highly enhanced compared to the SM Higgs case, 
due to the extra techni-quark loop contribution, which compensates the overall suppression by the direct 
VBF coupling $v_{\rm EW}/F_\phi \simeq 0.2$ to lift the event rate up to be comparable to the SM Higgs case. 
(The detailed estimate of the ggF contamination is given in Appendix~\ref{dijet}.) 
Note also the suppression of the VH-$b\bar{b}$-channel, which would be the characteristic signature of the TD 
to be distinguishable from the SM Higgs. More data from the upcoming LHC Run-II will draw a conclusive answer 
to whether or not the LHC Higgs is the SM Higgs, or the TD.

\begin{table} 
\begin{tabular}{|c|c|c|}
\hline 
\hspace{30pt} $N_C$ \hspace{30pt} &  
\hspace{30pt} $[v_{\rm EW}/F_\phi]_{\rm best}$ \hspace{30pt} & 
\hspace{30pt}$\chi^2_{\rm min}/{\rm d.o.f.}$ \hspace{30pt} \\ 
\hline\hline 
3 & 0.27 & $ 25/17\simeq 1.5$  \\ 
\hline 
4 & 0.23 & $ 16/17 \simeq 0.92$\\ 
\hline 
5 & 0.17 & $ 32/17 \simeq 2.0 $ \\ 
\hline  
\hline 
0\,[\rm SM Higgs] & 1 & 8.0/18 $\simeq 0.44$ \\ 
\hline 
\end{tabular}
\caption{The best fit values of $v_{\rm EW}/F_\phi$ for the one-family model with 
$N_C=3,4,5$ displayed together with 
the minimum of the $\chi^2$ ($\chi^2_{\rm min}$) normalized by the degree of freedom. 
Also has been shown in the last column the case of the SM Higgs corresponding to $N_C=0$ and $v_{\rm EW}/F_\phi=1$. }
\label{TDfit}
\end{table}

\begin{table}[ht] 
\begin{tabular}{|c|c|c|} 
 \hline 
\hspace{30pt} 
TD signal strengths $(v_{\rm EW}/F_\phi=0.23, N_C=4)$ 
\hspace{30pt} 
& 
\hspace{30pt} 
ATLAS 
\hspace{30pt} 
& 
\hspace{30pt} 
 CMS  
 \hspace{30pt} 
\\  
 \hline \hline 
$\mu_{\gamma\gamma}^{\rm ggF} \simeq 1.4$ & $1.32 \pm 0.38$ & $1.13 \pm 0.35$ \\ 
\hline 
$\mu_{ZZ}^{\rm ggF} \simeq 1.0$ & $1.7 \pm 0.5$ & $0.83 \pm 0.28$\\ 
\hline 
$\mu_{WW}^{\rm ggF} \simeq 1.0$ & $0.98 \pm 0.28$ & $0.72 \pm 0.37$ \\ 
\hline 
$\mu_{\tau\tau}^{\rm ggF} \simeq 1.0$ & $2.0 \pm 1.4$ & $1.1 \pm 0.46$ \\ 
\hline 
\hline 
$\mu_{\gamma\gamma}^{\rm VBF} \simeq 0.87$ (0.019) & $0.8 \pm 0.7$ & $1.16 \pm 0.59$ \\ 
\hline 
$\mu_{ZZ}^{\rm VBF} \simeq 0.61$ (0.014) & $0.3 \pm 1.3$  & $1.45 \pm 0.76$ \\ 
\hline 
$\mu_{WW}^{\rm VBF} \simeq 0.61$ (0.014) &  $1.28 \pm 0.51$ & $0.62 \pm 0.53$ \\ 
\hline 
$\mu_{\tau\tau}^{\rm VBF} \simeq 0.61$ (0.014) &  $1.24 \pm 0.57$ & $0.94 \pm 0.41$ \\ 
\hline 
$\mu_{bb}^{\rm VH} \simeq 0.014$ &  $0.52 \pm 0.40$ & $1.0 \pm 0.50$ \\ 
\hline 
\end{tabular}
\caption{
The predicted signal strengths of the TD with $v_{\rm EW}/F_\phi=0.23$ in the case of the one-family model with 
$N_C=4$. The numbers in the parentheses  correspond to the amount estimated without contamination from 
the ggF process. 
Also have been displayed the latest data on the Higgs coupling measurements reported from the 
ATLAS and CMS experiments~\cite{ATLAS:2015:Higgs,ATLAS:2013oma,Aad:2013wqa,ATLAS:2013wla,ATLAS-tautau,TheATLAScollaboration:2013lia,CMS:ril,CMS:xwa,Chatrchyan:2013iaa,CMS:2013yea,Chatrchyan:2014nva,Chatrchyan:2013zna}. }
\label{mu-vals}
\end{table}

The ATLAS and CMS have made a plot of the LHC Higgs couplings to the SM particles against the SM particle masses~\cite{SM-Higgs-Yukawa}, 
shown that the LHC Higgs couplings to fermions have aligned very well with the SM Higgs boson
properties. The plot has been made by assuming no contributions beyond the SM in loops, 
i.e., no contributions beyond SM to diphoton and digluon couplings. 
However, as explicitly seen from Eq.(\ref{betas}), the technidilaton couplings to 
diphoton and digluon significantly include the terms beyond the SM, technifermion contributions 
charged under the $U(1)_{\rm em}$ or QCD color. 
In this respect, such a plot cannot be applied to the technidilaton. 
In fact, the successful consistency with the LHC Higgs coupling measurement, as shown in Table~\ref{mu-vals}, 
is due to those beyond SM contributions, which especially enhance the ggF production cross section, 
balanced by the overall suppression due to the coupling $F_\phi$ larger than $v_{\rm EW}$ 
by a factor of 5.

 \section{Beyond technidilaton: Other technihadrons?}
 \label{BeyondTD}
 
 As we discussed in subsection \ref{ConformalPT} (see discussions below Eq.(\ref{Miranskyscaling2})),
 other techni-hadron (techni-$\rho$, techni-$a_1$, technibaryon, etc.) masses  also have masses of order, $M_A ={\cal O}(C_A(r) m_F)=C_A (r) \cdot \Lambda_{\rm TC}\cdot  f\left(\frac{\alpha}{\alpha_{\rm cr}} \right) (\ll \Lambda_{\rm TC})$, with the universal scaling of Miransky-BKT type, $f\left(\frac{\alpha}{\alpha_{\rm cr}} \right)\sim \tilde f(r) $, up to the non-universal coefficient $C_A(r)$ depending on the each techni-hadron $A$, with possible dependence on $r=N_F/N_C$ in the anti-Veneziano limit. This is in sharp contrast to the ordinary QCD where $F_\pi ={\cal O}(m_F)={\cal O} (\Lambda_{\rm QCD})$. The TD as a pseudo NG boson has the mass solely due to the explicit breaking of the scale symmetry via the PCDC just in the same way as the pion does. As mentioned above, the TD mass $M_\phi ={\cal O} (m_F/2)$ is independent of $N_C,N_F$ as the PCDC relation dictates. 
 
 In contrast,    
 all the non-NG boson techni-hadrons have {\it no constraints from the 
 PCDC as the explicit breaking} of the scale symmetry but do have {\it constraints from the SSB} of the scale symmetry, so that they
  should have masses on the scale of the SSB  of the scale symmetry, characterized by $F_\phi\sim \sqrt{N_FN_C} m_F$ which is much larger than $2 m_F$ of the naive nonrelativistic quark model picture, particularly 
 in the anti-Veneziano limit of the walking case, $N_C \rightarrow \infty$ with $\lambda\equiv N_C \alpha=$ constant ($\alpha>\alpha_{\rm cr})$),
 and with $r\equiv N_F/N_C=$ constant $(\gg 1)$. We naturally
expect that  their masses are generally of order of ${\cal O}({\rm TeV's})$:   
\beq
M_{A} ={\cal O} (C_A(r)\,  m_F)  = {\cal O} ({\rm TeV's}) 
\gg 2 m_F \gg M_\phi  
\eeq
with $C_A(r) \gg 1 $ for $r\rightarrow r_{\rm cr}$.
This  is consistent with the straightforward computation of large $N_F$ QCD based on the ladder BS
equation combined with the ladder SD equation, $M_\rho \simeq 4 m_F\simeq 12 F_\pi$ (for $N_C=3$) \cite{Harada:2003dc,Kurachi:2006ej} (This corresponds to $\simeq 6 v_{\rm EW}$ in the one-family model with $N_F=8, N_C=4$.)  
, which  is  somewhat larger than
the QCD case $M_\rho \sim 8 F_\pi$.   Being highly strong-coupling relativistic result, it is contrasted to the naive weak-coupling non-relativistic quark-model view 
  $M_A \sim 2 m_F$. This is also compared with the $N_C$ counting of the  bound state masses ${\cal O} ( \Lambda_{\rm QCD})$ in the ordinary QCD, where the
 gluon loop is dominant, while {\it the fermion loop dominates in the anti-Veneziano limit in the walking theory}.
Also the holographic calculations tend to give $M_{A} \gg M_\phi$, and so do the recent lattice calculations \cite{Appelquist:2014zsa,LatKMI}.

 As usual, the IR conformal physics of the WTC should  be described by the low-lying composite fields as effective fields, 
 in a way to realize all the symmetry structure of the underlying theory.
 Such an effective theory of WTC having higher resonances together with the 125 GeV TD is already constructed as a straightforward extension of sChPT \cite{Matsuzaki:2012vc,Matsuzaki:2013eva}, i.e, the scale-invariant version \cite{Kurachi:2014qma} of the Hidden Local Symmetry (HLS) model \cite{Bando:1984ej,Harada:2003jx},  (the ``sHLS model''),
  where  the technirho mass terms have the scale-invariance non-linearly realized  by the TD field $\chi=e^{\phi/F_\phi}$, with the SSB of the scale invariance  characterized by the scale of $F_\phi$,
 while the Higgs (TD) mass term in the TD potential,  on the  order of $m_F (\ll F_\phi)$,  is the only source of the explicit breaking of the scale symmetry related (via PCDC) to the nonperturbative trace anomaly 
 of the underlying theory.

Interesting  candidate for such techni-hadrons may account for the diboson excesses recently observed at LHC at 2 TeV~\cite{Aad:2015owa, Khachatryan:2014hpa}, which can be identified with the walking technirho \cite{Fukano:2015hga}. A smoking gun  of the walking technirho is the absence of the decay to the 125 GeV Higgs (TD), which is forbidden by the scale symmetry explicitly
broken 
only by the Higgs (TD) mass term  (corresponding to the nonperturbative trace anomaly in the underlying WTC)
\cite{Fukano:2015uga}.  Actually, the salient feature of the scale symmetry of the generic effective theory not just the sHLS model, containing the SM
 gauge bosons and the Higgs plus new vector bosons (any other massive particles as well), 
 is the absence of the decay of the new vector bosons  such as the technirho (and also other higher resonances)  into  the 125 GeV Higgs plus the SM gauge bosons \cite{Fukano:2015uga}. If such a decay of new particles is not found at LHC Run II, then the 125 GeV Higgs is nothing but the dilaton (TD in the case of the WTC) responsible for the {\it nonlinearly realized} scale symmetry, i.e., the {\it SSB} of the scale symmetry, no matter what underlying theory may be beyond the SM. This should be tested in the future LHC experiments.

\section{Summary and discussions}
\label{Summary}

In conclusion we have shown that the technidilaton in the walking technicolor, typically realized in the one-family model ($N_F=8, N_C=4$), is a naturally light composite Higgs to be identified with the 125 GeV Higgs, and is a weakly coupled composite state out of the strongly coupled conformal gauge dynamics, 
with its each coupling being even weaker than the SM Higgs.

In this paper, the walking technicolor with $\gamma_m=1$, originally based on the ladder SD equation,
 is reformulated in terms of the
Caswell-Banks-Zaks infrared fixed point $\alpha_*$ of the $SU(N_C)$ gauge theory for $N_F$ massless flavors, with the intrinsic scale $\Lambda_{\rm TC}$, in the anti-Veneziano limit Eq.(\ref{antiVeneziano}):
\beq
N_C  \rightarrow \infty\,,\, \lambda\equiv N_C \,\alpha = {\rm fixed}\,,  \,\, \,{\rm with} \,\, r\equiv N_F/N_C= {\rm fixed} \gg 1\,,
\eeq 
where the input coupling in the SD kernel is given by the constant coupling 
Eq.(\ref{ladderCBZ}) and Fig.~\ref{alpha-beta-a-V}:
\beq
\alpha(x) = \alpha_*\,\theta(\Lambda_{\rm TC}^2-x)\, ,\quad (x=-p^2>0) \,.
\eeq 
We have shown in the anti-Veneziano limit that the SSB of the chiral (electroweak) and scale symmetries takes place due to the technifermion condensate in Eq.(\ref{Rcondensate}), $\langle \bar F F\rangle_R \simeq -\frac{N_C}{\pi^2} m_F^3$, at strong coupling, $\alpha=\alpha_* >\alpha_{\rm cr}\, (r<r_{\rm cr})$, in such a way that it is essentially a continuous phase transition at criticality $r=r_{\rm cr}$ as 
the conformal phase transition characterized by the Miransky-BKT scaling of the essential singularity, Eq.(\ref{Miranskyscaling}):
\beq
m_F\sim \Lambda_{\rm TC}\cdot  \exp \left(-\frac{\pi}{\sqrt{\frac{\alpha}{\alpha_{\rm cr}} -1}}\right) \ll \Lambda_{\rm TC}\, \quad 
\left(0<\frac{\alpha}{\alpha_{\rm cr}}-1=\frac{\alpha_*}{\alpha_{\rm cr}}-1 \sim (r_{\rm cr} -r) \ll 1  \right)\,.
\eeq
Here the CBZ IR fixed point (viewed from $\mu>\Lambda_{\rm TC}$) is now regarded as the UV fixed point  (viewed from
$\mu<\Lambda_{\rm TC}$. The corresponding nonperturbative beta function has a multiple zero with the zero curvature at $\alpha(\mu)=\alpha_*=\alpha_{\rm cr}$ as in Eq. (\ref
{NPbeta2}), where the coupling turns over to the weak coupling region $\alpha(\mu) <\alpha_*=\alpha_{\rm cr}$. See Fig.\ref{beta:whole}.

Accordingly, while there are no bound states in the conformal window $\alpha<\alpha_{\rm cr}$ (unparticle phase), bound states exist only in the SSB phase, all of order $m_F$ up to the factors depending on $r=N_F/N_C$ in the anti-Veneziano limit.

First, the pseudo NG boson masses come only from explicit breakings of the internal symmetry, the chiral $SU(8)_L\times SU(8)_R$ in the car of one-family model,  
through the Dashen formula.    
The technipions (uneaten pseudo NG bosons of the chiral symmetry) 
pick up the explicit breaking of $SU(8)_L \times SU(8)_R$ by the SM gauge interactions and ETC gauge interactions, 
\begin{eqnarray}
M_\pi^2 &=& \langle \pi| {\cal H}_{breaking}|\pi\rangle =\frac{1}{F_\pi^2} \langle 0|\delta_\pi \delta_\pi  {\cal H}_{breaking}|0\rangle
\nonumber
\,,\\
\quad  {\cal H}_{breaking}&=&g^2_{(SM/ETC)} \int d^4x D_{\mu\nu}^{(SM/ETC)}(x) \,\,T \,\left(J^\mu_{(SM/ETC)}(x) J^\nu(0)_{(SM/ETC)}\right)\,.
\end{eqnarray}
with enhancement due to the large anomalous dimension, where $\delta_\pi O\equiv [i Q_{5\pi},O]$, with the broken generator charge $Q_{5\pi}$ corresponding to the $\pi$, and $D_{\mu\nu}^{(SM/ETC)}$ is the
SM/ETC gauge boson propagator coupled to the source current $J^\mu_{(SM/ETC)}$. They all have mass 
 of
$\gtrsim {\cal O} (m_F)$,  see Eq.(\ref{TCpionmass}). 

Similarly,  
the technidilaton, the pseudo NG boson of
the scale symmetry,  acquires
the mass from the explicit breaking of the scale symmetry, $m_F$, 
since the SSB of the scale symmetry due to the mass generation of $m_F$ also breaks the scale symmetry explicitly. The mass is also evaluated 
through the Dashen formula for the nonperturbative trace anomaly, 
Eq.(\ref{Traceanomalycontinuum}), 
this time the PCDC relation, Eq.(\ref{PCDCladder}):
\begin{eqnarray}
M_\phi^2 &=& \langle \phi| {\cal H}_{breaking}|\phi\rangle
= \langle \phi|  \frac{1}{4} \theta^\mu_\mu|\phi\rangle
=\frac{1}{4F_\phi^2} \langle 0|\delta_D \delta_D  
\theta^\mu_\mu|0\rangle \nonumber\\
&=&\frac{1}{F_\phi^2}\cdot
 \langle 0| 4\,  \theta^\mu_\mu|0\rangle 
  \sim \frac{1}{N_F N_C m_F^2} \cdot \left(N_F N_C\frac{16 \xi^2}{\pi^4} m_F^4\right)\simeq \left(\frac{m_F}{2} \right)^2
  \ll F_\phi^2 \,,
 \end{eqnarray}
 where $\delta_D \theta^\mu_\mu = d_\theta \theta^\mu_\mu\, (d_\theta=4) $ is the dilatation transformation.
 Note that the technidilaton decay constant $F_\phi$ is the order parameter of the SSB of the scale symmetry, $F_\phi^2 \sim N_F N_C m_F^2$ by definition,
 while the {\it explicit breaking scale is $m_F$ which is much smaller than the SSB scale $F_\phi$ of the scale symmetry}  in the anti-Veneziano limit (see text). 
 
 We have particularly seen that the nonperturbative trace anomaly is RG-invariant:
 \beq
 \langle \theta^\mu_\mu\rangle = \frac{\beta^{(NP)}(\alpha(\mu))}{4\alpha(\mu)} \langle G_{\mu\nu}^2 \rangle^{\rm NP}_{(\mu)}=\mu-{\rm independent}, 
 \eeq
 in a way that 
 the techni-gluon condensate is enhanced by the anomalous dimension as in Eq.(\ref{G2:final}), which is precisely compensated as the vanishing beta function   
Eq.(\ref{beta2}), finally to arrive at the RG-invariant finite result as 
in Eq.(\ref{theta}). Thus  the 
small beta function near the criticality is only operative for the large hierarchy $m_F \ll \Lambda_{\rm TC}$, while further hierarchy  
$M_\phi \ll m_F\ll F_\phi $ is due to the anti-Veneziano limit~\footnote{ 
Note that other
explicit breakings, quark/lepton mass $m_{q/l}$ (Eq.(\ref{qlmass})), technipion mass $M_\pi$ (Eq.(\ref{TCpionmass})), also scale like $m_{q/l}/F_\phi, M_\pi/F_\phi \sim 1/\sqrt{N_F  N_C} \rightarrow 0$ in the
anti-Veneziano limit.  They have an exactly massless point (switching off the ETC gauge interaction),  in contrast to the technidilaton, though. 
}: 
\beq 
\frac{M_\phi}{F_\phi} \sim \frac{m_F}{F_\phi} \sim \frac{1}{\sqrt{N_F N_C}} \rightarrow 0\,.
\eeq 
It is a salient feature of the anti-Veneziano limit that
the technidilaton 
 has a limit of vanishing mass in units of $F_\phi$, Eq.(\ref{TDscaling}), though not the exact massless point at the criticality of the conformal phase transition point where no bound states exist for the exactly zero explicit breaking $m_F\equiv 0$, i.e., no nonperturbative  trace anomaly (see text).

  This is somewhat similar to the $\eta^\prime$
 meson in ordinary QCD, which is regarded as the pseudo NG boson having mass from the $U(1)_A$ anomaly, $M_{\eta^\prime}/\Lambda_{\rm QCD}\sim \frac{N_F}{N_C} \ll 1$ and 
 $M_{\eta^\prime}/F_\pi \sim \frac{\sqrt{N_F}}{N_C} \rightarrow 0$ 
 in the original Veneziano limit ($r=N_F/N_C \ll 1 $ instead of the anti-Veneziano limit $r\gg 1$). Fate of the $\eta^\prime$ in the anti-Veneziano limit was discussed in the text, see Eq.(\ref{etaprime}). The exact
 massless point is also absent for $\eta^\prime$, since the anomaly cannot be identically zero in the quantum theory. Also note that the realistic value of the $\eta'$ meson is far from
 light in the real-life QCD, which corresponds to the technipions in our case.

 For the phenomenological issue of the technidilaton to be identified with the 125 GeV Higgs, 
 we first noticed that the Pagels-Stokar formula for the weak scale $v_{\rm EW}$ in Eq.(\ref{PSformulaEW}) implies
 $(246 \, {\rm GeV})^2= v_{\rm EW}^2= N_F N_C\frac{\xi^2}{4\pi^2} \, m_F^2$.  
 Then we have a conceptual feature of the technidilaton in the anti-Veneziano limit:
 \beq
 \frac{M_\phi}{v_{\rm EW}} \sim 
 \frac{M_\phi}{F_\phi}
  \sim \frac{1}{\sqrt{N_F N_C}} \rightarrow 0\,.
   \eeq
 More quantitatively, we showed the ladder estimate of the PCDC relation together with the Pagels-Stokar formula leads to Eq.(\ref{PCDCrel}):
 \beq
 M_\phi^2 \simeq \left(\frac{v_{\rm EW}}{2}\right)^2 \cdot \left(  \frac{5\, v_{\rm EW}}{F_\phi} \right)^2 \cdot \left[\frac{8}{N_F}\frac{4}{N_C}\right]\,. 
 \eeq
 Hence in the particular case, the one-family model with $N_F=8, N_C=4$, we  have $m_F \simeq v_{\rm EW}$ and in fact realize the reality of 125 GeV Higgs as in Eq.(\ref{Fphi}):
  \beq
 M_\phi \simeq \frac{v_{\rm EW}}{2} \,,\quad F_\phi \simeq 5\, v_{\rm EW}\, \quad (N_F=8\,,\, N_C=4)\,.
 \eeq

 The result yields in fact a best fit to the current LHC data for the 125 GeV Higgs as was explained in details in Section \ref{LHCpheno}. See Table \ref{TDfit} and \ref{mu-vals}.
The couplings of the technidilaton to the SM particles are small by a factor of $\frac{v_{\rm EW}}{F_\phi} \simeq \frac{1}{5}$, which is compensated by the enhancement by the extra contributions
from the charged/colored technifermions in the one-family model, 
see Eqs.(\ref{scaling}) -- (\ref{g-dip-dig}). Then the net results happened to be similar to that of the SM Higgs within the errors of the current data at LHC.

As to the non-NG boson technihadons, \{A\}, they all have the mass $M_{A}$ characterized by the coefficient $C_A(r)$ depending on the ratio $r=N_F/N_C$
in the anti-Veneziano limit:   
\beq
M_{A}\gtrsim {\cal O}(C_A(r) \, m_F) 
 \gg m_F\gtrsim M_\phi\,,
\eeq 
which takes the form of the universal scaling of essential singularity: $M_A \sim C_A (r)\,
\Lambda_{\rm TC}\cdot  e^{-\pi/\sqrt{\alpha/\alpha_{\rm cr}-1}} \rightarrow 0 \,\, (\alpha/\alpha_{\rm cr}  \rightarrow 1,
{\rm or}\, r \rightarrow r_{\rm cr} )$.
In the anti-Veneziano limit at $\alpha \rightarrow \alpha_{\rm cr}$ we have
\beq
\frac{M_\phi}{M_A}  \sim \frac{1}{C_A(r)} \ll 1\,,  
\quad \frac{M_A}{M_B} \rightarrow \frac{C_A(r)}{C_B(r)} \ne 0, \infty \,.
\eeq

Interesting  candidate for such techni-hadrons may explain the diboson excesses recently observed at LHC at 2 TeV~\cite{Aad:2015owa, Khachatryan:2014hpa}, which can be identified with the walking technirho with $M_\rho\simeq 2$ TeV \cite{Fukano:2015hga}. The excesses suggest a characteristically small  width $\Gamma_{\rm total} 
<100$ GeV \cite{Aad:2015owa}, which can be naturally realized in the anti-Veneziano limit: 
\beq
\frac{\Gamma_{\rm total}}{M_\rho}\simeq    \frac{\Gamma(\rho \rightarrow WW/WZ)}{M_\rho} \simeq \frac{1}{48\pi}  \frac{g_{\rho\pi\pi}^2}{N_D} \sim \frac{1}{N_F N_C}\rightarrow 0\,,
\eeq 
where $N_D=N_F/2$ is the number of the weak-doublets. In fact our one-family model $N_F=8, N_C=4$ can
reproduces the features of the excesses very well \cite{Fukano:2015hga}.  The fact that $\Gamma_{\rm total} \simeq \Gamma(\rho \rightarrow WW/WZ)  $ is related to   
a  smoking gun  of the walking technirho, namely the absence of the decay to the 125 GeV Higgs (TD), which is forbidden by the scale symmetry explicitly
broken 
only by the Higgs (TD) mass term  (corresponding to the nonperturbative trace anomaly in the underlying WTC)
\cite{Fukano:2015uga}.  Actually, it was shown \cite{Fukano:2015uga} that the salient feature of the scale symmetry of the generic effective theory not just the sHLS model, containing the SM
 gauge bosons and the Higgs plus new vector bosons (any other massive particles as well), 
 is the absence of the decay of the new vector bosons    
 into  the 125 GeV Higgs plus the SM gauge bosons, invalidating the so-called ``Equivalence Theorem''. It was further shown that if such a decay of new particles is not found at LHC Run II, then the 125 GeV Higgs is nothing but the dilaton (technidilaton in the case of the WTC) responsible for the {\it nonlinearly realized} scale symmetry, i.e., the {\it SSB} of the scale symmetry, no matter whatever underlying theory may be beyond the SM. This should definitely be tested in the future LHC experiments.  We will see.

Several comments are in order:

1. The $S$ parameter: The ladder BS calculation of the S parameter from the techni-sector alone was done near the criticality \cite{Harada:2005ru,Kurachi:2006mu}, suggesting a sizable
reduction, up till the 40\% reduction per one weak doublet compared with the QCD. Including a (weak) ETC effects among technifermions ($G_b$ terms in Eq.(\ref{four-fermions}))  for $\alpha/\alpha_{\rm cr} >1$ in the ladder BS calculation
further reduces it to 
$\frac{S}{N_F N_C} \simeq 0.03$, which would imply $S \simeq 1$ for the one-family model ($N_F=8,\, N_C=4$). This is still in conflict with the bound from electroweak precision experiments, 
$S<0.1$. 
The $S$ parameter from the TC sector, however,  is not necessarily in conflict with the experimental
value of the $S$ from the electroweak precision measurements, since the contributions from the TC sector 
can easily be cancelled by the strong mixing with the SM fermion contribution
through the 
ETC  
interactions \cite{Dimopoulos:1979es} ($G_c$ terms in Eq.(\ref{four-fermions})), as in the fermion delocalization of the Higgsless model~\cite{Cacciapaglia:2004rb}. 
the inclusion of the strong ETC effects plus the induced four-fermion effects of WTC, 
and/or the strong ETC mixing effects between the technicolor and SM sectors ($G_c$ terms in Eq.(\ref{four-fermions})).
This should be studied 
explicitly in the ladder BS equation. 
 Moreover,   even within the TC sector alone, 
 there exists a way to resolve this problem as demonstrated in the holographic model~\cite{Haba:2010hu,Matsuzaki:2012xx},  
 where we can reduce  $S$ 
 by tuning the holographic parameter of strength of the techni-gluon condensate $G$ through the $z_m$ (position of the infrared brane)   
in a way 
consistent with the TD mass of 125 GeV and all the current LHC data
for the 125 GeV Higgs~\cite{Matsuzaki:2012xx}.  
This approach can be constrained from the technipion mass bound from experiments \cite{Kurachi:2014xla}. 
Such a large gluonic effects cannot be incorporated into the ladder calculations in principle, and should be checked by the lattice calculations.
The more straightforward calculations on the lattice
are highly desired anyway.

2. The top quark mass: The top mass  is too small for the anomalous dimension $\gamma_m =1$. There are possible resolutions: First, the ETC breaking takes place in a step-wise, with the smallest scale for the third family 
$\Lambda_3 \ll \Lambda_2 \ll \Lambda_1$, and $\Lambda_3$ is less constrained by the FCNC limit than $\Lambda_{2,1} (>10^3 \,{\rm TeV})$, as commonly used in the ETC model building~\cite{Farhi:1980xs,Appelquist:2003uu}.  
Another way out
 \cite{Miransky:1988gk, Matumoto:1989hf} is the even larger anomalous dimension $1< \gamma_m =1+\sqrt{1-\frac{\alpha}{\alpha_{\rm cr}}}<2$ \cite{Miransky:1988gk} in the ladder calculation of the gauged NJL model, Eq.(\ref{gaugedNJLgamma}), due to the strong ETC coupling ($G_b$ terms in Eq.(\ref{four-fermions})).  Note that the $G_a$ term for the third family  in Eq.(\ref{four-fermions}) is much stronger than those for the first and second families, and comparable to $G_b, G_c$ terms at the scale $\Lambda_3$. The strong $G_a$ term for the top quark triggers the top quark condensate \cite{Miransky:1988xi} as well as the technifermion condensate, with $\gamma_m \simeq 2$, but would lead to a different picture than the
 ``top-colored assisted TC'' \cite{Hill:1994hp},
which had a serious problem of the light top pion not absorbed into $W/Z$ \cite{Fukano:2008iv}, since the the $W/Z$ mass is already generated by the TC condensate. In the case at hand, along the critical line of the system together with almost comparable $G_b, G_c$ terms, only a single combination of the top and technifermions may condense, so that no extra NG bosons would be formed.

3.  Straightforward ladder BS calculations: The walking techni-hadrons spectra by the BS and SD equations were calculated  for scalar, vector, axialvector mesons:
$M_S$, $M_\rho$, $M_{a_1}$, together with the decay constants $G_S, F_\rho, F_{a_1}$ \cite{Harada:2003dc}.  The result shows
 $M_S/f_\pi \sim 4$, $M_\rho/F_\pi \simeq M_{a_1}/F_\pi \simeq 12$, which is 
compared with  the real-life QCD, $m_{f_0(1370)}/f_\pi \simeq 15$ ($m_{f_0(500)}/f_\pi \simeq 5$: may not be $\bar q q$ bound state), $m_\rho/f_\pi \simeq 8, m_{a_1}/f_\pi \simeq 13$. The near degeneracy $M_\rho\simeq M_{a_1}$ is also consistent with the lattice
results for $N_F=8$ $N_C=3$~\cite{Appelquist:2014zsa,LatKMI}. Since the calculation does not distinguish between the flavor  singlet and nonsinglet states,
the scalar state $S$ does not corresponds to the technidilaton as a flavor singlet $\bar F F $ bound state. Nevertheless, it would be suggestive that the scalar state $S$ has the mass much smaller than in the QCD. It is well known \cite{Hatsuda:1994pi} that 
the singlet-nonsinglet splitting can be done by introducing the Kobayashi-Maskawa-'t Hooft determinant \cite{Kobayashi:1970ji} arising from the instanton in such a way as to push the flavor-singlet scalar down and nonsinglet up. 
It would be interesting to see the same thing near criticality of the walking theories in the ladder BS equation. 
Another interesting feature of the result of  \cite{Harada:2003dc} is that 
$F_\rho/F_\pi \simeq 2$ compared with the QCD value $F_\rho/f_\pi \simeq \sqrt{2}$, which could be relevant to the 2 TeV diboson excesses at the LHC~\cite{Fukano:2015hga}.

4. One-family model on the lattice:
  We have shown that the ladder results for the one-family model with $N_F=8, N_C=4$ give  the technidilaton as the 125 GeV Higgs the best fit to the current LHC data.
  The holographic estimate also yields a similar result as far as the realistic point is concerned \cite{Matsuzaki:2012xx}.  It was further shown 
  that a natural setting of the ETC model building prefers $N_C=4$.    Although many lattice results 
 indicate  the walking behavior with $\gamma_m \simeq 1 $ \cite{Aoki:2013xza,Appelquist:2014zsa,Hasenfratz:2014rna} and a light flavor-singlet as a candidate for the technidilaton \cite{Aoki:2014oha} in  the $N_F=8, N_C=4$  theory, so far
  no lattice studies for $N_F=8, N_C=4$ have been done. Lattice results for $N_F=8, N_C=4$ are highly desired.

\acknowledgments
We would like to thank Masafumi Kurachi for the collaboration in the early 
stage of this work and useful discussions, and   
Robert Shrock for very helpful discussions and suggestions, 
and great encouragements. We also thank Hidenori S. Fukano for useful discussions.
This work was supported in part by 
the JSPS Grant-in-Aid for Scientific Research  (S) \#22224003, 
(C) \#23540300 (K.Y.) 
and the JSPS Grant-in-Aid for Young Scientists (B) \#15K17645 (S.M.).

\appendix


\section{Ladder estimate of the chiral condensate, anomalous dimension, and OPE}

The bare chiral condensate can be directly estimated from the SSB solution 
Eq.(\ref{Sigma:sol}) at $x=\Lambda^2$ with the
  UV boundary condition Eq.(\ref{UVBCm0}), $\tilde \omega \ln \left( 16\Lambda^2/m_F^2\right) \simeq \pi$, for $m_0=0$.
  It follows:
  \beq
  \Lambda^2 \Sigma(\Lambda)=- \Lambda^4 \Sigma^\prime (\Lambda) 
  \simeq \frac{2\xi}{\pi \tilde  \omega} \Lambda\, m_F^2 \sin 
  \left( 
  \tilde \omega \ln (16\Lambda^2/m_F^2) -2\tilde \omega \right)
   \simeq  \frac{4\xi}{\pi} \Lambda\,  m_F^2 ,
  \eeq
  which is also consistent with Eq.(\ref{asym}).
  Hence 
  the chiral condensate
  can be evaluated through the formula Eq.(\ref{condensate2}) with $m_0=0$ (or Eq.(\ref{condensate3})):  
  \cite{Miransky:1989qc} 
  \beq
  \langle \bar F F\rangle_0 =- \frac{N_C}{\pi^2} \frac{\alpha_{\rm cr}}{\alpha(\Lambda^2)} \Lambda^2 \Sigma(\Lambda)
  \simeq -\frac{4\xi  N_C }{\pi^3} m_F^2\, \Lambda 
\label{condensate}
\,,  
  \eeq
  without logarithmic factor, 
  which in fact implies 
  $Z_m^{-1}\propto \frac{\Lambda}{m_F}$, and hence 
    \beq
  \gamma_m = \Lambda \frac{\partial  }{\partial \Lambda} \ln Z_m^{-1} 
  = 1\, 
  \label{gamma1}
  \eeq 
  consistently with Eq.(\ref{anomdim}) obtained in comparison with the OPE. 
  
 In the case of $m_0\ne 0$,  the result $Z_m\sim \frac{m_F}{\Lambda}$  
  was also noted  in the SSB phase \cite{Miransky:1984ef} before the advent of WTC, through the UV boundary condition. 
 More specifically,  using the expression for $m_0\ne 0$ in Eq.(\ref{UVBCm0}) with
  $\Sigma (x=m_P^2)=m_P\simeq m_F +m_R$ for $m_R \ll m_F$, we have \cite{Miransky:1989qc}    
  \begin{eqnarray}  m_0
  &\equiv& \varphi (m_P^2) =(x \Sigma(x))^\prime|_{x=\Lambda^2}=\frac{\xi}{\pi \tilde \omega}\frac{m_P^2}{\Lambda} \, 
  \sin \left(\tilde \omega \ln \frac{16\, \Lambda^2}{m_P^2}\right)
  \simeq  \varphi(m_F^2) +    m_R\, \frac{\partial \varphi (m_P)}{\partial m_P}\Bigg|_{m_P=m_F} \\
 &\simeq& m_R\, \left(\frac{2\xi}{\pi} \frac{m_F}{\Lambda} \right) \,,
   \\ 
  Z_m  
   &\simeq& \frac{2\xi}{\pi} \frac{m_F}{\Lambda} \,,  
  \label{massZm}
  \eeq 
which agrees with the result obtained directly performing the integral 
Eq.(\ref{condensate1}) \cite{Miransky:1989qc}.
This  again yields the same result as Eq.(\ref{gamma1}), $\gamma_m=1$. 
 From Eq.(\ref{massZm}) we have the renormalized condensate at $\mu=m_F$:
 \beq
 \langle \bar F F \rangle_R =  \langle \bar F F \rangle_0 \,Z_m  \simeq   - \frac{8\xi^2  N_C }{\pi^4} m_F^3\,,
  \eeq
so that the multiplicative renormalization follows, i.e.,  $m_0 \langle \bar F F \rangle_0 =m_R  \langle \bar F F \rangle_R$.

  This large anomalous dimension provides an enhancement of the quark/lepton mass through the ETC having the scale $\Lambda=\Lambda_{\rm ETC} (\sim \Lambda_{\rm TC})$:
 \begin{eqnarray}
 m_{q/l} &\sim& - c_i \frac{\langle\bar  F F \rangle_0}{N_C \Lambda_{\rm ETC}^2} 
\simeq   
\left( c_i
\frac{4\xi}{\pi^3} \frac{m_F^2}
{\Lambda_{\rm ETC}}
\right) 
= 
\left[ \frac{c_i}{N_F} \left(\frac{4}{\xi\pi} \frac{4 v_{\rm EW}}{N_C \Lambda_{\rm ETC} }\right)\right] v_{\rm EW} \equiv y^{\rm eff} \cdot v_{\rm EW},\\
y^{\rm eff} &\sim &\frac{ c_i }{N_F} \left(\frac{4}{\xi\pi}\frac{4 v_{\rm EW}}{N_C \Lambda_{\rm ETC} }\right)\,, 
  \end{eqnarray}
 where $c_i$ is a model-dependent numerical constant of ${\cal O} (1)$ and the Pagels-Stokar formula $m_F^2 \simeq  v_{\rm EW}^2 \frac{4\pi^2}{N_F N_C} $ in Eq.(\ref{PSv}) was used.
 This is roughly $0.1$ GeV for the typical quark/lepton mass (except for the top quark) with the effective Yukawa coupling $y^{\rm eff} \lesssim 
 10^{-3}$,  in accord with the FCNC constraints $\Lambda_{\rm ETC} \gtrsim 10^3$ TeV $\simeq 4 v_{\rm EW}$. 
  Then the WTC with this ladder solution provides a concrete dynamics for a solution of  the FCNC problem by the large anomalous dimension, the solution 
  \cite{Holdom:1981rm} based on a pure assumption of existence of a theory having UV fixed point (without concrete dynamics nor a concrete value of the anomalous dimension).

 The asymptotic form of $\Sigma(x)$ ($m_0\ne 0$) takes the same form as Eq.(\ref{Sigma:sol}), with replacement of $m_F$ to $m_P$.
 \begin{eqnarray}
\Sigma(x) &\sim& \frac{2\xi }{\pi}\frac{1}{ \tilde \omega} \frac{m_P^2} {\sqrt{x}} \sin \left( 
\tilde \omega \ln \left(\frac{16x}{m_P^2} \right) -2\tilde \omega
\right) \simeq \frac{2\xi }{\pi}\frac{1}{ \tilde \omega} \frac{m_P^2} {\sqrt{x}} \sin \left( 
\tilde \omega \ln \left(\frac{16x}{m_F^2} \right) -2\tilde \omega \ln (1-\frac{m_R}{m_F}) -2\tilde \omega
\right)\\
 &\simeq&  \frac{4\xi }{\pi}  \frac{(m_F+m_R)^2} {\sqrt{x}}  \left(1-\frac{m_R}{m_F}\right) \simeq  \frac{4\xi }{\pi} \left[ \frac{m_F\, m_R} {\sqrt{x}} 
 +  \frac{m_F^2} {\sqrt{x}}  \right]
 \end{eqnarray} 
 This is perfectly consistent with the OPE with $\gamma_m =1$: 
 \beq
  \Sigma(x) \sim \frac{4\xi}{\pi}  m_R \left(\frac{x}{m_F^2}\right)^{- \gamma_m/2} 
  - \frac{\pi^3}{2\xi N_C} \frac{\langle \bar F F
  \rangle_R} {x} \left(\frac{x}{m_F^2}\right)^{\gamma_m/2 }
  \sim  \frac{4\xi}{\pi}  \left[\frac{m_R m_F}{\sqrt{x}}  + \frac{m_F^2}{\sqrt{x}} \right]\,.
  \eeq 
   Such a large anomalous dimension is due to the
UV fixed point at $\alpha_{\rm cr}$ whose effective coupling $C_2 \alpha_{\rm cr}=\pi/3$  remains strong all the way up to
the scale $\Lambda_{\rm TC}$.

  \section{Pagels-Stokar Formula}
\label{PS:sec}

The Pagels-Stokar formula \cite{Pagels:1979hd} is given as 
\begin{equation} 
  \frac{4 \pi^2}{N_C} \frac{F_\pi^2}{m_F^2} 
  = \int_0^{\Lambda 
^2 \to \infty} 
dx x \frac{\Sigma(x)^2 - \frac{x}{4} \frac{d  (\Sigma(x)^2)}{d x} }{(x +  \Sigma( x)^2)^2} 
\,. \label{PS}
\end{equation}
The integral is dominated by the infrared region and converges quickly for the ladder
SSB solution with $\gamma_m =1$,
$\Sigma (x) \sim m_F^2/\sqrt{x}$, and hence is insensitive to the value of $\Lambda^2$ as far as it is very large, say $\Lambda^2/m_F^2 >10^6$. Rather, the integral depends on the precise form in the infrared region $x<\Lambda^2$. Here we take 
 the mass function 
of  the ladder SD solution, Eq.(\ref{Sigma:sol}): 
 \begin{equation} 
 \Sigma(x) \simeq \xi \frac{m_F^2}{\sqrt{x}} \sqrt{\frac{{\rm cth}\pi \tilde \omega}{\pi \tilde \omega ({\tilde \omega}^2 + 1/4)}} 
\, \sin \left( \tilde \omega  \ln \left( \frac{16 x}{m_F^2} \right)  - 2 \tilde \omega  \right) \,.
\end{equation}
Then we get 
\begin{equation} 
  \frac{4 \pi^2}{N_C} \frac{F_\pi^2}{m_F^2} 
  \simeq 2.00\, \xi^2
\,.
\label{PS:val}
\end{equation}
From this we get 
\beq
v_{\rm EW}^2= N_D F_\pi^2 =\frac{N_F}{2} \frac{\xi^2}{2\pi^2 } N_C m_F^2\simeq \frac{\xi^2}{4\pi^2} N_F N_C m_F^2\,.
\label{PSv}
\eeq
The result  is compared with that obtained by using a naive mass function, 
$\Sigma(x)=m_F^2 x^{-1/2}$ for $x>m_F^2$ and $\Sigma(x)=m_F$ for $x<m_F^2$: 
$  
\frac{4 \pi^2}{N_{C}} \frac{F_\pi^2}{m^2} 
  \simeq 1.00 
$.

 \section{Estimate of ggF contamination for technidilaton production with forward dijet at LHC}
\label{dijet}

The $h + 2j$ production at the LHC arises dominantly from two processes, i.e., VBF and ggF:   
\begin{equation} 
 \sigma(h+2j) = \sigma_{\rm VBF}(h + qq) + \sigma_{\rm ggF}(h + gg) 
 \,.  \label{dijet-cross}
\end{equation}
In Ref.~\cite{DelDuca:2001fn} the $h + gg$ cross section has been estimated, at $\sqrt{s}=14$ TeV 
with a kinematical cut set, as a function of the Higgs mass $m_h$. 
 At $m_h=125$ GeV it reads 
 \begin{equation} 
  \sigma_{\rm ggF}^{\rm cut}(h+gg) \simeq 10 \, {\rm pb}
\,, \label{hgg}
 \end{equation}
which is about 70\% - 80\% amount of the full phase space due to the kinematical cut. 
  Taking into account the phase space cut  
one can evaluate the full cross section as  
\begin{eqnarray}  
  \sigma_{\rm ggF}^{\rm full}(h +gg) 
&\sim& 
\left( \frac{10}{8} - \frac{10}{7} \right) \times 
\sigma_{\rm ggF}^{\rm cut}(h+gg) 
\nonumber \\ 
&\sim& 
13 - 14 \, {\rm pb}
\,. \label{ggFdijet:SM}
\end{eqnarray} 
On the other hand, the $h + 0j$ ggF production cross section at $\sqrt{s}=14$ TeV 
 can be read off from Ref.~\cite{LHCWiki} as 
\begin{equation} 
 \sigma_{\rm ggF}^{\rm full}(h+0j) \simeq 49 \, {\rm pb}  
 \,, \label{ggF0jet:SM}
\end{equation} 
 which corresponds to the number obtained by integrating the full phase space.  
 Eqs.(\ref{ggFdijet:SM}) and (\ref{ggF0jet:SM}) allow us to numerically write the ratio 
\begin{equation} 
r_{\rm ggF+jj} \equiv 
\frac{\sigma_{\rm ggF}^{\rm full}(h+gg)}{\sigma_{\rm ggF}^{\rm full}(h+0j)} \simeq 0.3 
 \,.\label{scaling:1}
\end{equation}
Note that the dependence of the gluon-gluon-Higgs coupling is canceled in this ratio, 
so the value of the ratio can be applied to any Higgs candidate including the technidilaton.

We shall assume that 
the 8 TeV cross sections are also applicable to Eq.(\ref{scaling:1}). 
Then the signal strength of the technidilaton decaying to $X_1X_2$ through the dijet production channel   
can be evaluated as 
\begin{eqnarray} 
\mu_{2j}^{X_1X_2} 
&=& 
\frac{\sigma(\phi+2j)\times {\rm BR}(\phi \to X_1X_2)}{\sigma(h+2j) \times {\rm BR}(h \to X_1X_2)} 
\nonumber \\ 
&\sim& 
\frac{[\sigma_{\rm VBF}(\phi + qq) + r_{\rm ggF+jj} \times \sigma_{\rm ggF}(\phi + 0j) ] \times {\rm BR}(\phi \to X_1X_2)}{[\sigma_{\rm VBF}(h + qq) + r_{\rm ggF+jj} \times \sigma_{\rm ggF}(h + 0j) ] \times {\rm BR}(h \to X_1X_2)} 
\nonumber \\ 
&=& 
R_{\rm VBF} \cdot \frac{1 + r_{\rm contam}(\phi)}{1 + r_{\rm contam}(h)} \cdot r_{\rm BR}^{X_1X_2} 
\,, 
\end{eqnarray}
where 
\begin{eqnarray}
 R_{\rm VBF} &=& \frac{\sigma_{\rm VBF}(\phi + qq)}{\sigma_{\rm VBF}(h + qq)} = \left( \frac{v_{\rm EW}}{F_\phi} \right)^2 
  \,, \nonumber \\ 
 r_{\rm contam}(\phi/h) &=& \frac{r_{\rm ggF+jj} \times \sigma_{\rm ggF}(\phi/h + 0j)}{\sigma_{\rm VBF}(\phi/h + qq)} 
 \,, \nonumber \\ 
 r_{\rm BR}^{X_1X_2} &=& \frac{{\rm BR}(\phi \to X_1X_2)}{{\rm BR}(h\to X_1X_2)} 
 \,. 
\end{eqnarray}
 Note that at the leading oder of perturbative computations 
the ratios $r_{\rm contam} (\phi)$ and $r_{\rm BR}^{X_1X_2}$ 
depend only on $N_C$ once the Higgs mass is fixed to be 125 GeV. 
At the 8 TeV LHC, for $N_C=4$ we have 
\begin{eqnarray} 
 r_{\rm contam} (\phi) &\simeq& 106 
\,, \nonumber \\ 
 r_{\rm BR}^{WW} &=& r_{\rm BR}^{ZZ} = r_{\rm BR}^{\tau\tau} = \simeq 0.26
 \,, \qquad 
 r_{\rm BR}^{\gamma\gamma} \simeq 0.37 
\end{eqnarray}
and for the SM Higgs, 
\begin{equation} 
r_{\rm contam} (h) \simeq 1.4
\,.  
\end{equation}  
Thus we estimate the signal strengths, 
\begin{eqnarray} 
 \mu_{VBF}^{WW/ZZ/\tau\tau} 
&\simeq& 0.6 \left(\frac{v_{\rm EW}/F_\phi}{ 0.23} \right)^2 
\,. \nonumber \\ 
 \mu_{VBF}^{\gamma\gamma} 
&\simeq& 0.8 
\left(\frac{v_{\rm EW}/F_\phi}{ 0.23} \right)^2 
\,. 
\end{eqnarray}


\end{document}